\documentstyle[10pt,epsf,epsfig,dp_delphititle]{dp_delphi}
\makeindex
\def\DpPaperGroup{EP}
\def\DpPaperRef{2000-087}
\def\DpDate{28 March 2000}
\def\DpAuthors{DELPHI Collaboration}
\def\DpSubmit{(Eur. Phys. J. C18(2000)229)}
\def\DpTitle{{
Study of ${\mathrm B^0_s}$-$\overline{\mathrm B^0_s}$ oscillations \\
 and ${\mathrm B^0_s}$ lifetimes
using \\ hadronic  decays of ${\mathrm B^0_s}$ mesons \\
}}
\def\DpComment{  }
\def\DpEMail{ }

\newcommand{\BC}{\begin{center}}
\newcommand{\EC}{\end{center}}
\newcommand{\BE}{\begin{equation}}
\newcommand{\EE}{\end{equation}}
\newcommand{\BEA}{\begin{eqnarray}}
\newcommand{\EEA}{\end{eqnarray}}
\newcommand{\BA}{\begin{array}}
\newcommand{\EA}{\end{array}}
\newcommand{\BI}{\begin{itemize}}
\newcommand{\EI}{\end{itemize}}
\newcommand{\BF}{\begin{figure}}
\newcommand{\EF}{\end{figure}}
\newcommand{\BT}{\begin{table}}
\newcommand{\ET}{\end{table}}
\newcommand{\BTB}{\begin{tabular}}
\newcommand{\ETB}{\end{tabular}}
\newcommand\BM{\begin{minipage}}
\newcommand\EM{\end{minipage}}

\newcommand{\Erg}[3]{\ifmmode{#1\pm#2_{stat.}\pm#3_{sys.}}\else{$#1\pm#2_{stat.}\pm#3_{sys.}$}\fi}
\newcommand{\erg}[3]{\ifmmode{\scriptstyle#1\,\pm\,#2\,\pm\,#3}\else{$\scriptstyle#1\,\pm\,#2\,\pm\,#3$}\fi}
\begin{document}
\makeatletter
\newcount\@tempcntc
\def\@citex[#1]#2{\if@filesw\immediate\write\@auxout{\string\citation{#2}}\fi
  \@tempcnta\z@\@tempcntb\m@ne\def\@citea{}\@cite{\@for\@citeb:=#2\do
    {\@ifundefined
       {b@\@citeb}{\@citeo\@tempcntb\m@ne\@citea\def\@citea{,}{\bf ?}\@warning
       {Citation `\@citeb' on page \thepage \space undefined}}%
    {\setbox\z@\hbox{\global\@tempcntc0\csname b@\@citeb\endcsname\relax}%
     \ifnum\@tempcntc=\z@ \@citeo\@tempcntb\m@ne
       \@citea\def\@citea{,}\hbox{\csname b@\@citeb\endcsname}%
     \else
      \advance\@tempcntb\@ne
      \ifnum\@tempcntb=\@tempcntc
      \else\advance\@tempcntb\m@ne\@citeo
      \@tempcnta\@tempcntc\@tempcntb\@tempcntc\fi\fi}}\@citeo}{#1}}
\def\@citeo{\ifnum\@tempcnta>\@tempcntb\else\@citea\def\@citea{,}%
  \ifnum\@tempcnta=\@tempcntb\the\@tempcnta\else
   {\advance\@tempcnta\@ne\ifnum\@tempcnta=\@tempcntb \else \def\@citea{--}\fi
    \advance\@tempcnta\m@ne\the\@tempcnta\@citea\the\@tempcntb}\fi\fi}
 
\makeatother
\begin{titlepage}
\pagenumbering{roman}
\CERNpreprint{\DpPaperGroup}{\DpPaperRef} 
\date{{\small\DpDate}} 
\title{\DpTitle} 
\address{\DpAuthors} 
\begin{shortabs} 
\noindent
%
\newcommand{\et}{\end{tabular}}
\newcommand{\Bs}{\mathrm{B^0_s}}
\newcommand{\dgs}{\Delta \Gamma_{\Bs}}
\newcommand{\dgbs}{\Delta \Gamma_{\Bs}/\Gamma_{\Bs}}
\newcommand{\tbs}{\tau_{\Bs}}
\newcommand{\bt}{\begin{tabular}}
\newcommand{\dms}{\Delta m_{\Bs}}
\newcommand{\Dsl}{{\mathrm{D_s}}\ell}
\newcommand{\Dsh}{{\rm{D}_{s}^{\pm}}\rm{h}^{\mp}}
\newcommand {\Gec}     {{\mathrm{GeV}/\it{c}}}
\newcommand {\bb}     {b \overline{b}}
\newcommand {\qq}     {q \overline{q}}
\newcommand {\cc}     {c \overline{c}}
\newcommand {\Bh}     {\overline{\mathrm{B}}_{h}}
\newcommand {\bfake}  {\overline{\mathrm{B}}_{\overline{h}}}
\newcommand {\Bns}    {\overline{\mathrm{B}}_{ns}}
\newcommand {\Gecqt}   {{$\mathrm{GeV}/{\it{c}}^2$}}
\newcommand {\Gecqm}   {{\mathrm{GeV}/{\it{c}}^2}}
\newcommand {\Me}      {{\mathrm{MeV}}}
\newcommand {\Mec}     {{\mathrm{MeV}/\it{c}}}
\newcommand {\Mecqt}   {{$\mathrm{MeV}/{\it{c}}^2$}}
\newcommand {\Mecqm}   {{\mathrm{MeV}/{\it{c}}^2}}
\newcommand{\gammas}{\mathrm{\Gamma_{\Bs}}}
\newcommand{\gammad}{\mathrm{\Gamma_{\Bd}}}
\newcommand{\gammasl}{\mathrm{\Gamma_{\Bs}^{L}}}
\newcommand{\gammash}{\mathrm{\Gamma_{\Bs}^{H}}}
\newcommand{\mbs}{m_{\Bs}}
\newcommand{\Bob}{\overline{\mathrm{B^0}}}
\newcommand{\dmd}{\Delta m_{\Bd}}
\newcommand{\dmq}{\Delta m_{\rm B^0_q}}
\newcommand{\BsDsm}{{\mathrm{B_s}^{0}} \rightarrow \mathrm{D_s} \mu X}
\newcommand{\BsDsX}{{\mathrm{B_s}^{0}} \rightarrow \mathrm{D_s} X}
\newcommand{\BDsX}{\mathrm{B} \rightarrow \mathrm{D_s} X}
\newcommand{\BDomX}{\mathrm{B} \rightarrow \mathrm{D}^0 \mu X}
\newcommand{\BDpmX}{\mathrm{B} \rightarrow \mathrm{D}^+ \mu X}
\newcommand{\Dsfmn}{\mathrm{D_s} \rightarrow \phi \mu \nu}
\newcommand{\Dsfipi}{\mathrm{D_s} \rightarrow \phi \pi}
\newcommand{\DsfX}{\mathrm{D_s} \rightarrow \phi X}
\newcommand{\DpfX}{\mathrm{D}^+ \rightarrow \phi X}
\newcommand{\DofX}{\mathrm{D}^0 \rightarrow \phi X}
\newcommand{\DfX}{\mathrm{D} \rightarrow \phi X}
\newcommand{\DsD}{\mathrm{B} \rightarrow \mathrm{D_s} \mathrm{D}}
\newcommand{\DsmX}{\mathrm{D_s} \rightarrow \mu X}
\newcommand{\DmX}{\mathrm{D} \rightarrow \mu X}
\newcommand{\Zbb}{\mathrm{Z} \rightarrow b \overline{b}}
\newcommand{\Zcc}{\mathrm{Z} \rightarrow c \overline{c}}
\newcommand{\Rbb}{\frac{\Gamma_{\mathrm{Z} \rightarrow b \overline{b}}}
{\Gamma_{Z \rightarrow \mathrm{hadrons}}}}
\newcommand{\Rcc}{\frac{\Gamma_{\mathrm{Z} \rightarrow c \overline{c}}}
{\Gamma_{Z \rightarrow \mathrm{hadrons}}}}
\newcommand{\str}{s \overline{s}}
\newcommand{\Bsb}{\overline{\mathrm{B^0_{\rm{s}}}}}
\newcommand{\Bp}{\mathrm{B^{+}}}
\newcommand{\Bm}{\mathrm{B^{-}}}
\newcommand{\Bo}{\mathrm{B^{0}}}
\newcommand{\Bd}{\mathrm{B^{0}_{d}}}
\newcommand{\tbd}{\tau_{\Bd}}
\newcommand{\B}{\mathrm{B^{\pm}_{u}}}
\newcommand{\Bdb}{\overline{\mathrm{B^{0}_{d}}}}
\newcommand{\Lb}{\Lambda^0_{\mathrm{b}}}
\newcommand{\Lbb}{\overline{\Lambda^0_{\mathrm{b}}}}
\newcommand{\Kstar}{\mathrm{K^{*0}}}
\newcommand{\Kstarr}{\mathrm{K^{*\pm}}}
\newcommand{\Kstarm}{\mathrm{K^{*-}}}
\newcommand{\Kstarmn}{\mathrm{K^{*-}}}
\newcommand{\phim}{\mathrm{\phi}}
\newcommand{\Ds}{\mathrm{D_s}}
\newcommand{\Dsp}{\mathrm{D_s}^+}
\newcommand{\Dsn}{\rm{D}_{s}^{-}}
\newcommand{\Dp}{\mathrm{D}^+}
\newcommand{\Dn}{\mathrm{D}^0}
\newcommand{\Dsb}{\overline{\mathrm{D_s}}}
\newcommand{\Dm}{\mathrm{D}^-}
\newcommand{\Dnb}{\overline{\mathrm{D}^0}}
\newcommand{\Lc}{\Lambda_{\mathrm{c}}}
\newcommand{\Lcb}{\overline{\Lambda_{\mathrm{c}}}}
\newcommand{\Dstarp}{\mathrm{D}^{*+}}
\newcommand{\Dstarm}{\mathrm{D}^{*-}}
\newcommand{\Dsstarp}{\mathrm{D_s}^{*+}}
\newcommand{\Pb}{P_{\mathrm{b-baryon}}}
\newcommand{\KKpi}{\mathrm{ K K \pi }}
\newcommand{\GeV}{\mathrm{GeV}}
\newcommand{\MeV}{\mathrm{MeV}}
\newcommand{\nb}{\mathrm{nb}}
\newcommand{\Zzero}{\mathrm{Z}}
\newcommand{\MZ}{M\mathrm{_Z}}
\newcommand{\MW}{M\mathrm{_W}}
\newcommand{\GF}{G_F}
\newcommand{\Gm}{G_{\mu}}
\newcommand{\MH}{M\mathrm{_H}}
\newcommand{\MT}{m_{\mathrm{top}}}
\newcommand{\GZ}{\Gamma_{\mathrm{Z}}}
\newcommand{\Afb}{A_{\mathrm{FB}}}
\newcommand{\Afbs}{A_{\mathrm{FB^s}}}
\newcommand{\sigmaf}{\sigma_{\mathrm{F}}}
\newcommand{\sigmab}{\sigma_{\mathrm{B}}}
\newcommand{\NF}{N_{F}}
\newcommand{\NB}{N_{B}}
\newcommand{\Nnu}{N_{\nu}}
\newcommand{\RZ}{R\mathrm{_Z}}
\newcommand{\rhob}{\rho_{eff}}
\newcommand{\Gammanz}{\mathrm{\Gamma_{Z}^{new}}}
\newcommand{\Gammani}{\mathrm{\Gamma_{inv}^{new}}}
\newcommand{\Gammasz}{\mathrm{\Gamma_{Z}^{SM}}}
\newcommand{\Gammasi}{\mathrm{\Gamma_{inv}^{SM}}}
\newcommand{\Gammaxz}{\mathrm{\Gamma_{Z}^{exp}}}
\newcommand{\Gammaxi}{\mathrm{\Gamma_{inv}^{exp}}}
\newcommand{\rhoZ}{\rho_{\mathrm{Z}}}
\newcommand{\thw}{\theta_{\mathrm{W}}}
\newcommand{\swsq}{\sin^2\!\thw}
\newcommand{\swsqmsb}{\sin^2\!\theta_{\mathrm{W}}^{\overline{\mathrm{MS}}}}
\newcommand{\swsqbar}{\sin^2\!\overline{\theta}_{\mathrm{W}}}
\newcommand{\cwsqbar}{\cos^2\!\overline{\theta}_{\mathrm{W}}}
\newcommand{\swsqb}{\sin^2\!\theta^{eff}_{\mathrm{W}}}
\newcommand{\ee}{{e^+e^-}}
\newcommand{\eeX}{{e^+e^-X}}
\newcommand{\gaga}{{\gamma\gamma}}
\newcommand{\mumu}{\ifmmode {\mu^+\mu^-} \else ${\mu^+\mu^-} $ \fi}
\newcommand{\eeg}{{e^+e^-\gamma}}
\newcommand{\mumug}{{\mu^+\mu^-\gamma}}
\newcommand{\tautau}{{\tau^+\tau^-}}
\newcommand{\qqb}{q\bar{q}}
\newcommand{\eegg}{e^+e^-\rightarrow \gamma\gamma}
\newcommand{\eeggg}{e^+e^-\rightarrow \gamma\gamma\gamma}
\newcommand{\eeee}{e^+e^-\rightarrow e^+e^-}
\newcommand{\eeeeee}{e^+e^-\rightarrow e^+e^-e^+e^-}
\newcommand{\eeeeg}{e^+e^-\rightarrow e^+e^-(\gamma)}
\newcommand{\eeeegg}{e^+e^-\rightarrow e^+e^-\gamma\gamma}
\newcommand{\eeeg}{e^+e^-\rightarrow (e^+)e^-\gamma}
\newcommand{\eemumu}{e^+e^-\rightarrow \mu^+\mu^-}
\newcommand{\eetautau}{e^+e^-\rightarrow \tau^+\tau^-}
\newcommand{\eehad}{e^+e^-\rightarrow \mathrm{hadrons}}
\newcommand{\eettg}{e^+e^-\rightarrow \tau^+\tau^-\gamma}
\newcommand{\eell}{e^+e^-\rightarrow l^+l^-}
\newcommand{\Ztopig}{{\mathrm{Z}}^0\rightarrow \pi^0\gamma}
\newcommand{\Ztogg}{{\mathrm{Z}}^0\rightarrow \gamma\gamma}
\newcommand{\Ztoee}{{\mathrm{Z}}^0\rightarrow e^+e^-}
\newcommand{\Ztoggg}{{\mathrm{Z}}^0\rightarrow \gamma\gamma\gamma}
\newcommand{\Ztomumu}{{\mathrm{Z}}^0\rightarrow \mu^+\mu^-}
\newcommand{\Ztotautau}{{\mathrm{Z}}^0\rightarrow \tau^+\tau^-}
\newcommand{\Ztoll}{{\mathrm{Z}}^0\rightarrow l^+l^-}
\newcommand{\Ztocc}{{\mathrm{Z}\rightarrow c \bar c}}
\newcommand{\Lamp}{\Lambda_{+}}
\newcommand{\Lamm}{\Lambda_{-}}
\newcommand{\Pt}{P_{t}}
\newcommand{\Gee}{\Gamma_{ee}}
\newcommand{\Gpig}{\Gamma_{\pi^0\gamma}}
\newcommand{\Ggg}{\Gamma_{\gamma\gamma}}
\newcommand{\Gggg}{\Gamma_{\gamma\gamma\gamma}}
\newcommand{\Gmumu}{\Gamma_{\mu\mu}}
\newcommand{\Gtautau}{\Gamma_{\tau\tau}}
\newcommand{\Ginv}{\Gamma_{\mathrm{inv}}}
\newcommand{\Ghad}{\Gamma_{\mathrm{had}}}
\newcommand{\Gnu}{\Gamma_{\nu}}
\newcommand{\GnuSM}{\Gamma_{\nu}^{\mathrm{SM}}}
\newcommand{\Gll}{\Gamma_{l^+l^-}}
\newcommand{\Gff}{\Gamma_{f\overline{f}}}
\newcommand{\Gtot}{\Gamma_{\mathrm{tot}}}
\newcommand{\Rb}{R\mathrm{_b}}
\newcommand{\Rc}{R\mathrm{_c}}
\newcommand{\al}{a_l}
\newcommand{\vl}{v_l}
\newcommand{\af}{a_f}
\newcommand{\vf}{v_f}
\newcommand{\ael}{a_e}
\newcommand{\ve}{v_e}
\newcommand{\amu}{a_\mu}
\newcommand{\vmu}{v_\mu}
\newcommand{\atau}{a_\tau}
\newcommand{\vtau}{v_\tau}
\newcommand{\ahatl}{\hat{a}_l}
\newcommand{\vhatl}{\hat{v}_l}
\newcommand{\ahate}{\hat{a}_e}
\newcommand{\vhate}{\hat{v}_e}
\newcommand{\ahatmu}{\hat{a}_\mu}
\newcommand{\vhatmu}{\hat{v}_\mu}
\newcommand{\ahattau}{\hat{a}_\tau}
\newcommand{\vhattau}{\hat{v}_\tau}
\newcommand{\vtildel}{\tilde{\mathrm{v}}_l}
\newcommand{\avsq}{\ahatl^2\vhatl^2}
\newcommand{\Ahatl}{\hat{A}_l}
\newcommand{\Vhatl}{\hat{V}_l}
\newcommand{\Afer}{A_f}
\newcommand{\Ael}{A_e}
\newcommand{\Aferb}{\bar{A_f}}
\newcommand{\Aelb}{\bar{A_e}}
\newcommand{\AVsq}{\Ahatl^2\Vhatl^2}
\newcommand{\Iwk}{I_{3l}}
\newcommand{\Qch}{|Q_{l}|}
\newcommand{\roots}{\sqrt{s}}
\newcommand{\pT}{p_{\mathrm{T}}}
\newcommand{\mt}{m\mathrm{_t}}
\newcommand{\Rechi}{{\mathrm{Re}} \left\{ \chi (s) \right\}}
\newcommand{\up}{^}
\newcommand{\abscosthe}{|cos\theta|}
\newcommand{\dsum}{\Sigma |d_\circ|}
\newcommand{\zsum}{\Sigma z_\circ}
\newcommand{\sint}{\mbox{$\sin\theta$}}
\newcommand{\cost}{\mbox{$\cos\theta$}}
\newcommand{\mcost}{|\cos\theta|}
\newcommand{\epair}{\mbox{$e^{+}e^{-}$}}
\newcommand{\mupair}{\mbox{$\mu^{+}\mu^{-}$}}
\newcommand{\taupair}{\mbox{$\tau^{+}\tau^{-}$}}
\newcommand{\gamgam}{\mbox{$e^{+}e^{-}\rightarrow e^{+}e^{-}\mu^{+}\mu^{-}$}}
\newcommand{\fullskip}{\vskip 16cm}
\newcommand{\halfskip}{\vskip  8cm}
\newcommand{\quarskip}{\vskip  6cm}
\newcommand{\abitskip}{\vskip 0.5cm}
\newcommand{\ba}{\begin{array}}
\newcommand{\ea}{\end{array}}
\newcommand{\bc}{\begin{center}}
\newcommand{\ec}{\end{center}}
\newcommand{\be}{\begin{eqnarray}}
\newcommand{\eeq}{\end{eqnarray}}
\newcommand{\bes}{\begin{eqnarray*}}
\newcommand{\ees}{\end{eqnarray*}}
\newcommand{\Kz}{\ifmmode {\mathrm{K^0_s}} \else ${\mathrm{K^0_s}} $ \fi}
\newcommand{\Zz} {\mathrm{Z}}
\newcommand{\qqbar}{\ifmmode {q\bar{q}} \else ${q\bar{q}} $ \fi}
\newcommand{\ccbar}{\ifmmode {c\bar{c}} \else ${c\bar{c}} $ \fi}
\newcommand{\bbbar}{\ifmmode {b\bar{b}} \else ${b\bar{b}} $ \fi}
\newcommand{\xxbar}{\ifmmode {x\bar{x}} \else ${x\bar{x}} $ \fi}
\newcommand{\rphi}{\ifmmode {\R\phi} \else ${R\phi} $ \fi}

\noindent
Oscillations of $\mbox{B}^0_{\mathrm{s}}$ mesons have been studied in samples  
selected from about 3.5 million hadronic $\Zz$ decays detected by DELPHI between 1992 and 1995.
One analysis uses events in the exclusive decay channels: 
$\Bs \rightarrow \rm{D}_{s}^{-} \pi^{+}$ or $\rm{D}_{s}^{-}\mathrm{a_1^+}$ 
and $\Bs \rightarrow \overline{\mathrm{D}}^0\mathrm{K}^- \pi^+ $ or $\overline{\mathrm{D}}^0\mathrm{K}^-\mathrm{a_1^+}$, 
where the ${\mathrm{D}}$ decays are completely reconstructed. 
In addition, $\mbox{B}^0_{\mathrm{s}}-\overline{\mbox{B}^0_{\mathrm{s}}}$ oscillations have been studied in events with an
exclusively reconstructed  $\Ds$ accompanied in the same hemisphere by a high momentum hadron of opposite charge. 
Combining the two analyses, a limit on the mass difference between the physical $\mbox{B}^0_{\mathrm{s}}$ states has been obtained:
\bes
   \ba{c}
 \dms > 4.0~\mbox{ps}^{-1}~\mbox{at the 95\% C.L.} \\
\mbox{$\mathrm{with~ a~ sensitivity~ of}$}~\dms = 3.2~\mbox{ps}^{-1}.
   \ea
 \ees
Using the latter sample of events, the $\Bs$ lifetime has been measured and an upper limit on the 
decay width difference 
between the two physical $\Bs$ states  has been obtained:
$$
\tau_{\Bs}  = 1.53^{+0.16}_{-0.15}(\rm{stat}.)\pm{0.07}(\rm{syst}.)~\mathrm{ps}
$$
$$
\dgbs  < 0.69~\mbox{at the 95\% C.L.}
$$
The combination of these results with those obtained using  
$\mathrm{\rm{D}_{s}^{\pm}} \ell^{\mp}$ sample gives: 
\bes
   \ba{c}
 \dms > 4.9~\mbox{ps}^{-1}~\mbox{at the 95\% C.L.} \\
\mbox{$\mathrm{with~ a~ sensitivity~ of}$}~\dms = 8.7~\mbox{ps}^{-1}.
   \ea
 \ees
\begin{center}
$\tau_{\Bs}  = 1.46\pm{0.11}~\mathrm{ps}~$ 
and
 $~\dgbs  < 0.45~\mbox{at the 95\% C.L.}$
\end{center}
\end{shortabs}
\vfill
\begin{center}
\DpSubmit \ \\ 
\DpComment \ \\
\DpEMail \ \\
\end{center}
\vfill
\clearpage
\headsep 10.0pt
\addtolength{\textheight}{10mm}
\addtolength{\footskip}{-5mm}
\begingroup
%
\newcommand{\DpName}[2]{\hbox{#1$^{\ref{#2}}$},\hfill}
\newcommand{\DpNameTwo}[3]{\hbox{#1$^{\ref{#2},\ref{#3}}$},\hfill}
\newcommand{\DpNameThree}[4]{\hbox{#1$^{\ref{#2},\ref{#3},\ref{#4}}$},\hfill}
\newskip\Bigfill \Bigfill = 0pt plus 1000fill
\newcommand{\DpNameLast}[2]{\hbox{#1$^{\ref{#2}}$}\hspace{\Bigfill}}
%
\footnotesize
\noindent
\DpName{P.Abreu}{LIP}
\DpName{W.Adam}{VIENNA}
\DpName{T.Adye}{RAL}
\DpName{P.Adzic}{DEMOKRITOS}
\DpName{I.Ajinenko}{SERPUKHOV}
\DpName{Z.Albrecht}{KARLSRUHE}
\DpName{T.Alderweireld}{AIM}
\DpName{G.D.Alekseev}{JINR}
\DpName{R.Alemany}{VALENCIA}
\DpName{T.Allmendinger}{KARLSRUHE}
% ...
\normalsize
\endgroup
\titlefoot{Department of Physics and Astronomy, Iowa State
     University, Ames IA 50011-3160, USA
    \label{AMES}}
\titlefoot{Physics Department, Univ. Instelling Antwerpen,
     Universiteitsplein 1, BE-2610 Wilrijk, Belgium \\
     \indent~~and IIHE, ULB-VUB,
     Pleinlaan 2, BE-1050 Brussels, Belgium \\
     \indent~~and Facult\'e des Sciences,
     Univ. de l'Etat Mons, Av. Maistriau 19, BE-7000 Mons, Belgium
    \label{AIM}}
\titlefoot{Physics Laboratory, University of Athens, Solonos Str.
     104, GR-10680 Athens, Greece
    \label{ATHENS}}
\titlefoot{Department of Physics, University of Bergen,
     All\'egaten 55, NO-5007 Bergen, Norway
    \label{BERGEN}}
\titlefoot{Dipartimento di Fisica, Universit\`a di Bologna and INFN,
     Via Irnerio 46, IT-40126 Bologna, Italy
    \label{BOLOGNA}}
\titlefoot{Centro Brasileiro de Pesquisas F\'{\i}sicas, rua Xavier Sigaud 150,
     BR-22290 Rio de Janeiro, Brazil \\
     \indent~~and Depto. de F\'{\i}sica, Pont. Univ. Cat\'olica,
     C.P. 38071 BR-22453 Rio de Janeiro, Brazil \\
     \indent~~and Inst. de F\'{\i}sica, Univ. Estadual do Rio de Janeiro,
     rua S\~{a}o Francisco Xavier 524, Rio de Janeiro, Brazil
    \label{BRASIL}}
\titlefoot{Comenius University, Faculty of Mathematics and Physics,
     Mlynska Dolina, SK-84215 Bratislava, Slovakia
    \label{BRATISLAVA}}
\titlefoot{Coll\`ege de France, Lab. de Physique Corpusculaire, IN2P3-CNRS,
     FR-75231 Paris Cedex 05, France
    \label{CDF}}
\titlefoot{CERN, CH-1211 Geneva 23, Switzerland
    \label{CERN}}
\titlefoot{Institut de Recherches Subatomiques, IN2P3 - CNRS/ULP - BP20,
     FR-67037 Strasbourg Cedex, France
    \label{CRN}}
\titlefoot{DESY-Zeuthen, Platanenallee 6, D-15735 Zeuthen, Germany
    \label{DESY}}
\titlefoot{Institute of Nuclear Physics, N.C.S.R. Demokritos,
     P.O. Box 60228, GR-15310 Athens, Greece
    \label{DEMOKRITOS}}
\titlefoot{FZU, Inst. of Phys. of the C.A.S. High Energy Physics Division,
     Na Slovance 2, CZ-180 40, Praha 8, Czech Republic
    \label{FZU}}
\titlefoot{Dipartimento di Fisica, Universit\`a di Genova and INFN,
     Via Dodecaneso 33, IT-16146 Genova, Italy
    \label{GENOVA}}
\titlefoot{Institut des Sciences Nucl\'eaires, IN2P3-CNRS, Universit\'e
     de Grenoble 1, FR-38026 Grenoble Cedex, France
    \label{GRENOBLE}}
\titlefoot{Helsinki Institute of Physics, HIP,
     P.O. Box 9, FI-00014 Helsinki, Finland
    \label{HELSINKI}}
\titlefoot{Joint Institute for Nuclear Research, Dubna, Head Post
     Office, P.O. Box 79, RU-101 000 Moscow, Russian Federation
    \label{JINR}}
\titlefoot{Institut f\"ur Experimentelle Kernphysik,
     Universit\"at Karlsruhe, Postfach 6980, DE-76128 Karlsruhe,
     Germany
    \label{KARLSRUHE}}
\titlefoot{Institute of Nuclear Physics and University of Mining and Metalurgy,
     Ul. Kawiory 26a, PL-30055 Krakow, Poland
    \label{KRAKOW}}
\titlefoot{Universit\'e de Paris-Sud, Lab. de l'Acc\'el\'erateur
     Lin\'eaire, IN2P3-CNRS, B\^{a}t. 200, FR-91405 Orsay Cedex, France
    \label{LAL}}
\titlefoot{School of Physics and Chemistry, University of Lancaster,
     Lancaster LA1 4YB, UK
    \label{LANCASTER}}
\titlefoot{LIP, IST, FCUL - Av. Elias Garcia, 14-$1^{o}$,
     PT-1000 Lisboa Codex, Portugal
    \label{LIP}}
\titlefoot{Department of Physics, University of Liverpool, P.O.
     Box 147, Liverpool L69 3BX, UK
    \label{LIVERPOOL}}
\titlefoot{LPNHE, IN2P3-CNRS, Univ.~Paris VI et VII, Tour 33 (RdC),
     4 place Jussieu, FR-75252 Paris Cedex 05, France
    \label{LPNHE}}
\titlefoot{Department of Physics, University of Lund,
     S\"olvegatan 14, SE-223 63 Lund, Sweden
    \label{LUND}}
\titlefoot{Universit\'e Claude Bernard de Lyon, IPNL, IN2P3-CNRS,
     FR-69622 Villeurbanne Cedex, France
    \label{LYON}}
\titlefoot{Univ. d'Aix - Marseille II - CPP, IN2P3-CNRS,
     FR-13288 Marseille Cedex 09, France
    \label{MARSEILLE}}
\titlefoot{Dipartimento di Fisica, Universit\`a di Milano and INFN,
     Via Celoria 16, IT-20133 Milan, Italy
    \label{MILANO}}
\titlefoot{Niels Bohr Institute, Blegdamsvej 17,
     DK-2100 Copenhagen {\O}, Denmark
    \label{NBI}}
\titlefoot{NC, Nuclear Centre of MFF, Charles University, Areal MFF,
     V Holesovickach 2, CZ-180 00, Praha 8, Czech Republic
    \label{NC}}
\titlefoot{NIKHEF, Postbus 41882, NL-1009 DB
     Amsterdam, The Netherlands
    \label{NIKHEF}}
\titlefoot{National Technical University, Physics Department,
     Zografou Campus, GR-15773 Athens, Greece
    \label{NTU-ATHENS}}
\titlefoot{Physics Department, University of Oslo, Blindern,
     NO-1000 Oslo 3, Norway
    \label{OSLO}}
\titlefoot{Dpto. Fisica, Univ. Oviedo, Avda. Calvo Sotelo
     s/n, ES-33007 Oviedo, Spain
    \label{OVIEDO}}
\titlefoot{Department of Physics, University of Oxford,
     Keble Road, Oxford OX1 3RH, UK
    \label{OXFORD}}
\titlefoot{Dipartimento di Fisica, Universit\`a di Padova and
     INFN, Via Marzolo 8, IT-35131 Padua, Italy
    \label{PADOVA}}
\titlefoot{Rutherford Appleton Laboratory, Chilton, Didcot
     OX11 OQX, UK
    \label{RAL}}
\titlefoot{Dipartimento di Fisica, Universit\`a di Roma II and
     INFN, Tor Vergata, IT-00173 Rome, Italy
    \label{ROMA2}}
\titlefoot{Dipartimento di Fisica, Universit\`a di Roma III and
     INFN, Via della Vasca Navale 84, IT-00146 Rome, Italy
    \label{ROMA3}}
\titlefoot{DAPNIA/Service de Physique des Particules,
     CEA-Saclay, FR-91191 Gif-sur-Yvette Cedex, France
    \label{SACLAY}}
\titlefoot{Instituto de Fisica de Cantabria (CSIC-UC), Avda.
     los Castros s/n, ES-39006 Santander, Spain
    \label{SANTANDER}}
\titlefoot{Dipartimento di Fisica, Universit\`a degli Studi di Roma
     La Sapienza, Piazzale Aldo Moro 2, IT-00185 Rome, Italy
    \label{SAPIENZA}}
\titlefoot{Inst. for High Energy Physics, Serpukov
     P.O. Box 35, Protvino, (Moscow Region), Russian Federation
    \label{SERPUKHOV}}
\titlefoot{J. Stefan Institute, Jamova 39, SI-1000 Ljubljana, Slovenia
     and Laboratory for Astroparticle Physics,\\
     \indent~~Nova Gorica Polytechnic, Kostanjeviska 16a, SI-5000 Nova Gorica, Slovenia, \\
     \indent~~and Department of Physics, University of Ljubljana,
     SI-1000 Ljubljana, Slovenia
    \label{SLOVENIJA}}
\titlefoot{Fysikum, Stockholm University,
     Box 6730, SE-113 85 Stockholm, Sweden
    \label{STOCKHOLM}}
\titlefoot{Dipartimento di Fisica Sperimentale, Universit\`a di
     Torino and INFN, Via P. Giuria 1, IT-10125 Turin, Italy
    \label{TORINO}}
\titlefoot{Dipartimento di Fisica, Universit\`a di Trieste and
     INFN, Via A. Valerio 2, IT-34127 Trieste, Italy \\
     \indent~~and Istituto di Fisica, Universit\`a di Udine,
     IT-33100 Udine, Italy
    \label{TU}}
\titlefoot{Univ. Federal do Rio de Janeiro, C.P. 68528
     Cidade Univ., Ilha do Fund\~ao
     BR-21945-970 Rio de Janeiro, Brazil
    \label{UFRJ}}
\titlefoot{Department of Radiation Sciences, University of
     Uppsala, P.O. Box 535, SE-751 21 Uppsala, Sweden
    \label{UPPSALA}}
\titlefoot{IFIC, Valencia-CSIC, and D.F.A.M.N., U. de Valencia,
     Avda. Dr. Moliner 50, ES-46100 Burjassot (Valencia), Spain
    \label{VALENCIA}}
\titlefoot{Institut f\"ur Hochenergiephysik, \"Osterr. Akad.
     d. Wissensch., Nikolsdorfergasse 18, AT-1050 Vienna, Austria
    \label{VIENNA}}
\titlefoot{Inst. Nuclear Studies and University of Warsaw, Ul.
     Hoza 69, PL-00681 Warsaw, Poland
    \label{WARSZAWA}}
\titlefoot{Fachbereich Physik, University of Wuppertal, Postfach
     100 127, DE-42097 Wuppertal, Germany
    \label{WUPPERTAL}}
\titlefoot{Now at University of Florida
    \label{FLORIDA}}
\addtolength{\textheight}{-10mm}
\addtolength{\footskip}{5mm}
\clearpage
\headsep 30.0pt
\end{titlepage}
%
\pagenumbering{arabic} 
\setcounter{footnote}{0} %
\large
%
\newcommand{\et}{\end{tabular}}
\newcommand{\Bs}{\mathrm{B^0_s}}
\newcommand{\dgs}{\Delta \Gamma_{\Bs}}
\newcommand{\dgbs}{\Delta \Gamma_{\Bs}/\Gamma_{\Bs}}
\newcommand{\tbs}{\tau_{\Bs}}
\newcommand{\bt}{\begin{tabular}}
\newcommand{\dms}{\Delta m_{\Bs}}
\newcommand{\Dsl}{{\mathrm{D_s}}\ell}
\newcommand{\Dsh}{{\rm{D}_{s}^{\pm}}\rm{h}^{\mp}}
\newcommand {\Gec}     {{\mathrm{GeV}/\it{c}}}
\newcommand {\bb}     {{\mathrm b} \overline{{\mathrm b}}}
\newcommand {\qq}     {{\mathrm q} \overline{{\mathrm q}}}
\newcommand {\cc}     {{\mathrm c} \overline{{\mathrm c}}}
\newcommand {\Bh}     {\overline{\mathrm{B}}_{h}}
\newcommand {\bfake}  {\overline{\mathrm{B}}_{\overline{h}}}
\newcommand {\Bns}    {\overline{\mathrm{B}}_{ns}}
\newcommand {\Gecqt}   {{$\mathrm{GeV}/{\it{c}}^2$}}
\newcommand {\Gecqm}   {{\mathrm{GeV}/{\it{c}}^2}}
\newcommand {\Me}      {{\mathrm{MeV}}}
\newcommand {\Mec}     {{\mathrm{MeV}/\it{c}}}
\newcommand {\Mecqt}   {{$\mathrm{MeV}/{\it{c}}^2$}}
\newcommand {\Mecqm}   {{\mathrm{MeV}/{\it{c}}^2}}
\newcommand{\gammas}{\mathrm{\Gamma_{\Bs}}}
\newcommand{\gammad}{\mathrm{\Gamma_{\Bd}}}
\newcommand{\gammasl}{\mathrm{\Gamma_{\Bs}^{L}}}
\newcommand{\gammash}{\mathrm{\Gamma_{\Bs}^{H}}}
\newcommand{\mbs}{m_{\Bs}}
\newcommand{\Bob}{\overline{\mathrm{B^0}}}
\newcommand{\dmd}{\Delta m_{\Bd}}
\newcommand{\dmq}{\Delta m_{\rm B^0_q}}
\newcommand{\BsDsm}{{\mathrm{B_s}^{0}} \rightarrow \mathrm{D_s} \mu X}
\newcommand{\BsDsX}{{\mathrm{B_s}^{0}} \rightarrow \mathrm{D_s} X}
\newcommand{\BDsX}{\mathrm{B} \rightarrow \mathrm{D_s} X}
\newcommand{\BDomX}{\mathrm{B} \rightarrow \mathrm{D}^0 \mu X}
\newcommand{\BDpmX}{\mathrm{B} \rightarrow \mathrm{D}^+ \mu X}
\newcommand{\Dsfmn}{\mathrm{D_s} \rightarrow \phi \mu \nu}
\newcommand{\Dsfipi}{\mathrm{D_s} \rightarrow \phi \pi}
\newcommand{\DsfX}{\mathrm{D_s} \rightarrow \phi X}
\newcommand{\DpfX}{\mathrm{D}^+ \rightarrow \phi X}
\newcommand{\DofX}{\mathrm{D}^0 \rightarrow \phi X}
\newcommand{\DfX}{\mathrm{D} \rightarrow \phi X}
\newcommand{\DsD}{\mathrm{B} \rightarrow \mathrm{D_s} \mathrm{D}}
\newcommand{\DsmX}{\mathrm{D_s} \rightarrow \mu X}
\newcommand{\DmX}{\mathrm{D} \rightarrow \mu X}
\newcommand{\Zbb}{\mathrm{Z} \rightarrow b \overline{b}}
\newcommand{\Zcc}{\mathrm{Z} \rightarrow c \overline{c}}
\newcommand{\Rbb}{\frac{\Gamma_{\mathrm{Z} \rightarrow b \overline{b}}}
{\Gamma_{Z \rightarrow \mathrm{hadrons}}}}
\newcommand{\Rcc}{\frac{\Gamma_{\mathrm{Z} \rightarrow c \overline{c}}}
{\Gamma_{Z \rightarrow \mathrm{hadrons}}}}
\newcommand{\str}{s \overline{s}}
\newcommand{\Bsb}{\overline{\mathrm{B^0_{\rm{s}}}}}
\newcommand{\Bp}{\mathrm{B^{+}}}
\newcommand{\Bm}{\mathrm{B^{-}}}
\newcommand{\Bo}{\mathrm{B^{0}}}
\newcommand{\Bd}{\mathrm{B^{0}_{d}}}
\newcommand{\tbd}{\tau_{\Bd}}
\newcommand{\B}{\mathrm{B^{\pm}_{u}}}
\newcommand{\Bdb}{\overline{\mathrm{B^{0}_{d}}}}
\newcommand{\Lb}{\Lambda^0_{\mathrm{b}}}
\newcommand{\Lbb}{\overline{\Lambda^0_{\mathrm{b}}}}
\newcommand{\Kstar}{\mathrm{K^{*0}}}
\newcommand{\Kstarr}{\mathrm{K^{*\pm}}}
\newcommand{\Kstarm}{\mathrm{K^{*-}}}
\newcommand{\Kstarmn}{\mathrm{K^{*-}}}
\newcommand{\phim}{\mathrm{\phi}}
\newcommand{\Ds}{\mathrm{D_s}}
\newcommand{\Dsp}{\mathrm{D_s}^+}
\newcommand{\Dsn}{\rm{D}_{s}^{-}}
\newcommand{\Dp}{\mathrm{D}^+}
\newcommand{\Dn}{\mathrm{D}^0}
\newcommand{\Dsb}{\overline{\mathrm{D_s}}}
\newcommand{\Dm}{\mathrm{D}^-}
\newcommand{\Dnb}{\overline{\mathrm{D}^0}}
\newcommand{\Lc}{\Lambda_{\mathrm{c}}}
\newcommand{\Lcb}{\overline{\Lambda_{\mathrm{c}}}}
\newcommand{\Dstarp}{\mathrm{D}^{*+}}
\newcommand{\Dstarm}{\mathrm{D}^{*-}}
\newcommand{\Dsstarp}{\mathrm{D_s}^{*+}}
\newcommand{\Pb}{P_{\mathrm{b-baryon}}}
\newcommand{\KKpi}{\mathrm{ K K \pi }}
\newcommand{\GeV}{\mathrm{GeV}}
\newcommand{\MeV}{\mathrm{MeV}}
\newcommand{\nb}{\mathrm{nb}}
\newcommand{\Zzero}{\mathrm{Z}}
\newcommand{\MZ}{M\mathrm{_Z}}
\newcommand{\MW}{M\mathrm{_W}}
\newcommand{\GF}{G_F}
\newcommand{\Gm}{G_{\mu}}
\newcommand{\MH}{M\mathrm{_H}}
\newcommand{\MT}{m_{\mathrm{top}}}
\newcommand{\GZ}{\Gamma_{\mathrm{Z}}}
\newcommand{\Afb}{A_{\mathrm{FB}}}
\newcommand{\Afbs}{A_{\mathrm{FB^s}}}
\newcommand{\sigmaf}{\sigma_{\mathrm{F}}}
\newcommand{\sigmab}{\sigma_{\mathrm{B}}}
\newcommand{\NF}{N_{F}}
\newcommand{\NB}{N_{B}}
\newcommand{\Nnu}{N_{\nu}}
\newcommand{\RZ}{R\mathrm{_Z}}
\newcommand{\rhob}{\rho_{eff}}
\newcommand{\Gammanz}{\mathrm{\Gamma_{Z}^{new}}}
\newcommand{\Gammani}{\mathrm{\Gamma_{inv}^{new}}}
\newcommand{\Gammasz}{\mathrm{\Gamma_{Z}^{SM}}}
\newcommand{\Gammasi}{\mathrm{\Gamma_{inv}^{SM}}}
\newcommand{\Gammaxz}{\mathrm{\Gamma_{Z}^{exp}}}
\newcommand{\Gammaxi}{\mathrm{\Gamma_{inv}^{exp}}}
\newcommand{\rhoZ}{\rho_{\mathrm{Z}}}
\newcommand{\thw}{\theta_{\mathrm{W}}}
\newcommand{\swsq}{\sin^2\!\thw}
\newcommand{\swsqmsb}{\sin^2\!\theta_{\mathrm{W}}^{\overline{\mathrm{MS}}}}
\newcommand{\swsqbar}{\sin^2\!\overline{\theta}_{\mathrm{W}}}
\newcommand{\cwsqbar}{\cos^2\!\overline{\theta}_{\mathrm{W}}}
\newcommand{\swsqb}{\sin^2\!\theta^{eff}_{\mathrm{W}}}
\newcommand{\ee}{{{\mathrm e}^+{\mathrm e}^-}}
\newcommand{\eeX}{{e^+e^-X}}
\newcommand{\gaga}{{\gamma\gamma}}
\newcommand{\mumu}{\ifmmode {\mu^+\mu^-} \else ${\mu^+\mu^-} $ \fi}
\newcommand{\eeg}{{e^+e^-\gamma}}
\newcommand{\mumug}{{\mu^+\mu^-\gamma}}
\newcommand{\tautau}{{\tau^+\tau^-}}
\newcommand{\qqb}{q\bar{q}}
\newcommand{\eegg}{e^+e^-\rightarrow \gamma\gamma}
\newcommand{\eeggg}{e^+e^-\rightarrow \gamma\gamma\gamma}
\newcommand{\eeee}{e^+e^-\rightarrow e^+e^-}
\newcommand{\eeeeee}{e^+e^-\rightarrow e^+e^-e^+e^-}
\newcommand{\eeeeg}{e^+e^-\rightarrow e^+e^-(\gamma)}
\newcommand{\eeeegg}{e^+e^-\rightarrow e^+e^-\gamma\gamma}
\newcommand{\eeeg}{e^+e^-\rightarrow (e^+)e^-\gamma}
\newcommand{\eemumu}{e^+e^-\rightarrow \mu^+\mu^-}
\newcommand{\eetautau}{e^+e^-\rightarrow \tau^+\tau^-}
\newcommand{\eehad}{e^+e^-\rightarrow \mathrm{hadrons}}
\newcommand{\eettg}{e^+e^-\rightarrow \tau^+\tau^-\gamma}
\newcommand{\eell}{e^+e^-\rightarrow l^+l^-}
\newcommand{\Ztopig}{{\mathrm{Z}}^0\rightarrow \pi^0\gamma}
\newcommand{\Ztogg}{{\mathrm{Z}}^0\rightarrow \gamma\gamma}
\newcommand{\Ztoee}{{\mathrm{Z}}^0\rightarrow e^+e^-}
\newcommand{\Ztoggg}{{\mathrm{Z}}^0\rightarrow \gamma\gamma\gamma}
\newcommand{\Ztomumu}{{\mathrm{Z}}^0\rightarrow \mu^+\mu^-}
\newcommand{\Ztotautau}{{\mathrm{Z}}^0\rightarrow \tau^+\tau^-}
\newcommand{\Ztoll}{{\mathrm{Z}}^0\rightarrow l^+l^-}
\newcommand{\Ztocc}{{\mathrm{Z}\rightarrow c \bar c}}
\newcommand{\Lamp}{\Lambda_{+}}
\newcommand{\Lamm}{\Lambda_{-}}
\newcommand{\Pt}{P_{t}}
\newcommand{\Gee}{\Gamma_{ee}}
\newcommand{\Gpig}{\Gamma_{\pi^0\gamma}}
\newcommand{\Ggg}{\Gamma_{\gamma\gamma}}
\newcommand{\Gggg}{\Gamma_{\gamma\gamma\gamma}}
\newcommand{\Gmumu}{\Gamma_{\mu\mu}}
\newcommand{\Gtautau}{\Gamma_{\tau\tau}}
\newcommand{\Ginv}{\Gamma_{\mathrm{inv}}}
\newcommand{\Ghad}{\Gamma_{\mathrm{had}}}
\newcommand{\Gnu}{\Gamma_{\nu}}
\newcommand{\GnuSM}{\Gamma_{\nu}^{\mathrm{SM}}}
\newcommand{\Gll}{\Gamma_{l^+l^-}}
\newcommand{\Gff}{\Gamma_{f\overline{f}}}
\newcommand{\Gtot}{\Gamma_{\mathrm{tot}}}
\newcommand{\Rb}{R\mathrm{_b}}
\newcommand{\Rc}{R\mathrm{_c}}
\newcommand{\al}{a_l}
\newcommand{\vl}{v_l}
\newcommand{\af}{a_f}
\newcommand{\vf}{v_f}
\newcommand{\ael}{a_e}
\newcommand{\ve}{v_e}
\newcommand{\amu}{a_\mu}
\newcommand{\vmu}{v_\mu}
\newcommand{\atau}{a_\tau}
\newcommand{\vtau}{v_\tau}
\newcommand{\ahatl}{\hat{a}_l}
\newcommand{\vhatl}{\hat{v}_l}
\newcommand{\ahate}{\hat{a}_e}
\newcommand{\vhate}{\hat{v}_e}
\newcommand{\ahatmu}{\hat{a}_\mu}
\newcommand{\vhatmu}{\hat{v}_\mu}
\newcommand{\ahattau}{\hat{a}_\tau}
\newcommand{\vhattau}{\hat{v}_\tau}
\newcommand{\vtildel}{\tilde{\mathrm{v}}_l}
\newcommand{\avsq}{\ahatl^2\vhatl^2}
\newcommand{\Ahatl}{\hat{A}_l}
\newcommand{\Vhatl}{\hat{V}_l}
\newcommand{\Afer}{A_f}
\newcommand{\Ael}{A_e}
\newcommand{\Aferb}{\bar{A_f}}
\newcommand{\Aelb}{\bar{A_e}}
\newcommand{\AVsq}{\Ahatl^2\Vhatl^2}
\newcommand{\Iwk}{I_{3l}}
\newcommand{\Qch}{|Q_{l}|}
\newcommand{\roots}{\sqrt{s}}
\newcommand{\pT}{p_{\mathrm{T}}}
\newcommand{\mt}{m\mathrm{_t}}
\newcommand{\Rechi}{{\mathrm{Re}} \left\{ \chi (s) \right\}}
\newcommand{\up}{^}
\newcommand{\abscosthe}{|cos\theta|}
\newcommand{\dsum}{\Sigma |d_\circ|}
\newcommand{\zsum}{\Sigma z_\circ}
\newcommand{\sint}{\mbox{$\sin\theta$}}
\newcommand{\cost}{\mbox{$\cos\theta$}}
\newcommand{\mcost}{|\cos\theta|}
\newcommand{\epair}{\mbox{$e^{+}e^{-}$}}
\newcommand{\mupair}{\mbox{$\mu^{+}\mu^{-}$}}
\newcommand{\taupair}{\mbox{$\tau^{+}\tau^{-}$}}
\newcommand{\gamgam}{\mbox{$e^{+}e^{-}\rightarrow e^{+}e^{-}\mu^{+}\mu^{-}$}}
\newcommand{\fullskip}{\vskip 16cm}
\newcommand{\halfskip}{\vskip  8cm}
\newcommand{\quarskip}{\vskip  6cm}
\newcommand{\abitskip}{\vskip 0.5cm}
\newcommand{\ba}{\begin{array}}
\newcommand{\ea}{\end{array}}
\newcommand{\bc}{\begin{center}}
\newcommand{\ec}{\end{center}}
\newcommand{\be}{\begin{eqnarray}}
\newcommand{\eeq}{\end{eqnarray}}
\newcommand{\bes}{\begin{eqnarray*}}
\newcommand{\ees}{\end{eqnarray*}}
\newcommand{\Kz}{\ifmmode {\mathrm{K^0_s}} \else ${\mathrm{K^0_s}} $ \fi}
\newcommand{\Zz} {\mathrm{Z}}
\newcommand{\qqbar}{\ifmmode {q\bar{q}} \else ${q\bar{q}} $ \fi}
\newcommand{\ccbar}{\ifmmode {c\bar{c}} \else ${c\bar{c}} $ \fi}
\newcommand{\bbbar}{\ifmmode {b\bar{b}} \else ${b\bar{b}} $ \fi}
\newcommand{\xxbar}{\ifmmode {x\bar{x}} \else ${x\bar{x}} $ \fi}
\newcommand{\rphi}{\ifmmode {\R\phi} \else ${R\phi} $ \fi}

\section{Introduction}
\label{sec:1}


In this paper, the average lifetime of the $\Bs$ meson is measured
and limits are derived on the oscillation frequency of the 
$\Bs$-$\Bsb$ system, $\dms$, and on the decay width difference, $\dgs$, 
between mass eigenstates of this system.

 Starting with a ${\mathrm B^0_s}$ meson produced at time $t$=0, 
the probability, ${\cal P}(t)$, to observe a $\Bs$ or a $\Bsb$ decaying at 
the proper time $t$ can be written, neglecting effects from CP violation:
\begin{equation}
{\cal P}(t)~=~\frac{\Gamma_{\mathrm B^0_s}}{2} e^{- \Gamma_{\mathrm B^0_s} t}
(\cosh (\frac{\dgs}{2} t)~\pm~\cos (\dms t) )
\label{eq:tot}
\end{equation}
where $\gammas~=~(\gammash + \gammasl)/2$, $\dgs~=~\gammasl-\gammash$ 
and $\dms~=~\mbs^{\mathrm H}~-~\mbs^{\mathrm L}$; L and H denote the light and heavy physical 
states, respectively; $\dgs$ and $\dms$ are defined to be positive~\cite{ref:bbd}
and the plus (minus) signs in equation~(\ref{eq:tot}) refer to $\Bs$ ($\Bsb$) decays. 
The oscillation period gives a direct measurement of the mass difference 
between the two physical states. 
The Standard Model predicts that $\dgs \ll \dms$ and the previous expression 
simplifies to :
\begin{equation}
{\cal P}_{\mathrm B^0_s}^{unmix}(t)~=~\Gamma_{\mathrm B^0_s} \; e^{- \Gamma_{\mathrm B^0_s} t} \cos^2 (\frac{\dms t}{2} )
\label{eq:osc1}
\end{equation}
for $\Bs \rightarrow \Bs$ and similarly:
\begin{equation}
{\cal P}_{\mathrm B^0_s}^{mix}(t)~=~\Gamma_{\mathrm B^0_s} \; e^{- \Gamma_{\mathrm B^0_s} t} \sin^2 (\frac{\dms t}{2} )
\label{eq:osc2}
\end{equation}
for  $\Bs \rightarrow {\Bsb}$.
The oscillation frequency, proportional to $\dms$, can be obtained from the 
fit of the  time distributions given in relations~(\ref{eq:osc1}) and 
~(\ref{eq:osc2}), whereas expression~(\ref{eq:tot}), without distinguishing 
between the $\Bs$ and the $\Bsb$, can be used to determine the average 
lifetime and the difference between the lifetimes of the heavy and light 
mass eigenstates\footnote{Throughout the paper the dimension of $\dms$ will be
expressed in ps$^{-1}$ units, because the argument $\dms t$ in the expression
~(\ref{eq:tot}) has no dimension. The conversion is given by ps$^{-1}$ = $6.58\times10^{-4}$eV/$c^2$. }.
\vspace{0.2cm}

B physics allows a precise determination of 
some of the parameters of the CKM matrix. All the nine elements
can be expressed in term of 4 parameters that are, in Wolfenstein
parameterisation~\cite{ref:wolf}, $\lambda$, A, $\rho$ and $\eta$.
The most uncertain parameters are $\rho$ and $\eta$.

Several quantities which depend on $\rho$ and $\eta$ can be measured
and, if the Standard Model is correct, they must define compatible values for 
the two parameters inside measurement errors and theoretical uncertainties.
These quantities are $\epsilon_{\mathrm K}$, the parameter introduced to measure CP 
violation in the K system, $|V_{\mathrm ub}|/|V_{\mathrm cb}|$, the ratio between the modulus 
of the CKM matrix elements corresponding to b$\rightarrow$u and 
b$\rightarrow$c transitions, and the mass difference $\dmd$. 

In the Standard Model, ${\mathrm B^0_q}$-$\overline{\mathrm B^0_q}$ 
(${\mathrm q} = {\mathrm d,s}$) mixing is a direct consequence of second order weak 
interactions. Having kept only the dominant top quark contribution, $\dmq$ 
can be expressed in terms of Standard Model parameters~\cite{ref:dmd_teo}:
\begin{equation}
\dmq= \frac{G_{\mathrm F}^2}{6 \pi^2}|V_{\mathrm tb}|^2|V_{\mathrm tq}|^2 m_{\mathrm t}^2 m_{\mathrm B_q} 
      f_{\mathrm B_q}^2 B_{\mathrm B_q} 
      \eta_{\mathrm B} F(x_{\mathrm t}).
\label{eq:dmth}
\end{equation}
In this expression $G_{\mathrm F}$ is the Fermi coupling constant; $F(x_{\mathrm t})$, with 
$x_{\mathrm t}=\frac{m_{\mathrm t}^2}{m_{\mathrm W}^2}$, results from the evaluation of 
the second order weak ``box'' diagram responsible for the mixing
and has a smooth dependence on $x_{\mathrm t}$;
$\eta_{\mathrm B}$ is a QCD correction factor obtained at next-to-leading order 
in perturbative QCD~\cite{ref:bubu}.
The dominant uncertainties in equation~(\ref{eq:dmth}) come from
the evaluation of the B meson decay constant $f_{\mathrm B_q}$ and of the ``bag'' 
parameter $B_{\mathrm B_q}$. The two elements of 
the $V_{\mathrm CKM}$ matrix are equal to:
\begin{equation}
 |V_{\mathrm td}| = A \lambda^3 \sqrt{( 1- \rho )^2 +  \eta^2}
  ~ ,~~~~~        |V_{\mathrm ts}| = A \lambda^2,
\end{equation}
neglecting terms of order ${\cal O}(\lambda^4)$.

In the Wolfenstein parameterisation, $|V_{\mathrm ts}|$ 
is independent of $\rho$ and $\eta$. A measurement of $\dms$ is  
thus a way to measure the value of the non-perturbative QCD
parameters.

Direct information on $V_{\mathrm td}$ can be inferred by measuring $\dmd$.
Several experiments have accurately measured $\dmd$, nevertheless this 
precision cannot be fully exploited to extract information on $\rho$ and 
$\eta$ because of the large uncertainty which originates in the 
evaluation of the non-perturbative QCD parameters.

An efficient constraint is the ratio between the Standard Model expectations 
for $\dmd$ and $\dms$, given by:
\begin{equation}
\frac{\dmd}{\dms}~=~\frac{ m_{\Bd} f^2_{\Bd} B_{\Bd} \eta_{\Bd}}
{ m_{\Bs} f^2_{\Bs} B_{\Bs} \eta_{\Bs}} \frac{\left | V_{\mathrm td} \right |^2}
{\left | V_{\mathrm ts} \right |^2} \; .
\label{eq:ratiodms}
\end{equation}
A measurement of the ratio ${\dmd}/{\dms}$ gives the same type of constraint, 
in the ${\rho}-{\eta}$ plane, as a measurement of $\dmd$, but is 
expected  to be better under control from theory, 
since the ratios of the decay constants 
$f_{\mathrm B_{q}}$ and of the bag parameters $\rm{B}_{\mathrm B_{q}}$ for
 $\Bd$ and $\Bs$ are better known  than their individual values \cite{ref:bbag1}. 

Using existing measurements which constrain $\rho$ and $\eta$, 
except those on $\dms$, the distribution for
the expected values of
$\dms$ can be obtained. It has been shown, for example, in the context of the
Electroweak Standard Model and QCD assumptions, 
that $\dms$ has to lie,
at 68$\%$ C.L., between 8~ps$^{-1}$ and 16~ps$^{-1}$ and is expected to be smaller than
21~ps$^{-1}$ at the 95\% C.L.~\cite{ref:bello}.
\vspace{0.2cm}

The search for ${\mathrm B^0_s}-\overline{\mathrm B^0_s}$ oscillations has 
been the subject of recent intense activity. No signal has been observed so far 
and the lower limit on the oscillation frequency comes from the 
combination of the results obtained at LEP, CDF and SLD experiments: 
$\Delta m_{\Bs} > 14.3~{\mathrm ps}^{-1}$ at the 95\%~C.L.~\cite{ref:osciwg}.  
The sensitivity of present measurements is estimated as $14.7~{\mathrm ps}^{-1}$.
These results have been obtained by combining analyses which select various $\Bs$
decay channels\footnote{Unless explicitly stated otherwise, charge conjugate
states are always implied.}. 
Some analyses~\cite{ALDMS_LQ,OPALDMS} use events containing simply a lepton emitted 
at large transverse momentum relative to the axis of the jet from which it emerges.
 In this case, the proper time is 
measured using an inclusive vertex algorithm to reconstruct the decay distance and 
the energy of the B hadron candidate.
In other analyses, like  ${\mathrm D_s^{\pm}} \ell^{\mp}$~\cite{ref:dsl,ADMSDSL}, the identified lepton is accompanied, 
in the same hemisphere, by an exclusively reconstructed $\Ds$.
In $\rm{D}_{s}^{\pm} {\mathrm h}^{\mp}$ analyses (see~\cite{ref:dsh_al} and the present paper),
such leptons are replaced by one or more charged hadrons.

Progress before the next generation accelerators is expected to come from improved analyses of 
these channels, but the sensitivity at high frequency is essentially limited 
by the damping of the reconstructed oscillation amplitude due to the limited resolution on the $\Bs$ proper 
time.  In general, the proper time resolution  $\sigma_t $ contains two terms and one of them is a time-dependent:
\begin{equation}
 \sigma_t  = \sqrt{ {\sigma_L}^2 + (\sigma_{p}/p)^2 \times t^2 }
\end{equation}
\label{eq:damp}
where ${\sigma_{L}}$ is related to the decay distance resolution,  
${\sigma_{p}/p}$ is the relative error on the momentum reconstruction and ${t}$ is the proper decay time.
The damping factor of the oscillation amplitude in this case  is given by 
the following expression~\cite{ref:amplitude}:

\begin{equation}
 damping = \exp({ -(\dms\sigma _{L})^{2}/2})\sqrt{\pi } ) ({\mathrm x}) \exp({\mathrm x}^{2}){\mathrm ERFC}({\mathrm x}),
\end{equation}
where x=1/\( \sqrt{2} \)/\( \dms \) 1/(\( \sigma _{p} \)/$p$) 
and ERFC(\( x \))=\( 2/\sqrt{\pi } \) \( \int ^{\infty }_{x} \)\( e^{-t^{2}} \)d\( t \). \\
Exclusive (completely reconstructed) decays have a better time resolution for two reasons. 
As there is no missing particle in the decay, the $\Bs$ momentum is known with a good precision and therefore 
the contribution of the momentum uncertainty to the proper time resolution 
is negligible. Hence, $\sigma_t={\sigma_L}$ and the previous expression is  
simplified: 
${damping} = \rm{exp}({-(\Delta m_{\Bs} \sigma_L)^2/2})$. 
In addition, the reconstructed channels are two-body or quasi two-body decays, with an opening 
angle of their decay products which is on average larger than in multi-body final states; 
this results in a better accuracy on their decay position determination.
\vspace{0.2cm}

The $\Bs$ meson lifetime is expected to be equal to the 
$\Bd$  lifetime~\cite{ref:bigi} within one percent. In the Standard Model, the ratio between the
decay width and the mass differences in the $\Bo$-$\Bob$ system is of the order 
of $(m_{\mathrm b}/m_{\mathrm t})^2$, although large QCD corrections are expected.
Explicit calculations to leading order in QCD correction, in the OPE 
formalism~\cite{ref:bbd} predict:
\bes
\dgbs  = 0.16 ^{+0.11}_{-0.09} 
\ees
where the quoted error is dominated by the uncertainty related to
hadronic matrix elements.
Recent calculations~\cite{ref:bbd_new} at next-to-leading order predict a 
lower value:
\bes
 \dgbs = 0.006^{+0.150}_{-0.063} \; .
\ees
An interesting approach consists in using the ratio between
$\dgs$ and $\dms$~\cite{ref:bbd_new}:
\begin{equation}
\frac{\dgs}{\dms} = (2.63 ^{+0.67}_{-1.36}) 10^{-3},
\end{equation}
to constrain the upper part of the $\dms$ spectrum with an upper
limit on $\dgbs$.
If, in future, the theoretical uncertainty can be reduced, this method 
can give an alternative way of determining $\dms$ via 
$\dgs$ and, in conjunction with the determination of $\dmd$,
 can provide an extra constraint on the $\rho$ and 
$\eta$ parameters.
\vspace{0.2cm}

After a description of the main features of the DELPHI detector, 
the event selection and the event simulation in Section~\ref{sec:delphi_det},
the two $\Bs$ analyses are described. 
 Section~\ref{sec:2} presents  an analysis of $\Bs$-$\Bsb$ oscillations
from a sample of  44 exclusively reconstructed $\Bs$ decays. 
Section~\ref{sec:3} is dedicated to the $\rm{D}_{s}^{\pm} {\mathrm h}^{\mp}$ analysis, 
in which $\Bs \rightarrow \Ds{\mathrm h} X$ decays with fully reconstructed $\Ds$ are selected. 
This analysis also includes a measurement of  
the $\Bs$ meson lifetime and  sets a limit on the difference between the decay widths of the physical $\Bs$ states, 
$\dgbs$. In Section~\ref{sec:4} the combined limit on $\Delta m_{\Bs}$ is given. 
Finally in Section~\ref{sec:5}, these three results are combined with those obtained using
a ${\mathrm D_s^{\pm}} \ell^{\mp}$ sample~\cite{ref:dsl}.

\section{The DELPHI detector}
\label{sec:delphi_det}
 
The events used in this analysis have been recorded with the DELPHI detector 
at LEP operating at energies close to the  $\Zz$ peak. The DELPHI detector and its performance 
were described in detail elsewhere~\cite{ref:perfo}. In this section,
 components of the detector and their characteristics, which are the most 
relevant for these analyses  are summarised.

\subsection{Charged particle reconstruction}

The detector elements used for tracking are the Vertex Detector (VD), the Inner Detector (ID), the Time Projection 
Chamber (TPC) and the Outer Detector (OD). The VD provides the high precision needed near the primary vertex. 

For the data taken from 1991 to 1993, the VD consisted of three cylindrical layers of silicon detectors 
(at radii of 6.3, 9.0 and 10.9~cm) measuring points in the plane transverse to the beam direction 
($r\phi$ coordinate\footnote{The DELPHI reference frame is defined with $z$ along the e$^-$ beam,
$x$ towards the centre of LEP and $y$ upwards. Angular coordinates are
$\theta$, measured from $z$, and the azimuth, $\phi$, from the $x$-axis,
while $r$ is the distance from the $z$-axis.})
in the polar angle range between $43^\circ$ and $137^\circ$.
In 1994, the first and the third layers were equipped with detector modules with double sided 
readout, providing a single hit precision of 7.6~$\mu$m in $r\phi$, 
similar to that obtained previously, and 9~$\mu$m in $z$~\cite{Delphi:VD}.
For high momentum particles with associated hits in the VD, the impact parameter 
precision close to the interaction region is 20~$\mu$m in the $r\phi$ plane and 34~$\mu$m in the $rz$ plane.
Charged particle tracks are reconstructed with 95\% efficiency and with a 
momentum resolution $\sigma_p/p < 1.5 \times 10^{-3} p$ ($p$ in~${\mathrm GeV}/\it{c}$) in 
the polar angle region between $25^\circ$ and $155^\circ$.

\subsection{Hadronic Z selection and event topology}

Hadronic events from Z  decays were selected by requiring a  
multiplicity of charged particles greater than four and a total energy of charged particles 
greater than 0.12$\sqrt s$, where $\sqrt s$ is the centre-of-mass energy and all particles were assumed to
be pions; charged particles were required to have a momentum greater than 
\mbox{0.4~${\mathrm GeV}/\it{c}$} and a polar angle between $20^\circ$ and $160^\circ$. 
The overall trigger and selection efficiency is (95.0$\pm$0.1)\%~\cite{delphi:Zshape}. 
A total of 
3.5 million hadronic events was obtained from the 1992-1995 data.

Each selected event was divided into two hemispheres separated by the plane
transverse to the sphericity axis.
A clustering analysis, based on the JETSET algorithm LUCLUS with default
parameters~\cite{ref:luclus}, was used to define jets using both charged and neutral particles.
These jets were used to compute the $p_t^{out}$
of each particle of the event, as the transverse momentum of this
particle with respect to the axis of the jet it belonged to,
after having removed this particle from its jet.

\subsection{Hadron identification}
\label{sec:hadid}

Hadron identification relied on the RICH detector and on the specific ionisation measurement performed by the TPC.

The RICH detector~\cite{ref:perfo} used two radiators.
A gas radiator separated kaons from pions between \mbox{3 and 9~${\mathrm GeV}/\it{c}$},
where kaons gave no Cherenkov light whereas pions did, and between 
\mbox{9 and 16~${\mathrm GeV}/\it{c}$}, using the measured Cherenkov angle. It also provided 
kaon/proton separation from \mbox{8 to 20~${\mathrm GeV}\it{/c}$}. 
A liquid radiator, which has been fully operational since 1994, 
 provided
${\mathrm p/K}/\pi$ separation in the momentum range between 0.7 and 8~${\mathrm GeV}/c$. 

The specific energy loss per unit length ($dE/dx$) was measured 
in the TPC by using up to 192 sense wires.
At least 30 contributing measurements were required to compute the truncated 
mean. In the momentum range between 3 and 25~${\mathrm GeV}/c$, this is fulfilled
for 55\% of the tracks and the $dE/dx$ measurement has a precision of $\pm7\%$.

The combination of the two measurements, $dE/dx$ and RICH, provides
four levels of pion, kaon and proton tagging~\cite{ref:idid}: ``very loose'', ``loose'', ``standard'' and ``tight'' 
corresponding to different purities. The efficiency and purity of the tagging
depend on the momentum of the particle.
In the typical momentum range of the B decay
products between 2 and 10~GeV$/c$, the average efficiency (purity) 
is about 75\% (50\%) for the ``loose'', 65\% (60\%), for the ``standard'' and 55\% (70\%) for the ``tight'' kaon tag. 
The ``very loose'' tag indicates that the particle is not identified as  a pion.
The purity is the proportion of genuine kaons in the selected sample. 

Very recently  a  neural network algorithm has been developed in order to combine the different RICH  
identification packages optimally. The result of the neural network gives a 
tagging variable $x_{net}$, which varies from $-1$ (pure background) to $+1$ (pure signal).  
In order to use the same definitions of the kaon tag  through all the paper,
the ``very loose'', ``loose'', ``standard'' and ``tight'' tags were defined 
for a tagging variable 
 $x_{net}$ larger than $-0.2$, $0.0$, $0.4$ and $0.6$, respectively.
This second algorithm has been used in the $\rm{D}_{s}^{\pm}{\mathrm h}^{\mp}$ analysis 
(Section~\ref{sec:3}). 


\subsection{Primary  and B vertices reconstruction }

The average beam spot position  evaluated using the events from  different periods of data
taking  has been used as a constraint  in the determination of the $\ee$ interaction 
point  on an event-by-event basis~\cite{ref:perfo}.
In 1994 and 1995 data, the position of the primary vertex
transverse to the beam is determined with a precision of about 40~$\mu$m in the horizontal direction,
and about 10~$\mu$m in the vertical direction. For 1992 and 1993 data, 
the uncertainties are larger by about 50\%.

For both 1992-1993 and 1994-1995 data, the B decay length was estimated as
$L=L_{xy}/\sin{\theta}_{\mathrm B}$, where $L_{xy}$ is the measured distance between 
the primary vertex and the B decay vertex in the plane transverse to 
the beam direction and ${\theta}_{\mathrm B}$ is the polar angle of the B flight direction, 
estimated from the B decay products. The transverse decay length  $L_{xy}$ was given the same 
sign as the scalar product of the B momentum with the vector joining the primary to 
the secondary vertex; this procedure gives  signed decay length $L$.

\subsection{${\mathrm \Lambda}$ and ${\mathrm K}^{0}_{\mathrm s}$ reconstruction}
\label{sec:vzero}
The ${\mathrm \Lambda} \rightarrow \mathrm{p}\pi^-$  and ${\mathrm K}^{0}_{\mathrm s} \rightarrow \pi^+\pi^-$ decays 
have been reconstructed using a kinematic fit. The distance in the $r\phi$~plane between the 
$V^0$ decay point and the primary vertex was required to be less than 90~cm. This 
condition meant that the decay products had track segments of at least 20~cm 
long in the TPC. The reconstruction of the $V^0$ vertex and selection cuts are
described in detail in reference~\cite{ref:perfo}.
Only ${\mathrm K^0_s}$ candidates passing the ``tight'' selection criteria
defined in~\cite{ref:perfo} were retained for this analysis.

\subsection{b-tagging}
\label{sec:btag}

The b-tagging package  is 
described in reference~\cite{btag}. The impact parameters of the charged particle tracks, with respect to 
the primary vertex, were used to build the probability 
that all tracks come from this vertex. Due to the long B hadron lifetimes, the probability distribution is peaked near zero 
for events which contain beauty quark whereas it is flat for events containing only light quarks.

\subsection{Event simulation}
\label{sec:ev_sim}

Simulated events were generated using the JETSET 7.3 program~\cite{ref:luclus}
with parameters tuned as in~\cite{ref:tuning} and with an updated model of  B decay 
branching fractions.
B hadron semileptonic decays were simulated using the ISGW model~\cite{ref:isgw}.
Generated events were passed through the full simulation of the DELPHI
detector~\cite{ref:perfo}, and the resulting simulated raw data 
were processed through the same reconstruction and analysis programs as the real data.

\section{Exclusively  reconstructed $\Bs$ decays }
\label{sec:2}

\subsection{Evaluation of $\Bs$ branching fractions}
\label{sec:21}
For this analysis $\Bs$ mesons were reconstructed in ${\mathrm D_s^-}
\pi^{+}$, ${\mathrm D_s^-}\mathrm{a_1^+}$,
$\overline{\mathrm D}^0 \mathrm{K^-} \pi^+$ and $\overline{\mathrm D}^0 
\mathrm{K^-}\mathrm{a_1^+}$ decay channels.
Contributions to the mass spectrum can come also from other decay channels 
such as ${\mathrm D_s^{*-}} \pi^+$,
${\mathrm D_s^{*-}} \mathrm{a}_{1}^{+}$, ${\mathrm \overline{D}}^{*}(2007)^{0} 
\mathrm{K}^{-} \pi^{+} $,
${\mathrm \overline{D}}^{*}(2007)^{0} \mathrm{K}^{-} \mathrm{a}_{1}^{+} $,
$\rm{D}_{s}^{-} \rho^{+}$ and ${\mathrm D_s^{*-}} \rho^{+}$, where the cascade photon
 or the
 neutral pion has not been reconstructed. All corresponding branching fractions are unmeasured.
An estimation will be used in the following to evaluate the number 
of the expected $\Bs$ events.

To evaluate the two-body
$\Bs \rightarrow {\mathrm D_s^{(*)-}} \pi^{+}(a_1^{+})$ branching fractions,
the equivalent decay channels for non-strange mesons were used.
This evaluation is based on the numerical application of
factorisation done in reference~\cite{ref:b-en-2-corps-factorisation}.

For each decay mode of the $\Bs \rightarrow {\mathrm D_s^{(*)-}}  {\mathrm M}^{+}$,
where ${\mathrm M}^{+}$ is a $\pi^+$ or a $\rho^{+}$,
the branching fraction is estimated using  the following formula: 
\begin{equation}
\label{eq:brds}
 {\mathrm Br}(\Bs \rightarrow {\mathrm D_s^{(*)-}}  {\mathrm M}^{+}) = 
 {\mathrm Br^{exp}}(\Bd \rightarrow {\mathrm D}^{(*)-}      {\mathrm M}^{+})  
 \frac{     {\mathrm Br^{th}}(\Bs \rightarrow {\mathrm D_s^{(*)-}}  {\mathrm M}^{+}) }
          { {\mathrm Br^{th}}(\Bd \rightarrow {\mathrm D}^{(*)-}      {\mathrm M}^{+}) },
\end{equation}
where the notation ${\mathrm D}^{(*)-}$ means ${\mathrm D}^{*}(2010)^{-}$ or 
${\mathrm D}^{-}$ meson.
The superscript ``exp'' designates the experimentally measured values~\cite{ref:book},
and ``th'' the predictions from reference~\cite{ref:b-en-2-corps-factorisation}.
The theoretical ratios in equation~(\ref{eq:brds}) are close to 1.

For the two-body decays containing D
mesons with a$_1$,
the theoretical branching fractions are not available, and 
the same factors as for the $\rho^{+}$ channels have been assumed.

\begin{table}[h]
\begin{center}
\begin{tabular}{|l|c||l|c|}
\hline 
\hline
  $\Bd$ decay channel                                                    & 
  Measured Br            &
  $\Bs$ decay channel                                                    & 
  Estimated Br            \\ \hline \hline
$ \Bd \rightarrow  \mathrm{D^{-}} \pi^+$                            & 
~$(3.0\pm0.4)\times 10^{-3} $  &
$ \Bs \rightarrow  {\mathrm D_s^-} \pi^+$                            & 
~$(3.4\pm0.4)\times 10^{-3} $  \\ \hline
$ \Bd \rightarrow  {\mathrm D}^{*}(2010)^{-} \pi^+$                            & 
~$(2.8\pm0.2)\times 10^{-3} $  &
$ \Bs \rightarrow  {\mathrm D_s^{*-}} \pi^+$                            & 
~$(3.0\pm0.2)\times 10^{-3} $  \\ \hline
$ \Bd \rightarrow \mathrm{D^{-}} \rho^{+}$                                 & 
~$(7.9\pm1.4)\times 10^{-3} $  &
$ \Bs \rightarrow {\mathrm D_s^-} \rho^{+}$                                  & 
~$(9.0\pm1.6)\times 10^{-3} $  \\ \hline
$ \Bd \rightarrow {\mathrm D}^{*}(2010)^{-} \rho^{+}$                              & 
~$(6.7\pm3.3)\times 10^{-3} $  &
$ \Bs \rightarrow {\mathrm D_s^{*-}} \rho^{+}$                                 & 
~$(6.9\pm3.4)\times 10^{-3} $  \\ \hline
$ \Bd \rightarrow  \mathrm{D^{-}}\mathrm{a}_1^+$                        & 
~$(6.0\pm3.3)\times 10^{-3} $  &
$ \Bs \rightarrow  {\mathrm D_s^-}\mathrm{a}_1^+$                        &
~$(6.8\pm3.6)\times 10^{-3} $  \\ \hline
$ \Bd \rightarrow  {\mathrm D}^{*}(2010)^{-}\mathrm{a}_1^+$                        & 
$(13.0\pm2.7)\times 10^{-3} $  &
$ \Bs \rightarrow  {\mathrm D_s^{*-}}\mathrm{a}_1^+$                        
& $(13.3\pm2.8)\times 10^{-3} $  \\ \hline
$ \Bd \rightarrow \mathrm{\overline{D}}^{(*)0} \pi^{-} \pi^{+}$     &   see
 Section~\ref{sec:24} \   &
$ \Bs \rightarrow \mathrm{\overline{D}}^{(*)0} \mathrm{K}^{-} \pi^{+}$     
& ~$(0.9\pm0.5)\times 10^{-3} $  \\ \hline
$ \Bd \rightarrow \mathrm{\overline{D}}^{(*)0} \pi^{-}\mathrm{a}_1^{+}$ &  
see Section~\ref{sec:24}                &
$ \Bs \rightarrow \mathrm{\overline{D}}^{(*)0} \mathrm{K}^{-}\mathrm{a}_1^{
+}$ & ~$(3.0\pm1.7)\times 10^{-3} $  \\ \hline
\end{tabular}
\caption[]{ \it {Values of the $\Bs$ branching fractions used in the current analysis.
 The notation
$\overline{\mathrm D}^{(*)0}$ defines $\overline{\mathrm D}^{*}(2007)^{0}
$ or $\overline{\mathrm D}^{0}$ charmed hadrons.} }
\label{tab:brachi}
\end{center}
\end{table}

Similar considerations are applied to the three-body decays
$\Bs \rightarrow \overline{\mathrm D}^{0}\mathrm{K}^- \pi^{+}(\mathrm{a}_{
1}^{+})$ and
$\Bs \rightarrow \overline{\mathrm D}^{*}(2007)^{0}\mathrm{K}^- \pi^{+}
(\mathrm{a}_{1}^{+})$ using the measurements of
${\mathrm Br}(\Bd \rightarrow \overline{\mathrm D}^{0} \pi^-\pi^+)$ and
${\mathrm Br}(\Bd \rightarrow \overline{\mathrm D}^{0} 
\pi^-\rm{a}_{1}^{+})$,
which are described in the Section~\ref{sec:24} of the current paper.
The ratio of $\Bs$ to $\Bd$ branching fractions was taken equal to 1.1 
(this factor corresponds to the average ratio 
of the four previous decay modes). Finally, the factor 3.3 was used as the ratio 
between  the $\Bs$ decay modes containing ${\mathrm a}_{1}^{+}$ and $\pi^{+}$ mesons. 
This factor was taken as an average between the experimental values for the 
following $\Bd$ two-body decays:
$$
 \frac{{\mathrm Br}(\Bd \rightarrow {\mathrm D}^- {\mathrm a}_{1}^{+}) + 
{\mathrm Br}(\Bd \rightarrow {\mathrm D}^{*}(2010)^{-} {\mathrm a}_{1}^{+}) }
 {{\mathrm Br}(\Bd \rightarrow {\mathrm D}^- \pi^+) + {\mathrm Br}(\Bd \rightarrow 
{\mathrm D}^{*}(2010)^{-} \pi^+) } =  3.3\pm 0.9 \; .
$$
 The theoretical calculation~\cite{ref:Neubert1997uc} gives for this ratio values
between 3.4 and 3.5. 
 Table~\ref{tab:brachi} presents
the evaluations of several branching fractions of interest for this paper.

\subsection{Event Sample}
\label{sec:22}
Events were selected using the following decay channels:
\begin{eqnarray*}
\begin{array}{ll}
\Bs \rightarrow {\mathrm D_s^-} \pi^{+}  & {\mathrm D_s^-} \rightarrow \phi \pi^{-}, \phi \pi^- \pi^+ \pi^- ,
 \mathrm{f}(980) \pi^- ,  \mathrm{K}^{0}_s \mathrm{K}^{-},
{{\mathrm{K}}}^{*0} \mathrm{K}^{-},  \Kstar \Kstarm             \\   
\Bs \rightarrow {\mathrm D_s^-}\mathrm{a_1^+} & {\mathrm D_s^-} \rightarrow \phi \pi^{-}, \mathrm{K}^{*0} \mathrm{K}^{-}  \\
\Bs \rightarrow \overline{\mathrm D}^0\mathrm{K}^- \pi^+        
                                    & \overline{\mathrm D}^0 \rightarrow \mathrm{K}^+ \pi^- ,\mathrm{K}^+\pi^-\pi^+\pi^- \\
\Bs \rightarrow \overline{\mathrm D}^0\mathrm{K}^-\mathrm{a_1^+}        
                                    & \overline{\mathrm D}^0 \rightarrow \mathrm{K}^+ \pi^- ,\mathrm{K}^+\pi^-\pi^+\pi^- 
\end{array}
\end{eqnarray*}
where the $\phi$, ${\mathrm K^0_s}$, ${\mathrm f}(980)$, ${\mathrm K}^{*0}$, $\Kstarm$ and ${\mathrm a_1}$ 
are reconstructed in their charged
decay channels: $\phi \rightarrow \mathrm{K}^{+}\mathrm{K}^{-}$, ${\mathrm K^0_s} \rightarrow \pi^{+}\pi^{-}$,
${\mathrm f}(980) \rightarrow \pi^{+}\pi^{-}$,
${\mathrm K}^{*0} \rightarrow \mathrm{K}^{+}\pi^{-}$, ${\mathrm K}^{*-} \rightarrow \mathrm{K}^{0}_s\pi^{-}$ 
and ${\mathrm a}_{1}^{+} \rightarrow \rho^0  \pi^+$, $\rho^0 \rightarrow \pi^+ \pi^-$. $\Ds$ and ${\mathrm D}^0$ mesons 
were reconstructed by considering charged particles  in the same hemisphere  with at least 
one VD hit.
 
D meson candidates were accepted
if their mass was within the intervals $1.93-2.01~\mathrm{GeV}/{\it{c}}^2$ for $\Ds$ and $1.83-1.90~\mathrm{GeV}/{\it{c}}^2$ for ${\mathrm D}^0$.
The ${\mathrm D}$ decay length was required to be positive and 
the $\chi^2$--probability of the fitted vertex to be larger 
than  10$^{-5}$ (10$^{-3}$ for the ${\mathrm D}^0 \rightarrow  \mathrm{K} \pi \pi \pi$ decay mode).
Different selection criteria were used for different decay channels according to the 
optimisations derived from dedicated simulated samples. They are described in the following: \\

$ \bullet$ \underline {${\mathrm D_s^-} \rightarrow \phi \pi^{-} $}\\
The $\phi$ meson was reconstructed in the decay mode $\phi \rightarrow \mathrm{K}^{+}\mathrm{K}^{-}$ by taking all 
possible pairs of oppositely charged particles if at least one of them was 
identified as a ``very loose'' kaon.  The invariant mass of these pairs 
had to be within $\pm$12~${\mathrm MeV}/{\it{c}}^2$ of the nominal $\phi$ mass value~\cite{ref:book}. 
The momenta of all  three particles were required to be larger than 1~${\mathrm GeV}/\it{c}$.
In the decay of the  $\Ds$ meson  into a vector ($\phi$) and a pseudoscalar meson ($\pi$),
 helicity conservation requires that the angle $\psi$, measured in the vector meson rest 
frame between the directions of its decay products 
and of the pseudoscalar meson, has a $\cos^{2}\psi$ distribution. The value of $|\cos\psi|$ was required to be larger than 0.3. \\

$\bullet\underline {{\mathrm D_s^-} \rightarrow \phi \pi^{-} \pi^{+} \pi^{-}}$\\
The $\phi$ meson was reconstructed as in the previous channel.
The momenta of all three pions had to be larger than 0.6~${\mathrm GeV}/\it{c}$. \\

$\bullet$\underline {${\mathrm D_s^-} \rightarrow  \mathrm{f}(980) \pi^{-}$}\\
The ${\mathrm f}(980)$ meson was reconstructed in the decay mode 
${\mathrm f}(980) \rightarrow \pi^{+} \pi^{-}$ by taking all 
possible pairs of oppositely charged particles classified as pions. This channel
suffers from a large combinatorial background which was reduced by selecting 
candidates having an  invariant $\pi^+\pi^-$ mass 
within 15~${\mathrm MeV}/{\it{c}}^2$ of the nominal ${\mathrm f}(980)$ mass~\cite{ref:book} and a total 
momentum larger than 8~${\mathrm GeV}/\it{c}$. The momenta of all three pions had to be larger than 1~${\mathrm GeV}/\it{c}$. \\

$\bullet$\underline {${\mathrm D_s^-} \rightarrow \mathrm{K}^{0}_s \mathrm{K}^{-}$}\\
${\mathrm K^0_s}$ meson reconstruction has been 
described in Section~\ref{sec:vzero}. In addition, 
the decay length of ${\mathrm K^0_s}$ candidates had to be positive and
their momentum to be larger than 2.5~${\mathrm GeV}/\it{c}$. The momentum of the K$^-$ candidate,
identified as  a ``very loose'' kaon, had to be larger than 1~GeV$/\it{c}$. \\

$\bullet$\underline {${\mathrm D_s^-} \longrightarrow{{\mathrm{K}}}^{*0} \mathrm{K}^{-} $}\\
The ${\mathrm{K}}^{*0} $ meson was reconstructed in the charged 
decay mode ${\mathrm{K}}^{*0} \rightarrow  \mathrm{K}^{+} \pi^{-}$.
The ${\mathrm K}^{+}$ candidate was required to be identified 
as a ``very loose'' kaon.
The momenta of both particles had to be larger than 1~${\mathrm GeV}/\it{c}$ and the invariant mass of the pair 
had to be within $\pm$40~${\mathrm MeV}/{\it{c}}^2$ of the nominal $\Kstar$ mass~\cite{ref:book}.  
The value of $|\cos\psi|$  (see the ${\mathrm D_s^-} \rightarrow \phi \pi^- $ selection) had to be larger than 0.3.
For the K$^-$ candidate, the  combinatorial background is higher than  
for the kaon  coming from  $\Kstar$ resonance. The momentum of the K$^-$ candidate from  ${\mathrm D_s^-}$  
had to exceed 2.5~{$\Gec$} and it had to be identified 
as a ``loose'' kaon.  \\

$\bullet$\underline {${\mathrm D_s^-} \longrightarrow\Kstar \Kstarmn$}\\
The  ${\mathrm{K}}^{*0} $ meson was reconstructed 
as previously, but with an invariant mass 
within $\pm$60~${\mathrm MeV}/{\it{c}}^2$ of the nominal 
${\mathrm{K}}^{*0}$ mass~\cite{ref:book}. This mass interval was chosen larger  
than in the previous ${\mathrm{K}}^{*0} $ selection,  because of 
the additional constraint from the ${\mathrm{K}}^{*- } $ mass.
 The ${\mathrm{K}}^{*- } $ meson was reconstructed in the  
decay mode ${\mathrm{K}}^{*- } \rightarrow \mathrm{K}^0_s \pi^-$.
The ${\mathrm K}^0_s$ meson reconstruction was discussed previously.
The invariant mass of ${\mathrm{K}}^{*- } $ candidates had to be within $\pm$60~${\mathrm MeV}/{\it{c}}^2$ of the nominal 
$\Kstarmn$ mass~\cite{ref:book}. \\

$\bullet$\underline {$\overline{\mathrm D}^0 \rightarrow  \mathrm{K}^+ \pi^- ,  \; \; 
 \overline{\mathrm D}^0 \rightarrow  \mathrm{K}^+ \pi^- \pi^+ \pi^-$} \\
 The $\overline{\mathrm D}^0$ meson decay to  $ \mathrm{K}^+ \pi^- $
  has been reconstructed by combining a kaon candidate,
 identified with the ``loose'' tag,
 with an oppositely charged pion with momentum larger than 1~${\mathrm GeV}/\it{c}$. 
 The $\overline{\mathrm D}^0$ meson decay to $  \mathrm{K}^+ \pi^- \pi^+ \pi^-$  has been 
reconstructed by combining a kaon candidate,
 identified with the ``standard'' tag, with three pions, each of them having a momentum larger than 0.5~${\mathrm GeV}/\it{c}$. 
In order to reduce the combinatorial background, for both decay modes,
 the kaon momentum  was required  to be larger than 
2.5~${\mathrm GeV}/\it{c}$ and the  $\overline{\mathrm D}^0$ momentum 
candidate to be larger than 10~${\mathrm GeV}/\it{c}$. \\

Selected  $\overline{\mathrm D}^{0}$ and ${\mathrm D_s^-}$  mesons were used to reconstruct 
$\Bs$ candidates
by fitting a common vertex for the ${\mathrm D_s^-} \pi^+ $, $ {\mathrm D_s^-}\mathrm{a}_1^{+} $,
$\overline{\mathrm D}^{0} \mathrm{K}^- \pi^+$ or $ \overline{\mathrm D}^{0} \mathrm{K}^-\mathrm{a}_1^{+}$ 
systems. The $ \pi$ momentum had to be larger than  4~${\mathrm GeV}/\it{c}$.
The ${\mathrm a_1}$ candidates were reconstructed using the combination of three pions with 
momenta larger than  0.8~${\mathrm GeV}/\it{c}$ and with an invariant mass situated 
within the interval $0.95$-$1.50$~GeV$/{\it{c}}^2$. 
At least one of the two $\pi^+\pi^-$ combinations was required to have an invariant
mass lying within $\pm$150~${\mathrm MeV}/{\it{c}}^2$ of the nominal $\rho$ mass~\cite{ref:book}. 
The ${\mathrm a_1}$ momentum had to be larger than  5~${\mathrm GeV}/\it{c}$ 
(6~${\mathrm GeV}/\it{c}$ for the $\overline{\mathrm D}^0 \rightarrow  \mathrm{K}^+ \pi^- \pi^+ \pi^+$ channel). 
For all candidates, the $\overline{\mathrm D}^{0}$ and ${\mathrm D_s^-}$ meson decay distance, 
relative to the B vertex had to be positive. Events with an estimated error on the
$\Bs$ decay distance larger than 250~$\mu m$ and those having a vertex $\chi^2$--probability 
smaller than 10$^{-3}$ were removed. 
In order to reduce the combinatorial background from charm and light quarks, the b-tagging probability
for the whole event and for the hemisphere opposite to the 
reconstructed B meson had to be smaller than 0.1.
Additional selections which depend on the $\Bs$ decay channel were applied, mainly 
for  $\overline{\mathrm D}^0$ decays which suffer from a larger combinatorial background than ${\mathrm D_s^-}$ candidates: \\
 
$\bullet\underline {\Bs \rightarrow {\mathrm D_s^-} \pi^+ \; , \;\; \Bs \rightarrow {\mathrm D_s^-}\mathrm{a}_1^+} $\\
The momentum of the $\Bs$ had to be larger than 22~${\mathrm GeV}/\it{c}$. For $\Bs \rightarrow {\mathrm D_s^-} \mathrm{a}_{1}^{+} $ candidates, only the 
combination with the largest $\Bs$ momentum has been kept.   \\  

$\bullet\underline{\Bs \rightarrow \overline{\mathrm D}^{0} \mathrm{K}^- \pi^+ \; , \; \; 
\Bs \rightarrow \overline{\mathrm D}^{0} \mathrm{K}^-\mathrm{a}_1^+} $ \\ 
For $\Bs \rightarrow \overline{\mathrm D}^{0} \mathrm{K}^- \pi^+$ decays, the $\Bs$  momentum had to be larger than 27~${\mathrm GeV}/\it{c}$.
Because of a high combinatorial background for $\Bs \rightarrow \overline{\mathrm D}^{0} 
\mathrm{K}^-\mathrm{a}_1^+$ decays, 
the $\Bs$ momentum had to be larger than 29~${\mathrm GeV}/\it{c}$
in the $\overline{ {\mathrm D}}^{0} \rightarrow  \mathrm{K}^+ \pi^- $ channel 
and larger than 33~${\mathrm GeV}/\it{c}$
in the $\overline{ {\mathrm D}}^{0} \rightarrow  \mathrm{K}^+ \pi^- \pi^+ \pi^- $ decay channel. 
In each event, only the candidate with the largest $\Bs$ momentum was kept.
For the K$^-$ candidate, identified 
as ``standard'', the momentum 
had to be larger than 2~${\mathrm GeV}/\it{c}$.  The source of the $\overline{ {\mathrm D}}^{0}$ and ${\mathrm K}^-$
meson  pair can be an excited D meson state. 
 The constraint on the mass value of such a state 
allows more soft K$^-$ mesons to be considered. The selection on the  K$^-$ momentum was reduced to 1.5~${\mathrm GeV}/\it{c}$, if 
the invariant mass of the  
$\overline{\mathrm D}^0 \mathrm{K}^-$ pair was compatible with the mass of the 
orbitally excited ${\mathrm D}_{{\mathrm s}J}(2573)$ state (the most probable value of $J$ is 2~\cite{ref:book}).
One  event was detected, which is compatible with the decay channel 
${\mathrm B^0_s} \rightarrow {\mathrm D}_{{\mathrm s}J}(2573)^{-} \pi^{+}$, 
${\mathrm D}_{{\mathrm s}J}(2573)^{-} \rightarrow \overline{\mathrm D}^0 \mathrm{K}^{-}$:
 the momentum of the K$^-$ candidate is 1.8  ${\mathrm GeV/\it{c}}$ and 
the mass difference 
$M(\overline{\mathrm D}^0\mathrm{K}^-) - M(\overline{\mathrm D}^0)$ is  $716.0\pm2.1$~MeV$/{\it{c}}^2$.
The expected mass difference of $708.9\pm1.8$~MeV$/{\it{c}}^2$ is in good agreement with the 
observed one, taking into account the full width of this state which is $15^{+5}_{-4}$~MeV$/{\it{c}}^2$~\cite{ref:book}.
\begin{table}[ht]
\begin{center}
\begin{tabular}{|c|c|c|c|c|c|}
\hline
\hline
                                 \multicolumn{6}{|c|}{Main peak}                                        \\ \hline
\hline
     Data set  &     Mass     &   Width ($\sigma$)  &  ${\mathrm N}_{signal}$ & Comb. bkg & Reflection      \\ 
           &     ($\mathrm{GeV}/{\it{c}}^2$)    &   ($\mathrm{GeV}/{\it{c}}^2$)  &    &  \% & \%     \\ \hline
Real Data   & $ 5.373\pm0.016 $  & $0.029\pm0.012 $  &  $8\pm4$ & $27\pm16$ & -      \\  \hline
Simulation & $5.370\pm0.002$  & $0.037\pm0.002 $  &  $5\pm1$ & - &   $7\pm3$     \\  \hline
\hline
                                         \multicolumn{6}{|c|}{Satellite peak}                                   \\ \hline
\hline
Real Data   & $ 5.050\pm0.054 $  & $0.111\pm0.049 $  &  $15\pm8$ & $55\pm13$ & -      \\  \hline
Simulation & $ 5.099\pm0.007  $  & $0.148\pm0.007 $  &  $11\pm2$ & - &   $12\pm6$     \\  \hline
\end{tabular}
\caption[]{ \it { Characteristics of the  $\Bs$ signals 
and comparison with the simulation. The expected numbers of events  were 
calculated taking into account the different branching fractions as given in Table~\ref{tab:brachi} and the corresponding 
reconstruction efficiencies. These efficiencies vary from  $(0.2\pm0.1)\%$ 
for the $\Bs \rightarrow \overline{\mathrm D}^{0} \mathrm{K}^-\mathrm{a}_1^{+}$
with $\overline{\mathrm D}^0 \rightarrow  \mathrm{K}^+ \pi^- \pi^+ \pi^-$, up to $(10.2\pm0.8)\%$
for the $\Bs \rightarrow {\mathrm D_s^-} \pi^+$ with ${\mathrm D_s^-} \rightarrow \phi \pi^{-}$.
The fitted or expected number of signal events and the fraction of combinatorial background are given inside a mass window
corresponding to $\pm3\sigma$($\pm2\sigma$) for the main(satellite) peaks. In real data
 the  number of
events in these mass windows are $11(33)$.
 The $\Bs$
mass in the simulation is  $5.370~ \mathrm{GeV}/{\it{c}}^2$. In the last column
the fraction of physical background, discussed in detail in Section~\ref{sec:23}, is given. }}
\label{tab:tab1}
\end{center}
\end{table}

The mass spectrum, obtained by summing up the contributions from the different channels is shown in Figure~\ref{fig:bsexcl1}. 
The data are indicated by points with error bars and the
     fit result is shown by the solid line.
The mass distribution in the signal region was fitted using two Gaussian functions of different widths to account for 
the exclusive $\Bs$ signal (main narrow peak) and for the presence of partially reconstructed  $\Bs$ decays (satellite wider peak).
According to the expected branching fractions, it was assumed that the dominant contributions  
for the satellite peak come from the following decay channels: ${\mathrm D_s^{*-}} \pi^+$, 
${\mathrm D_s^{*-}} \mathrm{a}_{1}^{+}$, $\overline{\mathrm D}^{*}(2007)^{0} \mathrm{K}^{-} \pi^{+} $, 
${\mathrm \overline{D}}^{*}(2007)^{0} \mathrm{K}^{-} 
{a}_{1}^{+} $, ${\mathrm D_s^-} \rho^{+}$ and ${\mathrm D_s^{*-}} \rho^{+}$, where the cascade photon or neutral pion (or both) were not 
reconstructed. In the simulation, it was verified that the mass distribution of the satellite peak can be 
described by a Gaussian function as shown in Figure~\ref{fig:bse_sat}. The central values and the widths of the Gaussian 
functions (main and satellite peaks) 
in Figure~\ref{fig:bsexcl1} were left free to vary in the fit. 
The combinatorial background was fitted using an exponential 
function with a slope fixed according to the simulation. This slope was verified in  several ways. Using 2.9 million 
$\bb$ and 5.3 million $\qq$ simulated events, the expected mass spectrum was obtained after having 
removed the $\Bs$ signal contribution in the mass region of the satellite and main peaks.
 The sum of the contributions from b events, charm events and  light-quark events is shown in Figure~\ref{fig:bsexcl1} 
and is in agreement with that obtained in data.
 Two further checks were performed, using 
events selected in the side-bands of the ${\mathrm D_s^-}$ 
and $\overline{\mathrm D}^0$ candidates 
and from wrong sign combinations. 
The slopes of both distributions are in agreement with
those obtained by fitting the data and the simulation.  

The fit to the mass distribution yielded a signal of $8\pm4\; \Bs$ decays in the main peak and $15\pm8$ events 
in the satellite peak. 
The probability that the background has fluctuated to give the observed number of events in the main peak is $3\times 10^{-4}$.
Table~\ref{tab:tab1} gives  the characteristics of the observed signals and
the comparison with simulation. 
Specific decay channels from ${\mathrm B^0_s}$, ${\mathrm B^0_d}$, ${\mathrm B}^+$ and $\overline{\Lb}$  were
simulated with statistics ranging from ten to several thousand times the expected rates in real data. 
These samples were used to determine efficiencies and reflection probabilities 
discussed below.

\subsection{Reflections from $\Bd$ and $\overline{\Lb}$ decays }
\label{sec:23}
For several ${\mathrm B^0_s}$ decay channels a possible physical background originates from non-strange B decays, when 
one of the pions or proton is misidentified as a kaon (kinematic reflections).  The main decay channels are
$\Bd \rightarrow  {\mathrm D}^{-} \pi^+(\mathrm{a}_1^+)$,  ${\mathrm D}^{-} \rightarrow  {{\mathrm{K}}}^{*0} \pi^{-}$ (which can
contribute to $\Bs \rightarrow {\mathrm D_s^-} \pi^+(\mathrm{a}_1^+)$,  
$ {\mathrm D_s^-} \rightarrow  {{\mathrm{K}}}^{*0} \mathrm{K}^{-}$ candidates) 
and from $\Bd \rightarrow  \overline{\mathrm D}^0 \pi^- \pi^+$, 
$\overline{\mathrm D}^{0} \rightarrow $ ${\mathrm K}^+ \pi^-$ or
${\mathrm K}^+ \pi^- \pi^+ \pi^-$ (which can contribute to $\Bs \rightarrow \overline{\mathrm D}^0 \mathrm{K}^- \pi^+$ 
candidates).
In the mass region of the main peak, $ 0.32\pm0.13$ events are expected to originate from kinematic reflections 
(with a 19\% contribution from $\overline{\Lb}$ decays).
These estimates  were obtained using dedicated Monte Carlo samples of $\Bd$ and $\overline{\Lb}$ events passed through 
the  full reconstruction algorithms and  satisfying the selection criteria. The corresponding uncertainties come from 
the limited knowledge of the assumed branching fractions and from the simulated events statistics.

To study the contribution in the satellite peak from kinematic reflections, the same decay processes were considered with channels in which the D meson 
is accompanied by a $\rho$. This gives $ 1.3\pm0.7 $ events.

In order to look for possible signals coming from kinematic reflections in real data, 
$\Bs$ candidates were considered in turn as $\Bd$ or $\overline{\Lb}$ hadrons 
by changing the kaon into a pion or an antiproton, respectively. 
The  mass distributions obtained are similar to those expected from genuine 
simulated $\Bs$ mesons and do not show
any accumulation of events in the $\Bd$ or $\overline{\Lb}$ mass regions.

\subsection{Reconstruction of non-strange ${\mathrm B}$ meson decays}
\label{sec:24}

Non-strange ${\mathrm B}$ mesons, decaying into a $\overline{\mathrm D}^0$ and a small
number of pions, were  reconstructed in order to verify the $\Bs$ reconstruction 
algorithms. The ${\mathrm B}^+ \rightarrow \overline{\mathrm D}^{0} \pi^{+}$, 
$\Bd \rightarrow  {\mathrm D}^{*}(2010)^{-} \pi^{+}$  and  
$\Bd \rightarrow {\mathrm D}^{*}(2010)^{-} \mathrm{a}_{1}^{+}$
decay channels were studied and
their mass distributions are shown in Figures~\ref{fig:bsexcl6}a and~\ref{fig:bsexcl6}b.
They were fitted using a Gaussian function for the signal
and using an exponential function for the combinatorial background.
An additional Gaussian function was used in Figure~\ref{fig:bsexcl6}a
 to account for the signal coming from the following decay channels: 
${\mathrm{B^+ \rightarrow \overline{\mathrm D}^{*}(2007)^{0} \pi^{+}}}$,
 ${\mathrm B}^+ \rightarrow \overline{\mathrm D}^{0} \rho^{+}$, 
 ${\mathrm{B^+ \rightarrow \overline{\mathrm D}^{*}(2007)^{0} \rho^{+}}}$,
where the $\pi^0$ and/or $\gamma$  from the $\overline{\mathrm D}^{*}(2007)^{0}$ and/or $\rho$
decays were not reconstructed.
The main selection criteria which were imposed are rather similar to those used in the $\Bs$ analysis. 
The information relevant to these reconstructed channels is given in Table~\ref{tab:tab3a}.
The numbers of observed events are in agreement with expectations. 

\begin{table}[h]
\begin{center}
\begin{tabular}{|l|c|c|c|c|}
\hline
\hline
                   Channel                         &   B meson mass      &  $N_{obs}$              & $ N_{exp}$  \\
                                                   &    $ \mathrm{(GeV/{\it{c}}^2)}$     &                         &             \\ 
\hline \hline
 B$^+ \rightarrow  \overline{\mathrm D}^{0} \pi^{+}$  &  $5.278\pm0.008$    &  $ 24\pm5 $           & $ 28\pm 5$ \\ \hline  
 $\Bd \rightarrow  {\mathrm D}^{*}(2010)^{-} \pi^{+}~+~ {\mathrm D}^{*}(2010)^{-} \mathrm{a}_{1}^{+}$        &  $5.277\pm0.011$    &  $ 11\pm5 $            & $10\pm 4$         \\ \hline \hline  
$\Bd \rightarrow \mathrm{\overline{D}}^{0} \mathrm{\pi}^{-} \pi^{+}~+~\mathrm{\overline{D}}^{0} \mathrm{\pi}^{-} \mathrm{a}_{1}^{+}$ & $5.287\pm0.024$ & $ 8\pm4$  &  $-$   \\ \hline  
\end{tabular}
\caption[]{ \it { Characteristics of the non-strange ${\mathrm B}$ decay mode
 and comparison with the simulation. The error on the expected number of events
comes from the errors in the branching fractions and reconstruction efficiencies. 
The branching fraction of the  decay mode in the last row is estimated in this paper. }}
\label{tab:tab3a}
\end{center}
\end{table}

Finally two $\Bd$ decay channels, $\Bd \rightarrow \overline{\mathrm D}^{0} \pi^- \pi^+ $ and 
$\Bd \rightarrow \overline{\mathrm D}^{0} \pi^- \mathrm{a}_{1}^{+}$ 
 which are not yet well established, were considered.

In the strange B sector (Section \ref{sec:21}) they correspond to the $\Bs \rightarrow  \overline{\mathrm D}^0\mathrm{K}^{-} \pi^{+} $ and
$\Bs \rightarrow  \overline{\mathrm D}^0\mathrm{K}^{-}  \mathrm{a}_{1}^{+} $ decays. 
$\Bd$ mesons were reconstructed using the same selection criteria as  for the corresponding $\Bs$ decay channel,
but replacing a K meson by a $\pi$. All particles, not explicitly identified  as protons or kaons 
were accepted as pions. In order to remove the $\Bd \rightarrow  {\mathrm D}^{*}(2010)^{-} \pi^{+}$ and         
$\Bd \rightarrow  {\mathrm D}^{*}(2010)^{-} \mathrm{a}_{1}^{+}$ contamination, candidates were required to have a mass difference
$M(\overline{\mathrm D}^{0}\pi^-) - M(\overline{\mathrm D}^{0})$ larger than $0.16~\mathrm{GeV}/{\it{c}}^2 $. 

Figure~\ref{fig:bsexcl_bd} shows the mass distribution for these two decay channels.
It was fitted with an exponential function for the background and two Gaussian functions for the signals corresponding 
to the main and satellite peaks, with respective widths equal to about
34~${\mathrm MeV}/{\it{c}}^2$ and 60~${\mathrm MeV}/{\it{c}}^2$, as obtained from the simulation. 
The position of the wider Gaussian,
 corresponding to $\Bd \rightarrow  \overline{\mathrm D}^{*}(2007)^{0} \pi^{-} \pi^{+}$ and $\Bd \rightarrow  
\overline{\mathrm D}^{*}(2007)^0 \pi^{-}\mathrm{a}_{1}^{+} $ signals with unreconstructed $\pi^0$ or $\gamma$, and the slope of the 
background exponential function were also taken from the simulation. The fit to the mass distribution yielded a signal 
of $8\pm4\; \Bd$ decays in the main peak. 
The probability that the background has fluctuated to give the observed number of events  is 
$3\times 10^{-2}$.

The number of observed events in the main peak can be translated into a branching fraction:
 ${\mathrm Br}(\Bd \rightarrow \mathrm{\overline{D}}^{0} \mathrm{\pi}^{-} \pi^{+}~+~\mathrm{\overline{D}}^{0} 
\mathrm{\pi}^{-} \mathrm{a}_{1}^{+} ) = (3.6\pm1.9)\times 10^{-3}.$ 
 The quoted error is completely dominated by statistics.
The CLEO Collaboration has given an upper limit~\cite{ref:book} only on 
the decay channel with a pion pair: 
${\mathrm Br}(\Bd \rightarrow \mathrm{\overline{D}}^{0} \mathrm{\pi}^{-} \pi^{+})~<~1.6\times 10^{-3}$ 
(at the 90\% C.L.). This limit is consistent with those which can be extracted from the measured  
channels
in this paper:
\begin{center}
 ${\mathrm Br}(\Bd \rightarrow \mathrm{\overline{D}}^{0} \mathrm{\pi}^{-} \pi^{+})~<~1.6\times 10^{-3}$ at  the $90\%$ C.L. \\ 
 ${\mathrm Br}(\Bd \rightarrow \mathrm{\overline{D}}^{0} \mathrm{\pi}^{-} {\mathrm a_{1}}^{+})~<~5.4\times 10^{-3}$ 
at the 90\% C.L. 
\end{center}
For this evaluation it was assumed (see Section~\ref{sec:21}) that the decay channel with ${\mathrm a_1^+}$ 
is produced with a rate which is ($3.3\pm0.9$) times larger than the channel with $\pi^+$. 


\subsection{ Study of ${\mathrm B^0_s}$-$\overline{\mathrm{B^0_s}}$ oscillations}
\label{sec:25}

\subsubsection{Algorithm for tagging b or $\overline{\mathrm b}$ quark at production time}
\label{sec:251}
The signature of the initial production of a b $(\overline{\mathrm b})$ quark
in the jet containing the ${\mathrm B^0_s}$ or $\overline{\mathrm B_s^0}$ candidate was determined using a combination of different variables
sensitive to the initial quark state. 
For each individual variable $X_i$, the probability density functions
$f_{\mathrm b}(X_i)$ $(f_{\overline{\mathrm b}}(X_i))$ for b ($\overline{\mathrm b}$) quarks were built
and the ratio $R_i = f_{\overline{\mathrm b}}(X_i)/f_{\mathrm b}(X_i)$ was computed. The combined
tagging variable is defined as:
\begin{equation}
\label{eq1}
x_{tag}  =  \frac{1 - R} {1 + R}~~, \: \mbox{where} \; \; R =  \prod R_i .
\end{equation}
The variable $x_{tag}$ varies between -1 and 1. 
Large and positive  values of $x_{tag}$ correspond 
to a high probability that 
the produced quark is a b rather than a $\overline{\mathrm b}$ in a given hemisphere. 

A set of nine discriminant variables has been selected for this
analysis. One set (three variables) is determined in the hemisphere which contains the
${\mathrm B_s^0}$ meson, the other set (five variables) in the hemisphere opposite
to the ${\mathrm B_s^0}$ meson, and one variable is common to both hemispheres.
Details concerning the definition of these variables and the method are given in reference~\cite{ref:dsl}. 

An event was classified as a mixed or as an unmixed candidate according to the
sign, $Q_D$, of the $\rm{D}_{s}$ electric charge for decay channels containing a $\rm{D}_{s}$ meson,
or according to the sign of the kaon for ${\mathrm{\overline{\mathrm D}^0\mathrm{K}^- \pi^+ }}$ decay channels, 
relative to the sign of the 
$x_{tag}$ variable. Mixed candidates are defined by requiring $x_{tag} \cdot Q_D < 0$, and unmixed
candidates correspond to $x_{tag} \cdot Q_D > 0$. 
The probability, ${\epsilon}_b^{tag}$, for tagging correctly 
the b or $\overline{\mathrm b}$ quark, from the measurement of $x_{tag}$, 
was evaluated by using a dedicated simulated event sample and was found to be (74.5$\pm0.5)\%$.

The corresponding probability for events in the combinatorial background was obtained using real
 data candidates selected in the side-bands of the ${\mathrm D}$ signal: the probabilities to classify 
these events as mixed or unmixed candidates are called ${\epsilon}^{mix}_{comb}$ 
and ${\epsilon}^{unmix}_{comb}$, respectively.
For the reflection events the analogous probabilities 
are called ${\epsilon}^{mix}_{ref}$, ${\epsilon}^{unmix}_{ref}$ and their values were taken from
the simulation.

\subsubsection{Proper time resolution}
\label{sec:252}
For each event, the $\Bs$ proper decay time was obtained from the measured decay length ($L_{\Bs}$) and the estimate 
of the $\Bs$ momentum ($p_{\Bs}$). The measured position of the ${\mathrm D_s^-}$ or $\overline{\mathrm D}^0$ decay vertex, 
the momentum, and the
corresponding measurement errors, were used to reconstruct a ${\mathrm D_s^-}$ or a $\overline{\mathrm D}^0$ particle.
A candidate $\Bs$ decay vertex was obtained by intercepting the trajectory of this particle with
the other charged particle tracks which are supposed to come from the $\Bs$ decay vertex. 

For the main peak, the ${\Bs}$ momentum is precisely known since all decay products are reconstructed.
For the satellite peak, this is no longer true, because there are one or two undetected neutral(s) ($\pi^0$ and/or $\gamma$) in 
the $\Bs$ decay. As discussed in Section~\ref{sec:22}, the satellite peak is 
assumed to be composed of ${\mathrm D_s^{*-}} \pi^+$,  ${\mathrm D_s^{*-}} \rm{a}_{1}^+$,
${\mathrm D_s^-} \rho^{+}$, ${\mathrm D_s^{*-}} \rho^{+}$ and ${\mathrm \overline{D}}^{*}(2007)^{0} \mathrm{K}^{-} \pi^{+} $,
${\mathrm \overline{D}}^{*}(2007)^{0} \mathrm{K}^{-}\mathrm{a}_{1}^{+} $ decays. The first four 
decay channels give ${\mathrm D_s^-} \pi^+(\rm{a}_{1}^+$)  and the last two channels give $  \mathrm{\overline{D}}^0 \mathrm{K}^{-} 
\pi^{+}(\mathrm{a}_{1}^{+}) $ final states. Kinematic mass constrained  fits (imposing the mass
of $\Bs$ and of the intermediate states with corresponding errors) were performed assuming 
always the presence of a $\gamma$ coming from the ${\mathrm{{\mathrm D_s^{*-}}}}$ or $\overline{\mathrm D}^{*}(2007)^{0}$ meson decays. 
This is a good approximation also for decay channels containing a $\rho^+$ meson. 
These mass constrained fits were performed on events lying in the mass region 
corresponding to $\pm 2\sigma$ of the fitted 
mass of the satellite peak. 

Except for the combinatorial background contribution, the predicted proper time
distributions were obtained by convoluting the theoretical functions with
resolution functions evaluated from simulated events. Due to the different decay length resolutions 
(for the different Vertex Detector configurations), proper time resolutions were used for data samples taken in 1992-1993 and 1994-1995 separately. 
The proper time resolution, ${\cal R}_{\mathrm{B_s}}(t-t_{i})$, was evaluated from
the difference between the generated
time ($t$) and the reconstructed time ($t_{i}$), fitting this distribution by
the sum of two Gaussian functions:
\be
{\cal R}_{\mathrm{B_s}}(t-t_{i}) = (1-f_2)G_1(t-t_i,\sigma_1)+f_2G_2(t-t_i,\sigma_2)
\eeq
where $G_1(t-t_i,\sigma_1)$ and $G_2(t-t_i,\sigma_2)$  are Gaussian functions 
with the corresponding  resolutions $\sigma_1$ and $\sigma_2$. In the general case, the $\sigma_i$ resolutions
 depend on the momentum uncertainty (see equation~(\ref{eq:damp})). But in this analysis, 
as the $\Bs$ momentum is known with 
a good precision,   the contribution of the momentum uncertainty to the proper 
time resolution is negligible. 
$f_{2}$ is the fraction of the second Gaussian function, 
which, by convention, is that with the worse resolution. 
The values of the corresponding parameters, obtained from simulated events, are given in Table~\ref{tab:tab4}.
The time resolution for events in the satellite peak 
is only slightly worse than for those belonging to the main peak.

\begin{table}[h]
\begin{center}
\begin{tabular}{|c|c|c|c|c|}
\hline
\hline
                                 \multicolumn{5}{|c|}{Main peak}                                        \\ \hline
\hline
        Decay channels  &     Years     &   $\sigma_{1}({\mathrm ps})$  &  $\sigma_{2}({\mathrm ps})$ &$ f_2 $      \\ \hline
all $\Bs$ channels   &   1992-1993    &    0.068            &       0.18        & 0.27        \\  \hline
all $\Bs$ channels   &   1994-1995     &    0.065            &       0.12        & 0.42        \\  \hline
\hline
                                         \multicolumn{5}{|c|}{Satellite peak}                                   \\ \hline
\hline
${\mathrm D_s^{*-}}\pi^+(\rm{a}_{1}^+)$, $\overline{\mathrm D}^{*}(2007)^{0} \mathrm{K}^-\pi^+(\rm{a}_{1}^+)$
                                        &  1992-1993     &    0.066            &       0.15        & 0.48        \\  \hline
${\mathrm D_s^{*-}}\pi^+(\rm{a}_{1}^+)$, $\overline{\mathrm D}^{*}(2007)^{0} \mathrm{K}^-\pi^+(\rm{a}_{1}^+)$
                                        &  1994-1995     &    0.081            &       0.17        & 0.30        \\  \hline
${\mathrm D_s^-}\rho^+ $,  ${\mathrm D_s^{*-}}\rho^+$     &  1992-1993     &    0.085            &       0.21        & 0.65         \\  \hline
${\mathrm D_s^-}\rho^+ $,  ${\mathrm D_s^{*-}}\rho^+ $    &  1994-1995     &    0.092            &       0.20        & 0.57        \\  \hline
\end{tabular}
\caption[]{ \it { Proper time resolution from simulation for 
${\mathrm 1992}$-${\mathrm 1993}$ and ${\mathrm 1994}$-${\mathrm 1995}$ data sets
and for the different decay channels. It has been parameterised by the sum of two Gaussian functions.  }}
\label{tab:tab4}
\end{center}
\end{table}

The time distribution ${\cal P}_{comb}(t_{i})$ for the combinatorial background under the main peak is obtained 
from real data by fitting the time distribution of wrong-sign and right-sign events lying in the side-bands of the $\Bs$ 
mass  distribution. It was verified, in the simulation, that the time distribution for these classes of events 
is similar to that obtained for events lying under the $\Bs$ mass peak. 
The time distribution ${\cal P}_{comb}(t_{i})$ for the combinatorial 
background under the satellite peak was taken directly from the simulation, 
since it has a  dependence on
the measured $\Bs$ mass, due to the procedure used to reconstruct the B momentum for these events.
The time distribution  ${\cal P}_{ref}(t_{i})$ for the reflection was also taken from 
the simulation.

\subsubsection{Fitting procedure}
\label{sec:253}

The oscillation analysis is  performed in the framework of the amplitude 
method~\cite{ref:amplitude}, which consists in measuring, for each value of 
the frequency $\Delta m_{\Bs}$, an amplitude $A$ and its error 
$\sigma(A)$. The parameter $A$ is introduced in the time evolution of pure ${\mathrm B^0_s}$ 
or $\overline{\mathrm B}^{0}_{\mathrm s}$ states, so that the value $A=1$ corresponds to 
a genuine signal for oscillation. The time-dependent probability that $\Bs$
is detected as a  $\Bs$ or $\overline{\mathrm B}^{0}_{\mathrm s}$ is then:
\begin{eqnarray}
{\cal P}^{unmix(mix)}(t)~
=~\frac{1}{2 \tau_{\Bs}}\times \exp({-t}/{\tau_{\Bs}}) \times
 (1 \pm A \; \cos(\Delta m_{\Bs} t))
\end{eqnarray}
where the  plus (minus) signs refer to $\Bs$ ($\overline{\Bs}$) decays.
The 95\%~C.L. excluded region for $\Delta m_{\Bs}$ is obtained by evaluating the probability that, in at 
most 5\% of the cases, a real signal having an amplitude equal to unity 
would give an observed amplitude 
smaller than the one measured. This corresponds to the condition:
$$A(\Delta m_{\Bs}) + 1.645~\sigma(A(\Delta m_{\Bs}))~<~1.$$
In the amplitude approach, the error $\sigma(A(\Delta m_{\Bs}))$ is related to
the probability to exclude a given value of $\Delta m_{\Bs}$.
The sensitivity is defined as the value of $\Delta m_{\Bs}$ which would just be
excluded, if the value of $A$, as measured in the experiment, was zero for all
values of $\Delta m_{\Bs}$, i.e. it is the expected 95\% confidence limit that an experiment of
the same statistics and resolution would be expected to achieve (a real
experiment would set a higher or lower limit, because of fluctuations in the
values of $A$ as a function of $\Delta m_{\Bs}$). 

The  probability distributions for mixed and unmixed\footnote{In the following, only the probability 
distribution for mixed events is written explicitly; the corresponding probability for unmixed events can be
obtained by changing $\epsilon$ into $(1-\epsilon)$.} 
events are:
\begin{eqnarray}
P^{mix}(t_i) = f_{\mathrm{B_s}} P_{\mathrm{B_s}}^{mix}(t_i)+ f_{ref} P_{ref}^{mix}(t_i) + f_{comb} P_{comb}^{mix}(t_i)
\end{eqnarray}
where $ f_{\mathrm{B_s}}$, $ f_{ref}$ and $f_{comb}$ are the relative fractions of the $\Bs$, 
reflection and combinatorial background events, respectively, which satisfy the condition  
 $ f_{\mathrm{B_s}}$+$ f_{ref}$+$f_{comb}$=1.
The expressions
for the different probability densities are:
\begin{itemize}
\item $\Bs$ mixing probability:
 \begin{eqnarray}
   \begin{array}{ll}
    P_{\mathrm{B_s}}^{mix}(t_i) &= \{~~{\epsilon}_b^{tag}  {\cal P}_{\mathrm{B_s}}^{mix}(t) +
(1-{\epsilon}_b^{tag})  {\cal P}_{\mathrm{B_s}}^{unmix}(t)~~\} \otimes {\cal R}_{\mathrm{B_s}}(t-t_i)
   \end{array}
 \end{eqnarray}

\item Reflection mixing probability:
 \begin{eqnarray}
     P_{ref}^{mix}(t_i) = {\epsilon}^{mix}_{ref}  {\cal P}_{ref}(t_i)
 \end{eqnarray}

\item Combinatorial background mixing probability:
 \begin{eqnarray}
     P_{comb}^{mix}(t_i) = {\epsilon}^{mix}_{comb}  {\cal P}_{comb}(t_i)
 \end{eqnarray}
\end{itemize}
The parameters ${\epsilon}_{\mathrm b}^{tag}$, ${\epsilon}^{mix}_{ref}$ 
and ${\epsilon}^{mix}_{comb}$ were defined in Section~\ref{sec:251}.  

The amplitude analysis is then performed using all events selected in a mass region between 4.83 and 5.46~GeV$/{\it{c}}^2$. 
The variation of the background level as a function of the reconstructed mass was included in this analysis on an event-by-event basis. 
Figure~\ref{fig:bsexcl_dms} shows the variation of the measured amplitude as a function of $\Delta m_{\Bs}$. 
With the present level of the statistics, this analysis  provides a  negligible low limit.
On the other hand, its most important feature, 
despite the low statistics, is the  relatively small error on the amplitude at high values of $\Delta m_{\Bs}$ with respect 
to the more inclusive analyses.
Due to the limited statistics, the error on $A$  ($\sigma(A)$) can be asymmetric. 
The error has been symmetrized by taking the larger value.
Figure~\ref{fig:bsexcl_err}a shows the variation of the measured error on the 
amplitude as a function of $\Delta m_{\Bs}$.
In Figure~\ref{fig:bsexcl_err}b the  results from the 
${\mathrm D_s^{\pm}} \ell^{\mp}$~\cite{ref:dsl} 
and from the ${\mathrm D_s^{\pm}} {\mathrm h}^{\mp}$ (see the second part
of the current paper) are compared with the exclusive $\Bs$ analysis. 
It should be noted that, due to the better proper time resolution, the resolution $\sigma(A)$  
of the $\Bs$ exclusive  analysis
increases more slowly. The ratio of the corresponding errors of  $\Bs$ exclusive
and ${\mathrm D_s^{\pm}} \ell^{\mp}$ analyses 
is about 5 (2) at the low (high) values of  $\Delta m_{\Bs}$.
%

The behaviour of $\sigma(A)$ was investigated using a ``toy'' Monte Carlo generated with the same characteristics as those measured
in real data. The individual toy experiments show a similar behaviour of $\sigma(A)$, as in the real data, 
as a function of $\Delta m_{\Bs}$.
For each value of $\Delta m_{\Bs}$, the distribution of $\sigma(A)$ and its variance was obtained. The central value and the 
region corresponding to a $\pm2 \sigma(A)$ variation are also shown in Figure~\ref{fig:bsexcl_err}a.

\subsubsection{Study of systematic uncertainties}
\label{sec:254}

Systematic uncertainties were evaluated by varying the parameters, which 
were kept constant in the fit according to their measured or expected errors.

\begin{itemize}
\item { Systematics from the tagging purity: } \\
a variation of $\pm$3\% on the expected tagging purity for the signal was considered, 
following the results given in~\cite{ref:dsl}.

\item { Systematics from the background level and kinematic reflection:} \\
the levels of background and of kinematic reflection  were varied separately
for the main peak and for the satellite peak according to the statistical uncertainties  given in Table~\ref{tab:tab1}. 
Since the level of the combinatorial background was used as a function of the reconstructed 
mass on an event-by-event basis, the measured central mass positions and the corresponding widths were also
varied by $\pm1\sigma$ around their fitted values.

\item { Systematics from the expected resolution on the B decay proper time: } \\
the widths  $\sigma_{1}$ and $\sigma_{2}$ described in Section~\ref{sec:252}
and given in Table~\ref{tab:tab4} were varied by $\pm10\%$~\cite{ref:dsl,ref:bspapertau,ref:oldbs}.

\item {Finally, uncertainties in the values of the $\Bs$ meson lifetime ($\tbs=1.46\pm0.06$~ps~\cite{ref:osciwg})
and in the parameterisation of the proper time of 
the combinatorial background  were taken into account.}
\end{itemize} 
In Figure~\ref{fig:bsexcl_dms}, which presents the variation of the measured amplitude as a function of $\Delta m_{\Bs}$,
the shaded area shows the contribution from systematics.

\section{ $\rm{D}_{s}^{\pm} {\mathrm h}^{\mp}$ analysis with fully reconstructed $\Ds$}
\label{sec:3}

Events with an exclusively reconstructed $\Ds$ accompanied by one or more large momentum 
hadrons were used to perform a second oscillation analysis. 
This channel is similar to the ${\mathrm D_s^{\pm}} \ell^{\mp}$ final state~\cite{ref:dsl} but, instead of a charged lepton, 
it uses charged hadrons. It provides a larger number of events but suffers
 from an ambiguity in the choice of the hadrons and from a reduced $\Bs$ purity of the selected sample.
This approach has already been  used in DELPHI to measure the $\Bs$ lifetime~\cite{ref:bspapertau}.

\subsection{Event selection }
\label{sec:31}
The $\Ds$ meson is selected in two decay channels:
\begin{eqnarray*}
\begin{array}{ll}
\Dsn \rightarrow \phi \pi^-, & \phi \rightarrow \mathrm{K}^+\mathrm{K}^-;
\\
\Dsn \rightarrow \mathrm{K}^{*0} \mathrm{K}^-, &
\mathrm{K}^{*0} \rightarrow \mathrm{K}^{+}\pi^-.
\\
\end{array}
\end{eqnarray*}

$\Ds$ candidates were reconstructed by making combinations of three charged particles located in the 
same event hemisphere,
 with a momentum larger than 1~{$\Gec$} and with reconstructed tracks associated to at least one VD hit.  
The value of $|\cos\psi|$ (see Section~\ref{sec:22})
was required to be larger than 0.4 and 0.6 in the $\Dsn \rightarrow \phi \pi^-$ and $\Dsn \rightarrow \mathrm{K}^{*0} \mathrm{K}^-$
decay channels, respectively. 
In order to reduce the combinatorial background from charm and light quarks, the b-tagging probability
for the whole event was required to be smaller than 0.5.

In this analysis the neural network algorithm  for hadron identification
described in Section~\ref{sec:hadid} was used.
For the $\Dsn \rightarrow \phi \pi^-$ decay mode, the invariant mass of $\phi$ candidates was required to be within $\pm12$~{\Mecqt} 
of the nominal $\phi$ mass and the $\phi$ momentum was required to be larger than 5~{$\Gec$}.
If the ${\mathrm K}^+ \mathrm{K}^-$ invariant mass was within  $\pm4${~\Mecqt} of the 
nominal $\phi $ mass,  both kaon candidates were required  to be identified 
as ``very loose'' kaons,  otherwise to be identified  
as ``loose'' kaons. 

For the $\Dsn \rightarrow \mathrm{K}^{*0} \mathrm{K}^-$ decay mode, the invariant mass of the 
${\mathrm K}^{*0}$ candidates was required to be within $\pm60$~{\Mecqt} of the nominal 
${\mathrm K}^{*0}$ mass value and the ${\mathrm K}^{*0}$ momentum was required to exceed 5~{$\Gec$}.
The momentum of the K$^-$ candidate from  ${\mathrm D_s^-}$ was required to exceed 3~{$\Gec$}. 
To suppress the physical
background from the ${\mathrm D}^{-} \rightarrow \mathrm{K}^{+} \pi^{-} \pi^{-}$ kinematic reflection,
the K$^-$ candidate was required to be identified as a ``standard'' kaon. 
For ${\mathrm K}^{*0} \rightarrow \mathrm{K}^{+}\pi^{-}$ decays, ``loose'' identified ${\mathrm K}^+$ were also accepted. 
In order to suppress the combinatorial background,
the K$^-$ candidate was required to be identified as ``tight'' kaon, if
the invariant mass of the 
${\mathrm K}^{*0}$ candidates was out of $\pm20$~{\Mecqt} of the nominal 
${\mathrm K}^{*0}$ mass value, and the value of $|\cos\psi|$  was required to be larger than 0.8.  
 
The selection of a hadron accompanying the $\Ds$ candidate is based on an impact parameter technique.
A sample of tracks coming predominantly from B hadron decays was preselected 
by using their impact parameters and the corresponding errors, both with respect to 
the primary vertex ($Im_{p}$, $\sigma_{{p}}$)  and to the secondary $\Ds$ decay 
vertex ($Im_s$, $\sigma_{{s}})$.
The hadron was then searched for amongst the preselected 
particles in the event, 
by requiring that its charge was opposite to the $\Ds$ charge and that it had the largest momentum. 
The efficiency of the hadron selection was about 80\% and,
among the selected hadrons, $(84\pm4)\%$ came from a B vertex.  
Details on the track preselection as well as on the hadron selection are given in reference~\cite{ref:bspapertau}. 

The B decay vertex was reconstructed by fitting the selected hadron and the $\Ds$ candidate to a common vertex.  
The $\chi^2$--probability of the fitted $\Bs$  vertex has been required to be larger than  10$^{-3}$. 
In order to increase the resolution on the measured decay length, only reconstructed events with 
a decay length error smaller than 0.07~cm were kept. 

If the previous procedure failed, a new attempt was made, 
using an inclusive algorithm which allowed several hadrons to be attached to the $\Ds$ candidate.
This algorithm is based on the difference in the rapidity distributions for particles coming from
fragmentation and from B decays.
The fragmentation particles, on average, have  lower rapidity~\cite{ref:mf} than B decay products.
The charged particles were ordered in increasing values of the rapidity and were attached in turn to the secondary 
$\Ds$ vertex. Up to three particles were accepted  with their total charge equal to $\pm$1 or 0.
The rapidity was calculated as $0.5 \log{((E+P_{L})/(E-P_{L}))}$, where $E$ is the energy of the particle
(assumed to be a pion)
and $P_{L}$ its longitudinal momentum with respect to the thrust axis of the event.
Only particles with
a momentum greater than 1 GeV/$\it{c}$ were accepted.
In addition,   
for the tracks satisfying the condition $Im_{p}/\sigma_{{p}}<3$, 
it was required that $ Im_s/\sigma_{{s}}< Im_{p}/\sigma_{{p}}$. Events
with a decay length error smaller than 0.07~cm and a  $\chi^2$--probability of the fitted $\Bs$  vertex 
larger than  10$^{-3}$ were kept. 

\begin{table}[h] 
\begin{center}
  \begin{tabular}{|l|c|c|} \hline
 $\Ds$ decay channels & $\Ds$ signal in 1992-1993 data & $\Ds$ signal in 1994-1995 data \\ \hline\hline 
\hbox{${\mathrm D_s^-}\rightarrow \phi\pi^-$}       
                                                 & $322\pm30~(0.60\pm0.04)$  & $468\pm42~(0.53\pm0.04)$    \\ \hline
\hbox{${\mathrm D_s^-} \rightarrow \mathrm{K}^{*0} \mathrm{K}^-$}  
                                                 & $152\pm28~~(0.70\pm0.06)$  & $324\pm35~(0.58\pm0.05) $  \\ \hline 
\end{tabular}
\end{center}
\caption{ \it  Number of $\Ds$ mesons reconstructed in the $\phi \pi^{-}$ and ${\mathrm K}^{*0}\mathrm{K}^-$
decay channels after selection of the accompanying hadron(s). 
The fraction of combinatorial background, given in parentheses, 
has been evaluated using a mass interval of 
$\pm2\sigma$ centred on the fitted $\Ds$ mass. }
\label{tab2}
\end{table}

The selected sample, specified as $\rm{D}_{s}^{\pm} {\mathrm h}^\mp$ in
the following, contained about $30\%$ of such ``multi-hadron''  events. Among them,
about $20\%$ have one,  $50\%$ two and $30\%$  three selected hadrons.  
The multi-hadron and single-hadron events  were treated in a similar way. 
Figure~\ref{fig:dsh1} shows the $\Ds$ signals after selection of the accompanying hadron(s).
The mass distributions  were fitted with two Gaussian functions of
equal widths to account for the ${\mathrm D_s^-}$ and D$^-$ signals and with an exponential 
function for the combinatorial background.
All parameters were allowed to vary in the fit.
Table~\ref{tab2} gives the number of observed events in the $\Ds$ signal, after background subtraction,  
and the fraction of combinatorial background. 
The $\Ds$ signal region corresponds to a mass interval of $\pm$2$\sigma$ centred on the 
fitted $\Ds$ mass.

The reconstructed number of ${\mathrm D_s^-}\rightarrow \phi\pi^-$ events
using 1994-1995 data  is about 1.5 larger  than those obtained using 1992-1993 data.
This factor  reflects the difference in statistics between the two data sets. 
Such ``statistical scaling'' does not apply to the ${\mathrm D_s^-} \rightarrow \mathrm{K}^{*0} \mathrm{K}^-$ 
decay mode, where the particle identification, which was better for the 1994-1995 data
set (see Section~\ref{sec:delphi_det}),
 plays a more important role.


Four events are in common with the exclusively
reconstructed $\Bs$ sample: one event in the main peak and three in the satellite peak.
These events were removed from the $\rm{D}_{s}^{\pm} {\mathrm h}^\mp$ sample
for the oscillation analysis.

\subsection {Sample composition}
\label{sec:32}

The  $\rm{D}_{s}^{\pm} {\mathrm h}^\mp$ sample contained a large physical background 
due to $\Ds$ from non-strange B hadron decays and from $\cc$ fragmentation. 
Four sources of events, originating from $\rm{B}$ decays,
 were considered: two from $\Bs$  and two from non-$\Bs$ mesons,
namely, $\rm{B}$ decaying to one  or two charmed mesons comprising at least one $\Ds$.
The relative fractions of these sources were calculated using five input parameters:
\begin{itemize}
\item
   Br(b $\rightarrow \rm{D}_{s}^{\pm}$X) at LEP~\cite{lepds};
\item
   Br(b $\rightarrow \rm{D}_{s}^{\pm}$X) at $\Upsilon(4S)$~\cite{cleodl};
\item
   Br(b $\rightarrow \Bsb$) at LEP~\cite{ref:book};
\item
    the probability in the non-strange $\rm{B}$ meson, that a $\rm{D}_{s}^{-}$ is produced at the lower vertex:
     Br($\mathrm{B_{u,d}}$ $\rightarrow \rm{D}_{s}^{-}$X)~\cite{ref:book1};
\item
   the probability to have two charmed hadrons in a b-decay: Br(b $\rightarrow $D$\overline{\rm{D}}$X)~\cite{ref:book2}.
\end{itemize}
The last  two probabilities were assumed to be the same for all $\rm{B}$ species.
To estimate the first two branching fractions ,
the averaged production rate of $\rm{D}_{s}$ from all B species,

${\mathrm Br}({\mathrm b} \rightarrow  {\mathrm D_s^{\pm}} X)\times {\mathrm Br}({\mathrm D_s^{\pm}} \rightarrow  \phi \pi^{\pm})$,
measured by the ALEPH, DELPHI and OPAL collaborations~\cite{lepds}, and the equivalent quantity 
${\mathrm Br}(\mathrm{B_{u,d}} \rightarrow  {\mathrm D_s^{\pm}} X) \times {\mathrm Br}({\mathrm D_s^{\pm}} \rightarrow  \phi \pi^{\pm})$, 
measured at the $\Upsilon(4S)$ by the CLEO and ARGUS collaborations \cite{cleodl}, were  used. 
The following  fractions for the different ${\mathrm B}$ decays contributing to the $\rm{D}_{s}^{\pm} {\mathrm h}^\mp$ 
signal were evaluated:
\begin{itemize}
\item
Fraction of $\Bs$ decaying to a  $\Ds$ and  no other charmed meson: $F_{\mathrm{B_s},{1{\mathrm D}}}$=($39\pm7$)\%.
\item
Fraction of $\Bs$ decaying to a $\Ds$ and  another charmed meson: $F_{\mathrm{B_s},{2{\mathrm D}}}$=($11\pm3$)\%.
\item
Fraction of B mesons (non-$\Bs$) decaying to a  $\Ds$ and  no other charmed meson: $F_{{\mathrm B},{1{\mathrm D}}}$=($9\pm5$)\%.
\item
Fraction of B mesons (non-$\Bs$) decaying to  a $\Ds$ and another charmed meson: $F_{{\mathrm B},{2{\mathrm D}}}$=($41\pm7$)\%.
\end{itemize}

The contribution $F_{\mathrm cc}$ from direct charm was estimated from 
the measurement 
of $\Ds$ production in charm events at LEP~\cite{lepds},  
taking into account  the $\Zz$ partial  widths into b  and c 
quark pairs: $F_{\mathrm cc}$=($27\pm5$)\%. 

Finally, the relative proportion  of the combinatorial background $f_{bkg}$ 
was taken directly from the fit of the real data  mass distributions
(see Table~\ref{tab2}).

\subsection{ Discriminant variables to increase the $\Bs$ purity}
\label{sec:33}
To increase the effective purity in $\Bs$ of the selected sample, five variables 
were used which allow the separation of the signal and background components.
These variables are: the $\Ds$ mass, the $\Ds$ momentum, the value of $|\cos\psi|$, 
the  $\chi^2$--probability of the fitted $\Ds$ decay vertex and the value of the b-tagging 
variable measured in the 
hemisphere opposite to that of the $\Ds$ meson. 

The relative components in the selected sample, defined in the previous 
section,  were calculated on an event-by-event basis:
\bes
  f^{eff}_{\mathrm{B_s},{1{\mathrm D}}}            & = &   
  F_{\mathrm{B_s},{1{\mathrm D}}}(1-f_{\mathrm cc}-f_{bkg}){\cal F}_{\mathrm b}({\mathrm b-tag})
\prod_{i=1,4} {\cal F}_{\mathrm bc}(v_{i})/tot          \\
  f^{eff}_{\mathrm{B_s},{2{\mathrm D}}}            & = &   
  F_{\mathrm{B_s},{2{\mathrm D}}}(1-f_{\mathrm cc}-f_{bkg}){\cal F}_{\mathrm b}({\mathrm b-tag}) 
\prod_{i=1,4} {\cal F}_{\mathrm bc}(v_{i})/tot          \\
  f^{eff}_{{\mathrm B},{1{\mathrm D}}}            & = &   
  F_{{\mathrm B},{1{\mathrm D}}}(1-f_{\mathrm cc}-f_{bkg}){\cal F}_{\mathrm b}({\mathrm b-tag}) 
\prod_{i=1,4} {\cal F}_{\mathrm bc}(v_{i})/tot  \\ 
  f^{eff}_{{\mathrm B},{2{\mathrm D}}}            & = &   
  F_{{\mathrm B},{2{\mathrm D}}}(1-f_{\mathrm cc}-f_{bkg}){\cal F}_{\mathrm b}({\mathrm b-tag}) 
\prod_{i=1,4} {\cal F}_{\mathrm bc}(v_{i})/tot  \\  
  f_{\mathrm cc}^{eff}            & = &   
  f_{\mathrm cc}{\cal F}_{\mathrm c}({\mathrm b-tag}) \prod_{i=1,4} {\cal F}_{\mathrm bc}(v_{i})/tot          \\  
  f_{bkg}^{eff}            & = &   
  f_{bkg} \prod_{i=1,5} {\cal F}_{bkg}(v_{i})/tot          \nonumber
\ees
where $v_{i}$ indicates the $i$-th discriminant variable, ${\cal F}_{\mathrm bc}$, ${\cal F}_{\mathrm b}$, 
${\cal F}_{\mathrm c}$ and
${\cal F}_{bkg}$ are the probability density functions for the b and c together, b, c 
and the combinatorial background  events, respectively. The relative charm contribution is 
$f_{\mathrm cc}=F_{\mathrm cc}(1-f_{bkg})$.
In these expressions,  the total  normalisation factor is:
$$
tot=f^{eff}_{\mathrm{B_s},{1{\mathrm D}}}+f^{eff}_{\mathrm{B_s},{2{\mathrm D}}}
+f^{eff}_{{\mathrm B},{1{\mathrm D}}}+f^{eff}_{{\mathrm B},{2{\mathrm D}}}+f^{eff}_{\mathrm cc}+f^{eff}_{bkg}.
$$
All discriminant variables, except for the b-tagging, were used to separate 
b and c events together from the combinatorial background ($bkg$). 
The  b-tagging variable was used to distinguish separately the b from c  and from 
combinatorial background events.  
The distributions of discriminant variables are shown in Figure~\ref{fig:dsh_disc} for events selected in the
signal region, after having subtracted the corresponding distributions of background events  obtained using 
events within the side-bands of the ${\mathrm{D_s}}$ signal. 
The corresponding distributions for background events are also shown in Figure~\ref{fig:dsh_disc}.
For the b-tagging variable, comparison with simulation is shown for the sum of the b and c events.   
The agreement between real data and simulated distributions is satisfactory.

The use of this procedure is equivalent to increase the $\Bs$ purity by 20\%. 

\subsection{ Measurement of the ${\mathrm B^0_s}$ lifetime}
\label{sec:34}

\subsubsection{Proper time resolution}
\label{sec:341}

For each event, the $\Bs$ proper decay time was obtained from the measured
decay length ($L_{\Bs}$) and the estimate of the $\Bs$ momentum ($p_{\Bs}$).
The technique used is described in~\cite{ref:bspapertau}. 
The predicted decay time distributions were obtained by convoluting the theoretical 
distributions with resolution functions evaluated from simulated events.
Different parameterisations were used for the two Vertex
Detector configurations installed in 1992-1993 and 1994-1995.
The proper time resolution was obtained from the ($t$-$t_i$) distribution of the difference 
between the generated ($t$) and the reconstructed ($t_{i}$) proper times.
The following distributions were considered:
\begin{itemize}
\item ${\cal R}_{{\mathrm B},{1{\mathrm D}}}(t-t_{i})$ is the resolution
 function for ${\mathrm B}$ decays
with only one charmed meson ($\rm{D}_{s}$) in the final state. It is parameterised as 
the sum of three Gaussian functions. The width of the second Gaussian is taken
to be proportional to the width of the first one. The third Gaussian
describes the component with a selected hadron coming from the primary
vertex; the fraction ($f_3$) of these events decreases exponentially as a function of the proper time:
\be
\ba{l}
  {\cal R}_{{\mathrm B},{1{\mathrm D}}}(t-t_{i}) = (1-f_2-f_3)G_1(t-t_i,\sigma_1)+ 
               f_2G_2(t-t_i,\sigma_2)+ 
               f_3G_3(t-t_i,\sigma_3)
\ea
\eeq
where $G_1(t-t_i,\sigma_1)$, $G_2(t-t_i,\sigma_2)$ and $G_3(t-t_i,\sigma_3)$ are Gaussian functions with
  $\sigma_1 = \sqrt{\sigma_{L1}^2+(\sigma_{p1}/p)^2 \times t^2}$,
$\sigma_2 = s_1\sigma_1$, $\sigma_3 = \sqrt{\sigma_{L3}^2+(\sigma_{p3}/p)^2 \times t^2}$;
$f_{2}$ and $f_{3}$ are the fractions of the second and third Gaussian functions,
respectively;  $f_3$ is defined as $f_3 = \exp(s_2-s_3t)$.  

The values for the decay length resolutions, $\sigma_{Lj}$, the  momentum
resolutions, $\sigma_{pj}/p$, the relative fractions, $f_j$, and the coefficients $s_j$, 
are given in Table~\ref{tab:b}.

\item ${\cal R}_{{\mathrm B},{2{\mathrm D}}}(t-t_{i})$ is the resolution function for ${\mathrm B}$ decays
with two ${\mathrm D}$ mesons in the final state. In this case, the selected hadron often
does not originate directly from the B vertex, but from the second D vertex, hence
 resulting in a worse resolution.
This resolution function is parameterised in a similar way as ${\cal R}_{{\mathrm B},{1{\mathrm D}}}(t-t_{i})$
and the values of the corresponding parameters are shown in Table~\ref{tab:b}.
\end{itemize}
\begin{table}[hbt]
\bc
\bt{|l|c|c|c|c|c|c|c|c|} \hline 
  \multicolumn{9}{|c|} {$\rm{D}_{s}^{\pm}{\mathrm h}^\mp$ sample} \\ \hline
  \mbox{Resol. function  (years) } &$\sigma_{L1}({\mathrm ps})$&$\sigma_{p1}/p$
  &$\sigma_{L3}({\mathrm ps})$&$\sigma_{p3}/p$&$f_2$&$s_1$&$s_2$ &$s_3$ (ps$^{-1}$) 
   \\ \hline \hline
   ${\cal R}^{\mathrm B}_{1{\mathrm D}}(t-t_{i})$ (1992-1993)   & 0.149 & 0.140  & 0.144 & 0.386 & 0.15 & 3.5 
   & -1.54 & 0.14 \\  \hline
   ${\cal R}^{\mathrm B}_{1{\mathrm D}}(t-t_{i})$ (1994-1995)   & 0.145 & 0.104  & 0.169 & 0.256 &0.10 & 2.5 
   & -1.87 & 0.17 \\  \hline \hline
   ${\cal R}^{\mathrm B}_{2{\mathrm D}}(t-t_{i})$ (1992-1993)    & 0.236 & 0.095 & 0.144 & 0.386 & 0.30 & 3.5 
   & -1.21 & 0.14 \\  \hline
   ${\cal R}^{\mathrm B}_{2{\mathrm D}}(t-t_{i})$ (1994-1995)    & 0.214 & 0.094 & 0.169 & 0.256 & 0.25 & 3.5 
   & -1.30 & 0.17 \\   \hline
\et
\ec
\caption{ \it Fitted values of the parameters used in the resolution functions 
${\cal R}_{{\mathrm B},{1{\mathrm D}}}(t-t_{i})$ and ${\cal R}_{{\mathrm B},{2{\mathrm D}}}(t-t_{i})$.}
\label{tab:b}
\end{table}

\subsubsection{ Likelihood fit}
\label{sec:342}
The $\Bs$ mean lifetime and  the time distribution of the combinatorial background 
were fitted simultaneously, 
using selected events lying within a mass interval of $\pm2\sigma$
centred on the measured $\Ds$ mass (2953 events)  and  events lying in the
$\Ds$ mass side-band (3373 events) between 2.1 and 2.3 \Gecqt.
The probability density function for the measured proper time, $t_i$, can be written as:
\be
P(t_i) =   f^{eff}_{\mathrm{B_s},{1{\mathrm D}}}     P_{\mathrm{B_s},{1{\mathrm D}}}(t_i) 
         + f^{eff}_{\mathrm{B_s},{2{\mathrm D}}}     P_{\mathrm{B_s},{2{\mathrm D}}}(t_i) 
         + f^{eff}_{{\mathrm B},{1{\mathrm D}}}       P_{{\mathrm B},{1{\mathrm D}}}(t_i)
         + f^{eff}_{{\mathrm B},{2{\mathrm D}}}       P_{{\mathrm B},{2{\mathrm D}}}(t_i) \\  
         + f_{\mathrm cc}^{eff}                P_{\mathrm cc}(t_i) 
         + f_{bkg}^{eff}               P_{bkg}(t_i). \nonumber  
\eeq
The different probability densities are expressed as convolutions of the 
physical probability densities with the appropriate resolution functions:

\begin{itemize}
\item 
For the signal:
\bes
   P_{\mathrm{B_s},{1{\mathrm D}}}(t_i) &=& \frac {1}{\tau_{\mathrm{B^0_s}}} \exp(-t/\tau_{\mathrm{B^0_s}}) \otimes 
   {\cal R}_{{\mathrm B},{1{\mathrm D}}}(t-t_i) \\
   P_{\mathrm{B_s},{2{\mathrm D}}}(t_i) &=& \frac {1}{\tau_{\mathrm{B^0_s}}}
 \exp(-t/\tau_{\mathrm{B^0_s}}) \otimes 
   {\cal R}_{{\mathrm B},{2{\mathrm D}}}(t-t_i),
\ees
where $t$ is the true proper time.
\item For the physical background coming from all B meson decays: 
\bes
   P_{{\mathrm B},{1{\mathrm D}}}(t_i)   &=&  \sum_{q \neq {\mathrm s}} \frac {1}
{\tau_{\mathrm{\mathrm B_q}}} 
f_{{\mathrm {\mathrm B_q}},{1{\mathrm D}}}~\exp(-t/\tau_{\mathrm B_q}) \otimes 
   {\cal R}_{{\mathrm B},{1{\mathrm D}}}(t-t_i) \\
   P_{{\mathrm B},{2{\mathrm D}}}(t_i)   &=&  \sum_{q \neq {\mathrm s}} \frac {1}
{\tau_{\mathrm{\mathrm B_q}}}f_{{\mathrm {\mathrm B_q}},{2{\mathrm D}}}~\exp(-t/\tau_{\mathrm B_q}) \otimes 
  {\cal R}_{{\mathrm B},{2{\mathrm D}}}(t-t_i). 
\ees
where $q$ runs over the various b-hadron species contributing to this background and  $f_{{\mathrm B_q},{1{\mathrm D}}}$, 
$f_{\mathrm {\mathrm B_q},{2{\mathrm D}}}$ are their corresponding fractions.

\item 
For the combinatorial background, the following function was used:
  \bes
  \ba{ll}
  P_{bkg}(t_i) = &  f^-   \exp(t/\tau^-)\otimes G(t-t_i,\sigma_-) +
                    f^+   \exp(-t/\tau^+)\otimes G(t-t_i,\sigma_+) + \\
                 &   (1-f^--f^+)                G(t-t_i,\sigma_0)
  \ea
  \ees
 Three distributions were considered: an exponential for poorly measured 
 events having negative $t$ (with lifetime $\tau^-$), an exponential for the flying background 
 (with lifetime $\tau^+$) and a central Gaussian for the non-flying one. 
 The seven parameters ($f^-$, $f^+$, $\tau^+$, $\tau^-$, $\sigma_-$, $\sigma_+$, $\sigma_0$) 
were allowed to vary in the fit.

\item 
For ``charm'' candidates, the function $ P_{\mathrm cc}(t_i)$ has the same form as $ P_{bkg}(t_i)$ and has been parameterised 
using simulated events. In this case all the parameters were fixed in the fit.
\end{itemize}

The $\Bs$ lifetime fit was performed in the proper time 
interval between -4~ps and 12~ps 
and the  result of the fit, shown in Figure~\ref{fig:dsh_tau}, is:
\begin{eqnarray*}
\tau_{\Bs} = \;\;1.53 \;^{+0.16}_{-0.15} \; ({\mathrm stat.})\;{\mathrm ps}.
\end{eqnarray*} 

\subsubsection{Systematic uncertainties on the $\Bs$ lifetime}
\label{sec:343}

The contributions to systematic uncertainties on the $\Bs$ lifetime measurement are summarised 
in Table~\ref{tab2_ol}.

\renewcommand\arraystretch{1.4}
\begin{table}[t]
\begin{center}
  \begin{tabular}{|c|c|} \hline
 Source of systematic uncertainty & $\tau_{\Bs}$ variation (${\mathrm ps}$)  
\\ \hline  \hline
   \hbox{Sample composition}                & $^{+0.013}_{-0.016}$    \\
   \hbox{$f_{bkg}$ }                        & $^{+0.046}_{-0.050}$    \\
   \hbox{${\mathrm{B_s}}$ purity}               & $^{+0.005}_{-0.015}$    \\
   \hbox{$t$ resolution              }      & $\pm$0.019    \\ 
   \mbox{$\tau_{{\mathrm B}^+}(1.65\pm0.03~{\mathrm ps})~\cite{ref:osciwg}$}  & $\pm$0.016 \\
   \mbox{$\tau_{\Bd} (1.56\pm0.03~{\mathrm ps})~\cite{ref:osciwg}$}  & $\pm$0.014 \\  \hline
   \hbox{Analysis bias correction}          & $\pm$0.040              \\ \hline  \hline
   \hbox{Total}                             & $\pm{0.07} $   \\ \hline  
  \end{tabular}
\end{center}
\caption{\it{ Sources of systematic uncertainties on the $\Bs$ lifetime.}}
\label{tab2_ol}
\end{table}
\renewcommand\arraystretch{1.0}

The systematic error due to the uncertainties in the relative fractions of the different $\Ds$ sources corresponds to
a $\pm1 \sigma $ variation of the fractions $f$ used in the
likelihood fit, excluding $f_{bkg}$ which is studied separately. 

The estimate of systematics related to the evaluation of the
$\Bs$ purity, on an event-by-event basis, was obtained in the following way.
The distributions of the different quantities contributing to the discriminant variable 
(Figure~\ref{fig:dsh_disc}) for signal and background events were re-weighted with a linear 
function in order to maximise the  agreement between data and the simulation.
The  ${\mathrm{B_s}}$ lifetime distribution was refitted with new probability distributions and the 
difference between the corresponding values of the fitted lifetime  taken as a systematic error.

Uncertainties on the determination of the resolution of the proper time receive 
two contributions: one from errors on the decay distance evaluation and the other
from errors on the measurement of the $\Bs$ momentum.
The systematic error coming from uncertainties on the time resolution 
was evaluated by varying  the widths 
$\sigma_{L} $ and $\sigma_{p}$   of the resolution 
function by $\pm$10\%~\cite{ref:dsl,ref:bspapertau,ref:oldbs}.
Finally, simulated $\Bs$ events, generated with a lifetime of $1.6 \; {\mathrm ps}$
and satisfying the same selection criteria as the real data, have a fitted
lifetime of $1.64\pm0.04 \; {\mathrm ps}$. The  $\Bs$ lifetime value obtained 
in the Section~\ref{sec:342}  was not corrected but
the statistical error of this comparison ($\pm$0.04~ps) was included in the systematic error.
The measured $\Bs$ lifetime  was found to be:
$$    \tau_{\Bs}  =  1.53^{+0.16}_{-0.15}({\mathrm stat.}) \pm{0.07}({\mathrm syst.})~{\mathrm ps}. $$ 

A similar analysis  by the ALEPH collaboration~\cite{ref:dsh_al} gave a 
consistent result: $\tau_{\Bs}  =  1.47\pm{0.14}({\mathrm stat.}) \pm{0.08}
({\mathrm syst.})~{\mathrm ps}. $

\subsection{Lifetime difference between ${\mathrm B^0_s}$ mass eigenstates}
\label{sec:35}

The $\Bs$ (or $\Bsb$) mesons are superpositions of the two mass eigenstates:
\bes
|\mathrm{B^0_s} \rangle = {\frac{1}{ \sqrt{2}}}(| {\mathrm B}^0_H \rangle + | {\mathrm B}^0_L \rangle)~~;~~
|\mathrm{\overline{B}^0_s} \rangle = {\frac{1}{ \sqrt{2}}}(| {\mathrm B}^0_H \rangle - | {\mathrm B}^0_L \rangle).
\ees
Neglecting CP violation, the time probability density is then given by:
\begin{eqnarray}
   {\cal P}(t) = (1-x_{\mathrm cp})\frac{\Gamma_{\mathrm H}\Gamma_{\mathrm L}}{\Gamma_{\mathrm H}+\Gamma_{\mathrm L}} \; 
(e^{-\Gamma_{\mathrm H}t}+e^{-\Gamma_{\mathrm L}t}) +{x_{\mathrm cp}}\Gamma_{\mathrm L} e^{-\Gamma_{\mathrm L}t}
\label{dG_Form}
\end{eqnarray}
where
$\Gamma_{\mathrm L} = \Gamma_{\Bs} + \Delta \Gamma_{\Bs}/2$, 
$\Gamma_{\mathrm H} = \Gamma_{\Bs} - \Delta \Gamma_{\Bs}/2$.          

The first term of equation  (\ref{dG_Form}) refers to final states
of identified beauty flavor, as in the $\Dsl$ case~\cite{ref:dsl}, it does include 
$\Ds$+light mesons final states. 
The second term corresponds to ${\mathrm D_s^{(*)+}} {\mathrm D_s^{(*)-}}$ final
states, which are dominantly ($ 98 \%$) CP even eigenstates~\cite{ref:aleksan}.
The value of $x_{\mathrm cp}$ was taken as the ratio of $F_{\mathrm{B_s},{2{\mathrm D}}}$ to
($F_{\mathrm{B_s},{1{\mathrm D}}}$+$F_{\mathrm{B_s},{2{\mathrm D}}}$): 
$x_{\mathrm cp}=0.22\pm0.07$ (see Section~\ref{sec:32}).

Two 
variables are then considered: $\tau_{\Bs}$ (${1/\Gamma_{\Bs}}$)
and $\dgbs$.
As the statistics in the sample is not sufficient to fit
simultaneously $\tbs$ and $\dgbs$,
the method used to evaluate $\dgbs$ consists in calculating 
the log-likelihood 
for the time distribution measured with the 
${\mathrm D_s^{\pm}} {\mathrm h}^\mp$ sample
and deriving the probability density function for $\dgbs$ by
constraining $\tau_{\Bs}$ to be equal to the $\Bd$ lifetime 
($\tau_{\Bd} = 1.56\pm0.03$~ps~\cite{ref:osciwg} and
$\tau_{\Bs}/\tau_{\Bd} = 1\pm{\cal O}(0.01)$ is predicted in \cite{ref:bigi}).

The log-likelihood function described in Section~\ref{sec:342}
was modified by replacing the probability density for $\Bs$
by (\ref{dG_Form}) convoluted with the appropriate resolution functions.
It was minimized in the $(\tau_{\Bs},\dgbs)$ plane 
and the likelihood difference with respect to its minimum  $\Delta {\cal L}$ 
(Figure~\ref{fig:dsh_dg}-a) was computed:
\bes
   \Delta{\cal L} = -\log{\cal L}_{tot}(\tbs,\dgbs) + 
   \log{\cal L}_{tot}((\tbs)^{min},(\dgbs)^{min}) \; .
\ees
The probability density function for the variables $\tbs$ and $\dgbs$ is proportional to:
\begin{eqnarray*}
   {\cal P}(\tbs,\dgbs) \propto e^{-\Delta {\cal L}}.
\end{eqnarray*}
The $\dgbs$ probability distribution was then obtained by convoluting
${\cal P }(\tbs,\dgbs)$ with the probability density function
$f_{\scriptsize({\tau_{\Bs}=\tau_{\Bd}})}(\tbs)$, expressing
the constraint $\tbs = \tbd$, and normalising the result to unity:
\begin{eqnarray*}
   {\cal P}(\dgbs) = { \int {{\cal P}(\tbs,\dgbs) f_{\scriptsize({\tau_{\Bs}=\tau_{\Bd}})}(\tbs) d\tbs}  \over 
    \int {{\cal P}(\tbs,\dgbs) f_{\scriptsize({\tau_{\Bs}=\tau_{\Bd}})}(\tbs) d\tbs d\dgbs}}  
\end{eqnarray*}
where
\begin{eqnarray*}
   f_{\scriptsize({\tau_{\Bs}=\tau_{\Bd}})}(\tbs)  = 
   {{1}/({\sqrt{2\pi}\sigma_{\tiny\tbd}})}\exp({-{{(\tbs-\tbd)^2}/{2\sigma^{2}_{\tiny\tbd}}}}) .
\end{eqnarray*}
The upper limit on $\dgbs$, calculated from ${\cal P}(\dgbs)$, is:
\bes
   \dgbs < 0.67~\mbox{at the 95\% C.L.}
\ees
This limit takes into account both statistical uncertainties and the systematic 
coming from the uncertainty\footnote{The uncertainty due to
the theoretical prediction of the equality of the $\tau_{\Bs}$ and
 $\tau_{\Bd}$ lifetimes is negligible with respect to the
present error on $\tau_{\Bd}$.} on the $\Bd$ lifetime.

The systematic uncertainty originating from other sources was evaluated
by convoluting the probability function ${\cal P}(\tbs,\dgbs)$
with the probability function of the corresponding parameters: $  {\cal P}(\dgbs) =$ 
\bes
\small
 {
  { \int {\cal P}(\tbs,\dgbs,x^1_{sys}, ...,x^n_{sys})
   f_{\scriptsize({\tau_{\Bs}=\tau_{\Bd}})}(\tbs) 
   f({x^1_{sys}}) ...f_({x^n_{sys}})
   d\tbs dx^1_{sys} ...dx^n_{sys}}
  \over
  { \int {\cal P}(\tbs,\dgbs,x^1_{sys}, ...,x^n_{sys})
   f_{\scriptsize({\tau_{\Bs}=\tau_{\Bd}})}(\tbs) 
   f({x^1_{sys}}) ...f_({x^n_{sys}})
   d\tbs dx^1_{sys} ...dx^n_{sys}} d\dgbs }
\ees
where $x^i_{sys}$ are the $n$ parameters considered in the systematic uncertainty and
$f(x^i_{sys})$ are the corresponding probability densities.

Only three systematics were considered here: the relative fraction of combinatorial
background $f_{bkg}$, the $\Bs$ purity of the selected
sample and the  $x_{\mathrm cp}$ fraction.
Other systematic uncertainties are expected to be small 
as they are in the lifetime measurement.

The $\dgbs$ probability distribution, obtained with the inclusion of the systematics,
is shown in Figure~\ref{fig:dsh_dg}-c,
the most probable value for $\dgbs$ is $0.35$ and  the upper limit at the 95\% confidence 
level is:
\bes
   \dgbs < 0.69~\mbox{at the 95\% C.L.}
\ees
It should be noted that the world average of the $\Bs$ lifetime 
cannot be used as a constraint in such an analysis,
since it depends on $\Delta\Gamma_{\mathrm B^0_s}$ and on
$\Gamma_{\mathrm B^0_s}$. Moreover, this dependence is also  different for
different decay channels.
In the ${\mathrm D_s^{\pm}} {\mathrm h}^\mp$  case the expression of the average $\Bs$ lifetime
is given by:
\begin{equation}
   \tau_{\Bs}({\mathrm D_s^{\pm}} {\mathrm h}^\mp)  = 
(1-x_{\mathrm cp})\frac{1 + (\frac{1}{2}\dgbs)^2}{\Gamma_{\Bs} (1 - (\frac{1}{2}\dgbs)^2)}
    + x_{\mathrm cp}/\Gamma_{\mathrm L} \; .
\end{equation}

\subsection{Study of ${\mathrm B^0_s}$-$\overline{\mathrm{B^0_s}}$ oscillations}
\label{sec:36}

\subsubsection{Tagging procedure}
\label{sec:361}
The tagging algorithm was explained in Section~\ref{sec:251}. 
An event is classified as a mixed or an unmixed candidate according to the relative
signs  of the $\rm{D}_{s}$ electric charge, $Q_{D}$, and of the tagging purity variable, $x_{tag}$. 

%
%

Mixed candidates have $x_{tag} \cdot Q_D < 0$, and unmixed
ones $x_{tag} \cdot Q_D > 0$. 
The probability, ${\epsilon}_b^{tag}$, of tagging the b or the 
$\overline{\mathrm b}$ quark correctly from the measurement of $x_{tag}$ was 
evaluated using a dedicated simulated event sample. The average tagging purity of the $x_{tag}$ variable, given by the
 simulation for true ${\mathrm B_s^0} \rightarrow  \mathrm{D^-_s h}^+ X$ decays, is $(71.4\pm0.4)\%$.
The purity is lower than that obtained in the ${\mathrm{D_s}} \ell$ sample~\cite{ref:dsl} 
because not all ${\mathrm{B_s}}$ charged decay
products are reconstructed in the present analysis. It was verified that the tagging purity is the same for different B
hadron species and varies by less than about $\pm 2\%$ whether the $\Bs$ has  oscillated or not. This effect is taken 
into account in the systematics. The corresponding probability distribution for events  in the 
combinatorial background was obtained using data candidates 
selected in the side-bands of the $\Ds$ signal: the probabilities of classifying 
these events as mixed or as unmixed candidates are called 
${\epsilon}^{mix}_{bkg}$ and ${\epsilon}^{unmix}_{bkg}$, respectively.
For the charm events, the analogous probabilities 
are called ${\epsilon}^{mix}_{\mathrm cc}$, ${\epsilon}^{unmix}_{\mathrm cc}$ and their values were taken from the simulation.

\subsubsection{Fitting procedure}
\label{sec:362}

From the expected proper time distributions and the tagging probabilities,
the probability functions for mixed and unmixed event candidates are 
\be
P^{mix}(t_i) =   f^{eff}_{\mathrm{B_s},{1{\mathrm D}}}     P^{mix}_{\mathrm{B_s},{1{\mathrm D}}}(t_i) 
         + f^{eff}_{\mathrm{B_s},{2{\mathrm D}}}     P^{mix}_{\mathrm{B_s},{2{\mathrm D}}}(t_i) 
         + f^{eff}_{{\mathrm B},{1{\mathrm D}}}       P^{mix}_{{\mathrm B},{1{\mathrm D}}}(t_i)
         + f^{eff}_{{\mathrm B},{2{\mathrm D}}}       P^{mix}_{{\mathrm B},{2{\mathrm D}}}(t_i) \\  
         + f_{\mathrm cc}^{eff}                P^{mix}_{\mathrm cc}(t_i) 
         + f_{bkg}^{eff}               P^{mix}_{bkg}(t_i), \nonumber  
\eeq
where $t_i$ is the reconstructed proper time. 
The analytical expressions for the different probability densities are given in the following, with $t$ being the true 
proper time:
\begin{itemize}
\item ${\mathrm B^0_s}$ signal mixing probability:
 \begin{eqnarray}
   \begin{array}{ll}
   P^{mix}_{\mathrm{B_s},{1{\mathrm D}}}(t_i)   &= \{~~{\epsilon}_b^{tag}  
    {\cal P}^{mix}_{\mathrm{B_s},{1{\mathrm D}}}(t) +
(1-{\epsilon}_b^{tag})  {\cal P}^{unmix}_{\mathrm{B_s},{1{\mathrm D}}}(t)~~\} \otimes 
   {\cal R}_{{\mathrm B},{1{\mathrm D}}}(t-t_i)
   \end{array}
 \end{eqnarray}
\item Physical background mixing probabilities:
\renewcommand\arraystretch{1.1}
 \begin{eqnarray}
  \begin{array}{ll}{
  P^{mix}_{\mathrm{B_s},{2{\mathrm D}}}(t_i) = \{  
     f_{\mathrm{B_s},{2{\mathrm D}}}{\epsilon}_b^{tag}/\tau_{\mathrm{B^0_s}}
     \exp(-t/\tau_{\mathrm{B^0_s}}) \}
\otimes    {\cal R}_{{\mathrm B},{2{\mathrm D}}}(t-t_i)} \\
   \end{array}
 \end{eqnarray}
\renewcommand\arraystretch{1.0}
\renewcommand\arraystretch{1.1}
 \begin{eqnarray}
   \begin{array}{ll}
 P^{mix}_{{\mathrm B},{1{\mathrm D}}}(t_i) = \{ & 
  f_{{\mathrm{B_d}},{1{\mathrm D}}} (~{\epsilon}_b^{tag} 
  {\cal P}^{mix}_{\mathrm{B_d},{1{\mathrm D}}}(t) + 
   (1-{\epsilon}_b^{tag})  {\cal P}^{unmix}_{\mathrm{B_d},{1{\mathrm D}}}(t)~) + \\ 
&    f_{{\mathrm B}^{+},{1{\mathrm D}}}(1-{\epsilon}_b^{tag})/\tau_{{\mathrm B}^+}
     \exp(-t/\tau_{{\mathrm B}^+}) +  \\
&    f_{\Lambda_{\mathrm b},{1{\mathrm D}}}(1-{\epsilon}_b^{tag})/\tau_{\Lambda_{\mathrm b}}
     \exp(-t/\tau_{\Lambda_{\mathrm b}}) \hspace{1cm} \}
\otimes    {\cal R}_{{\mathrm B},{1{\mathrm D}}}(t-t_i) 
   \end{array}
 \end{eqnarray}
\renewcommand\arraystretch{1.0}
\renewcommand\arraystretch{1.1}
 \begin{eqnarray}
   \begin{array}{ll}
 P^{mix}_{{\mathrm B},{2{\mathrm D}}}(t_i) = \{ & 
  f_{{\mathrm{B_d}},{2{\mathrm D}}} (~{\epsilon}_b^{tag} 
  {\cal P}^{unmix}_{\mathrm{B_d},{2{\mathrm D}}}(t) + 
   (1-{\epsilon}_b^{tag})  {\cal P}^{mix}_{\mathrm{B_d},{2{\mathrm D}}}(t)~) + \\ 
&    f_{{\mathrm B}^{+},{2{\mathrm D}}}(1-{\epsilon}_b^{tag})/\tau_{{\mathrm B}^+}
     \exp(-t/\tau_{{\mathrm B}^+}) +  \\
&    f_{\Lambda_{\mathrm b},{2{\mathrm D}}}(1-{\epsilon}_b^{tag})/\tau_{\Lambda_{\mathrm b}}
     \exp(-t/\tau_{\Lambda_{\mathrm b}}) \hspace{1cm} \}
\otimes    {\cal R}_{{\mathrm B},{2{\mathrm D}}}(t-t_i) 
   \end{array}
 \end{eqnarray}
\renewcommand\arraystretch{1.0}
\item Mixing probability for charm component:
 \begin{eqnarray}
     P_{\mathrm cc}^{mix}(t_i) = {\epsilon}^{mix}_{\mathrm cc}  
  {\cal P}_{\mathrm cc}(t_i) 
 \end{eqnarray}
\item Combinatorial background mixing probability:
 \begin{eqnarray}
     P_{bkg}^{mix}(t_i) = {\epsilon}^{mix}_{bkg}  
  {\cal P}_{bkg}(t_i) 
 \end{eqnarray}
\end{itemize}

The oscillation analysis was performed in the framework of the amplitude
method~\cite{ref:amplitude} as described in Section~\ref{sec:253}.
Considering only statistical uncertainties, the limit is: 
\begin{eqnarray}
   \begin{array}{ll}
~~~~~~~~~~\dms > 4.2~\mbox{ps}^{-1}~\mbox{at the 95\% C.L.} & \\
   \end{array}
 \end{eqnarray}
with a corresponding sensitivity equal to  $ 3.1~\mbox{ps}^{-1}$.
At $\dms$ = 10 ${\mathrm ps}^{-1}$, the error on the amplitude is 2.3 (see Figure~\ref{fig:dsh_dms}).

\subsubsection{Study of systematic uncertainties}
\label{sec:363}

Systematic uncertainties were evaluated by varying the parameters which 
were  kept constant in the fit, according to their measured or 
expected errors.

\begin{itemize}

\item { Systematics from the tagging probability: } \\
a conservative variation of $\pm$3\% on the expected tagging probability for the signal 
and for the other three processes contributing to the $\Dsh$ candidates was used.
 The same variation is assumed for the tagging purity
for the charm and combinatorial background events.
 The central values of these purities were fixed to the simulated ones.
%

\item { Systematics from the $\Bs$ purity:}\\
the same procedure already applied to the lifetime measurement has been used.

\item { Systematics from the resolution on the B decay proper time: } \\
the same procedure already applied to the lifetime measurement was used.
In addition, the systematic error due to the variation of the proper 
time distribution of the combinatorial background
was considered: the parameters used to define the background shape
in the lifetime fit were varied according to their fitted errors.
\end{itemize}

The inclusion of systematic uncertainties lowers the sensitivity 
to $2.7~{\mathrm ps}^{-1}$ and the 95\%~C.L limit becomes $\Delta m_{\Bs} > 4.1~{\mathrm ps}^{-1}$.

 The analogous analysis has been performed by the ALEPH collaboration~\cite{ref:dsh_al}, which
set a limit at  $\Delta m_{\Bs} > 3.9~{\mathrm ps}^{-1}$ at the 95\%~C.L., with a
 better sensitivity of $4.1~{\mathrm ps}^{-1}$.

\section{ Combined limit on $\Delta m_{\Bs}$ using exclusive $\Bs$ and 
 ${\mathrm D_s^{\pm}} \rm{h}^{\mp}$ events}
\label{sec:4}
This paper presents two analyses on $\Delta m_{\Bs}$ using   
 exclusively reconstructed $\Bs$ mesons and  $\rm{D}_{s}^{\pm} {\mathrm h}^{\mp}$ events.
Their results have been  combined (Figure~\ref{fig:bsdsh_dms}),
 taking into account correlations between systematic uncertainties in the two 
amplitude measurements~\cite{ref:osciwg}.  
A limit at the 95\% confidence level is obtained: 
\begin{eqnarray}
   \begin{array}{ll}
~~~~~~~~~~\dms > 4.0~\mbox{ps}^{-1}~\mbox{at the 95\% C.L.} & \\
   \end{array}
 \end{eqnarray}
with a corresponding sensitivity equal to $ 3.2~\mbox{ps}^{-1}$
(with statistical errors only, the limit would be $~\dms > 4.0~\mbox{ps}^{-1}$ at the 95\%~C.L. 
with a sensitivity of $4.4~\mbox{ps}^{-1}$).
 Figure~\ref{fig:bsdsh_dms} also shows the error on the amplitude for different values of $\dms$.
%

\section{ Combination of the DELPHI  $\Delta m_{\Bs}$, $\tau_{\Bs}$ and $ \dgbs$ analyses}
\label{sec:5}
DELPHI has performed three analyses on  $\dms$ using  ${\mathrm D_s^{\pm}} \ell^{\mp}$
 candidates~\cite{ref:dsl}, 
 ${\mathrm D_s^{\pm}} \rm{h}^{\mp}$ events and exclusively reconstructed $\Bs$ mesons.
They were combined, taking into account correlations between the
event samples and between systematic uncertainties 
in the different amplitude measurements (Figure~\ref{fig:delphi_dms}a).
This gives the following limit for $\dms$:
\bes
   \ba{c}
     \dms~>~4.9~{\mathrm ps}^{-1}~\mbox{at the 95\% C.L.} \\
\mbox{with a sensitivity of}~~~ \dms = 8.7 ~ \mbox{ps}^{-1}
   \ea
 \ees
The exclusion probability for this limit is $88\%$.
The variation, with $\dms$  of the error on the amplitude is given in Figure~\ref{fig:delphi_dms}b. 

The results of  two  analyses on  $\tau_{\Bs}$ and $ \dgbs$  using  ${\mathrm D_s^{\pm}} \ell^{\mp}$
and ${\mathrm D_s^{\pm}} \rm{h}^{\mp}$ events provide the following DELPHI results:

$$
\tau_{\Bs}  = 1.46\pm{0.11}~{\mathrm ps}
$$
$$
 \dgbs < 0.45~\mbox{at the 95\% C.L.}
$$

\section{ Conclusion }
\label{sec:6}
Using about 3.5 million hadronic $\Zz$ decays registered by
DELPHI between 1992 and 1995, two samples of events have been selected.
The first one consists of 44 reconstructed $\Bs$
events: 11 candidates (including 30\% of background) are completely reconstructed and 33 candidates
(including 55\% of background) are partially reconstructed ($\pi^0$ and/or $\gamma$ are not detected).
This analysis used twelve different decay channels of the $\Bs$ meson and is a first attempt to use such events
for the oscillation studies. Due to the excellent proper time resolution, this sample gives some contribution 
in the high $\Delta m_{\Bs}$ region. 

The second sample contains 2953 D$_s^{\pm} {\mathrm h}^\mp$ candidates (including 60\% of background)
with completely reconstructed ${\mathrm D_s^-}$ mesons in the $\phi \pi^{-}$ and ${\mathrm K}^{*0} \mathrm{K}^-$
decay channels. Using the D$_s^{\pm} {\mathrm h}^\mp$ sample, three studies have been performed.
The $\Bs$ lifetime has been measured and a limit on the fractional width
difference between the two physical $\Bs$ states has been obtained:
$$
\tau_{\Bs}  = 1.53^{+0.16}_{-0.15}({\mathrm stat.})\pm{0.07}({\mathrm syst.})~{\mathrm ps}
$$
$$
 \dgbs < 0.69~\mbox{at the 95\% C.L.}
$$
This last result has been obtained under the hypothesis that 
$\tau_{\Bs}=\tau_{\Bd}$.

Combining the two studies on $\Bs-\Bsb$ oscillations, a limit at the 95\%~C.L. on the mass 
difference between the physical ${\mathrm B^0_s}$ states has been set:
\begin{eqnarray}
   \begin{array}{ll}
~~~~~~~~~~\dms > 4.0~\mbox{ps}^{-1}~\mbox{at the 95\% C.L.} & \\
   \end{array}
\end{eqnarray}
with a corresponding sensitivity equal to $3.2~\mbox{ps}^{-1}$.

Previous DELPHI results on $\Bs$ lifetime obtained with the $\Dsh$ sample~\cite{ref:bspapertau} are superseded by the 
analysis presented in this paper.

Combination of the DELPHI $\Delta m_{\Bs}$, $\tau_{\Bs}$ and $ \dgbs$ analyses gives:
\bes
   \ba{c}
      \dms~>~4.9~{\mathrm ps}^{-1}~\mbox{at the 95\% C.L.} \\
\mbox{with a sensitivity of}~~~ \dms = 8.7~ \mbox{ps}^{-1}
   \ea
 \ees
$$
\tau_{\Bs}  = 1.46\pm{0.11}~{\mathrm ps}
$$
$$
 \dgbs < 0.45~\mbox{at the 95\% C.L.}
$$

\subsection*{Acknowledgements}
\vskip 3 mm
 We are greatly indebted to our technical 
collaborators, to the members of the CERN-SL Division for the excellent 
performance of the LEP collider, and to the funding agencies for their
support in building and operating the DELPHI detector.\\
We acknowledge in particular the support of \\
Austrian Federal Ministry of Science and Traffics, GZ 616.364/2-III/2a/98, \\
FNRS--FWO, Belgium,  \\
FINEP, CNPq, CAPES, FUJB and FAPERJ, Brazil, \\
Czech Ministry of Industry and Trade, GA CR 202/96/0450 and GA AVCR A1010521,\\
Danish Natural Research Council, \\
Commission of the European Communities (DG XII), \\
Direction des Sciences de la Mati$\grave{\mbox{\rm e}}$re, CEA, France, \\
Bundesministerium f$\ddot{\mbox{\rm u}}$r Bildung, Wissenschaft, Forschung 
und Technologie, Germany,\\
General Secretariat for Research and Technology, Greece, \\
National Science Foundation (NWO) and Foundation for Research on Matter (FOM),
The Netherlands, \\
Norwegian Research Council,  \\
State Committee for Scientific Research, Poland, 2P03B06015, 2P03B1116 and
SPUB/P03/178/98, \\
JNICT--Junta Nacional de Investiga\c{c}\~{a}o Cient\'{\i}fica 
e Tecnol$\acute{\mbox{\rm o}}$gica, Portugal, \\
Vedecka grantova agentura MS SR, Slovakia, Nr. 95/5195/134, \\
Ministry of Science and Technology of the Republic of Slovenia, \\
CICYT, Spain, AEN96--1661 and AEN96-1681,  \\
The Swedish Natural Science Research Council,      \\
Particle Physics and Astronomy Research Council, UK, \\
Department of Energy, USA, DE--FG02--94ER40817. \\

\newpage

\newpage
\begin{figure}
\begin{center}
 \epsfig{file=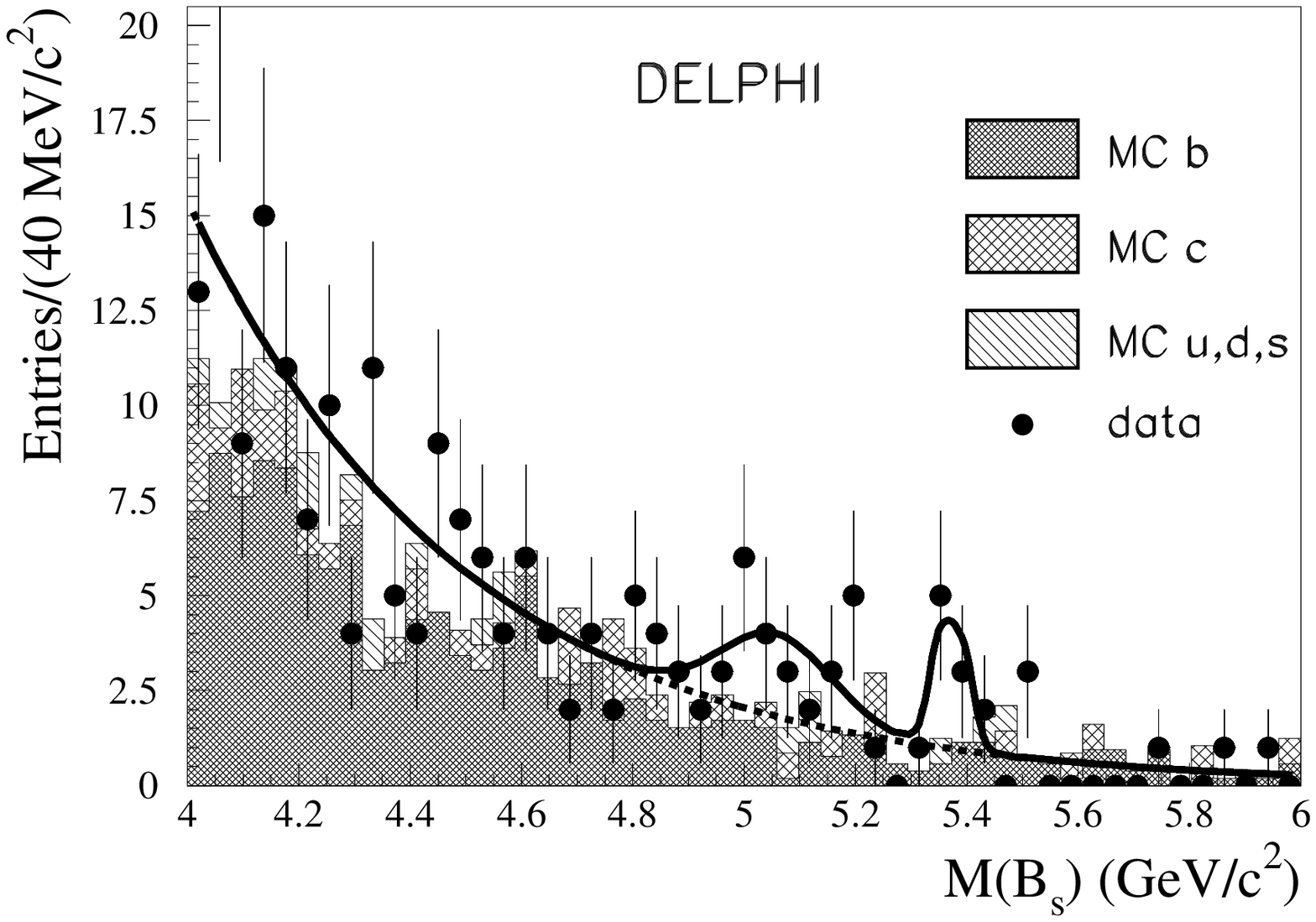,width=17cm,bbllx=0pt,bburx=595pt,bblly=210pt,bbury=580} 
\caption{ \it  $\Bs$ mass spectrum for the candidates selected in the twelve
   decay channels described in Section~\ref{sec:22}. The data are indicated 
   by the points with error bars and the result of the fit has been superimposed. 
   Details on this fit are given in the text. The histograms represent the 
   expected contribution from beauty events (after having removed the 
exclusively reconstructed $\Bs$ decay channels), 
from charm events  and from light quark events. 
}
\label{fig:bsexcl1}
\end{center}
\end{figure}

\newpage
\begin{figure}
\begin{center}
{\epsfig{file=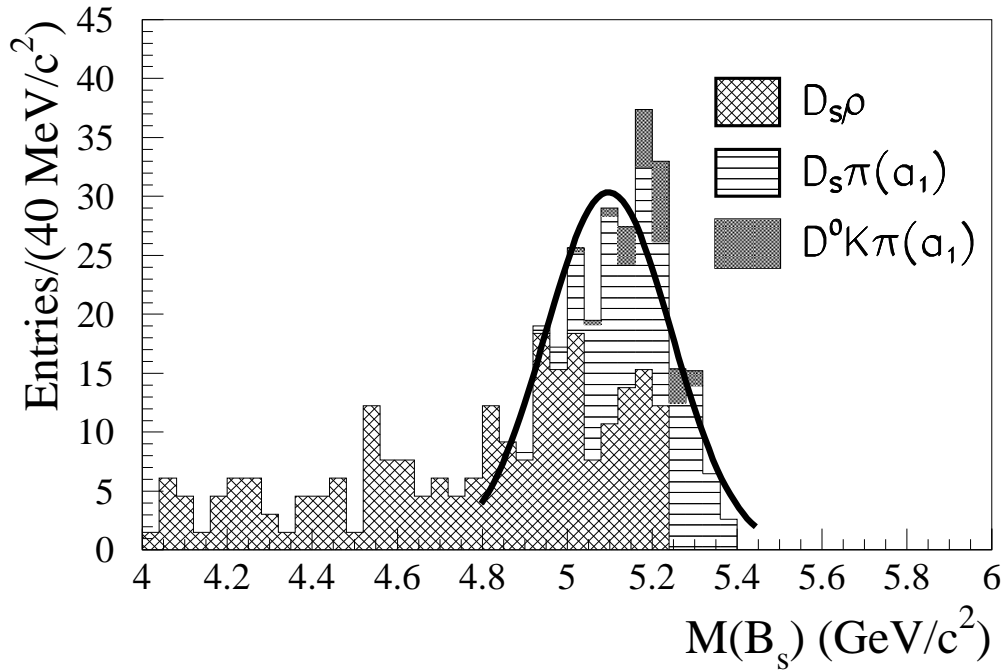,width=16cm}}
\caption{ \it The Monte Carlo composition of the satellite peak including the effects of
experimental resolution. The notations
in the plot are the following:  $\Ds \rho$ corresponds to ${\mathrm D_s^-} \rho^{+}$ and ${\mathrm D_s^{*-}} \rho^{+}$
 decay channels of the $\Bs$; $\Ds \pi({\mathrm a}_1)$ corresponds to ${\mathrm D_s^{*-}} \pi^+$ 
and ${\mathrm D_s^{*-}}  \rm{a}_{1}^+$;
 ${\mathrm D}^0 \mathrm{K} \pi({\mathrm a}_1)$ shows the contribution from $  \overline{\mathrm D}^{*}(2007)^{0} \mathrm{K}^{-} \pi^{+} $ and 
 $\overline{\mathrm D}^{*}(2007)^{0} \mathrm{K}^{-}\mathrm{a}_{1}^{+} $ decay channels. All contributions have been normalised 
according to the evaluation of the branching fractions discussed in Section~\ref{sec:21} and including reconstruction 
efficiencies evaluated using simulated events. }
\label{fig:bse_sat}
\end{center}
\end{figure}

\newpage
\begin{figure}
\begin{center}
{\epsfig{file=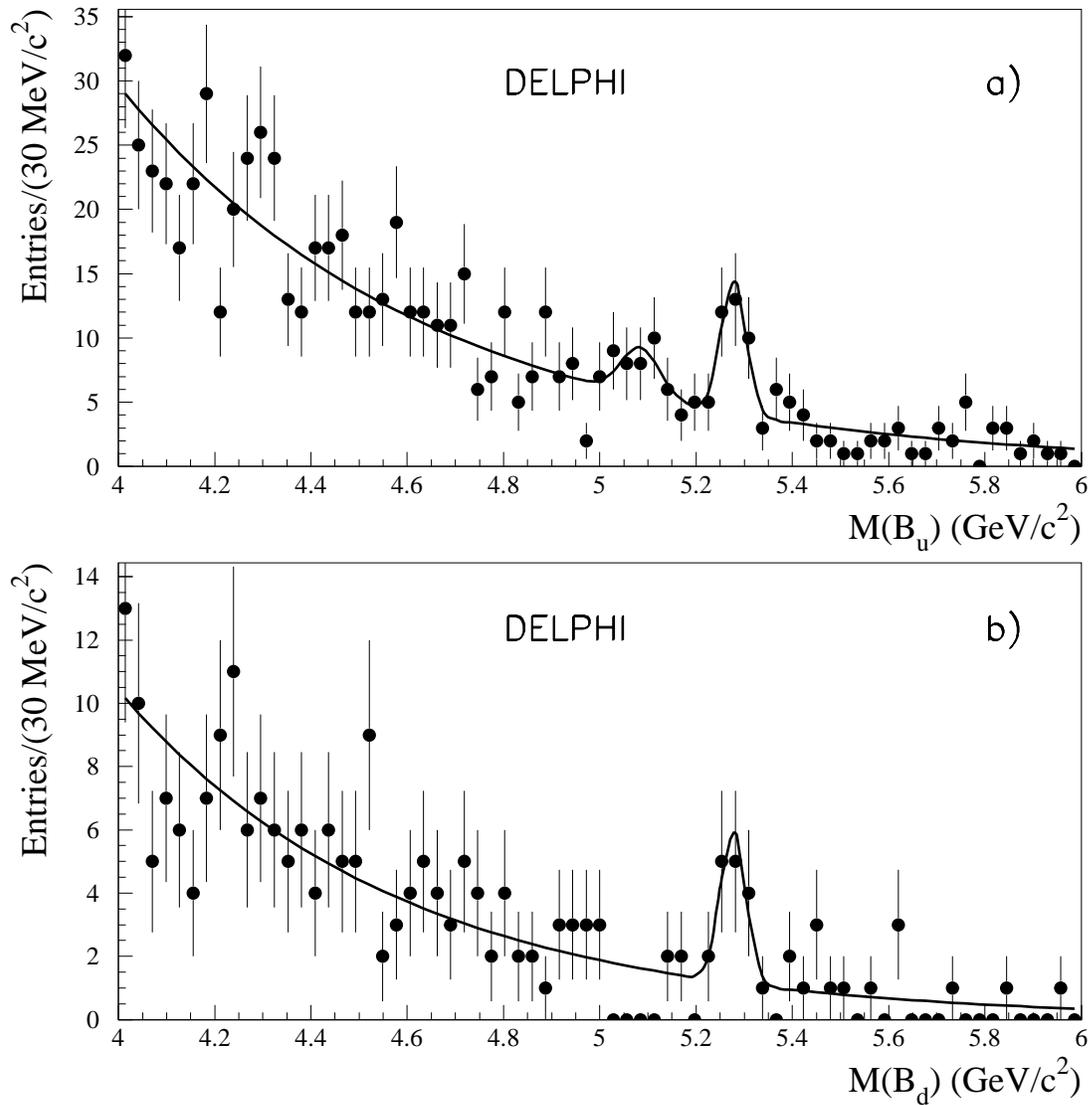,width=16cm}}
\caption{ \it Mass spectra for a) B$^+ \rightarrow  \overline{\mathrm D}^{0} \pi^{+}$,
 b) $\Bd \rightarrow  {\mathrm D}^{*}(2010)^{-} \pi^{+}$   and
  $\Bd \rightarrow {\mathrm D}^{*}(2010)^{-}\mathrm{a}_{1}^{+}$ decays.
 In the first plot a signal from B$^+ \rightarrow  {\mathrm D}^{*}(2007)^{0} \pi^{+}$,
 ${\mathrm B}^+ \rightarrow \overline{\mathrm D}^{0} \rho^{+}$ and 
 ${\mathrm{B^+ \rightarrow \overline{\mathrm D}^{*}(2007)^{0} \rho^{+}}}$ 
(satellite peak) is also visible.
Details on the fit are given in the text. 
}
\label{fig:bsexcl6}
\end{center}
\end{figure}

\newpage
\begin{figure}
\begin{center}
 \epsfig{file=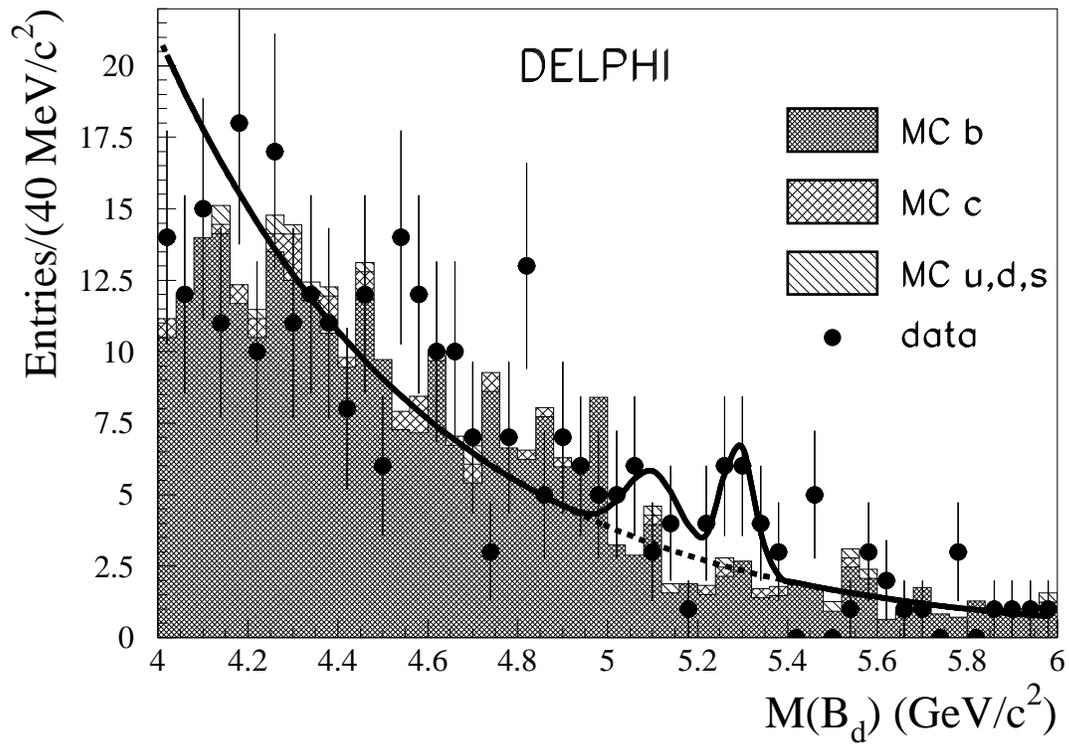,width=17cm,bbllx=0pt,bburx=595pt,bblly=210pt,bbury=580} 
\caption{ \it  $\Bd$ mass spectrum for the sum of $\Bd \rightarrow \overline{\mathrm D}^{0} \pi^{-} \pi^{+}$  
and  $\Bd \rightarrow \overline{\mathrm D}^0 \pi^{-}\mathrm{a}_{1}^{+} $ decays  selected in the four
decay channels described in Section~\ref{sec:24}. The data are indicated by the points with error bars and the result of the 
fit has been superimposed. The histograms represent the expected contribution from beauty events
 (after having removed  the
exclusively reconstructed $\Bd$ decay channels), 
  from charm events  and from light quark events.
 The widths of the signals have been fixed according to the values found in the simulation. Details on the fit 
are given in the text.} 
\label{fig:bsexcl_bd}
\end{center}
\end{figure}

\newpage
\begin{figure}
\begin{center}
{\epsfig{file=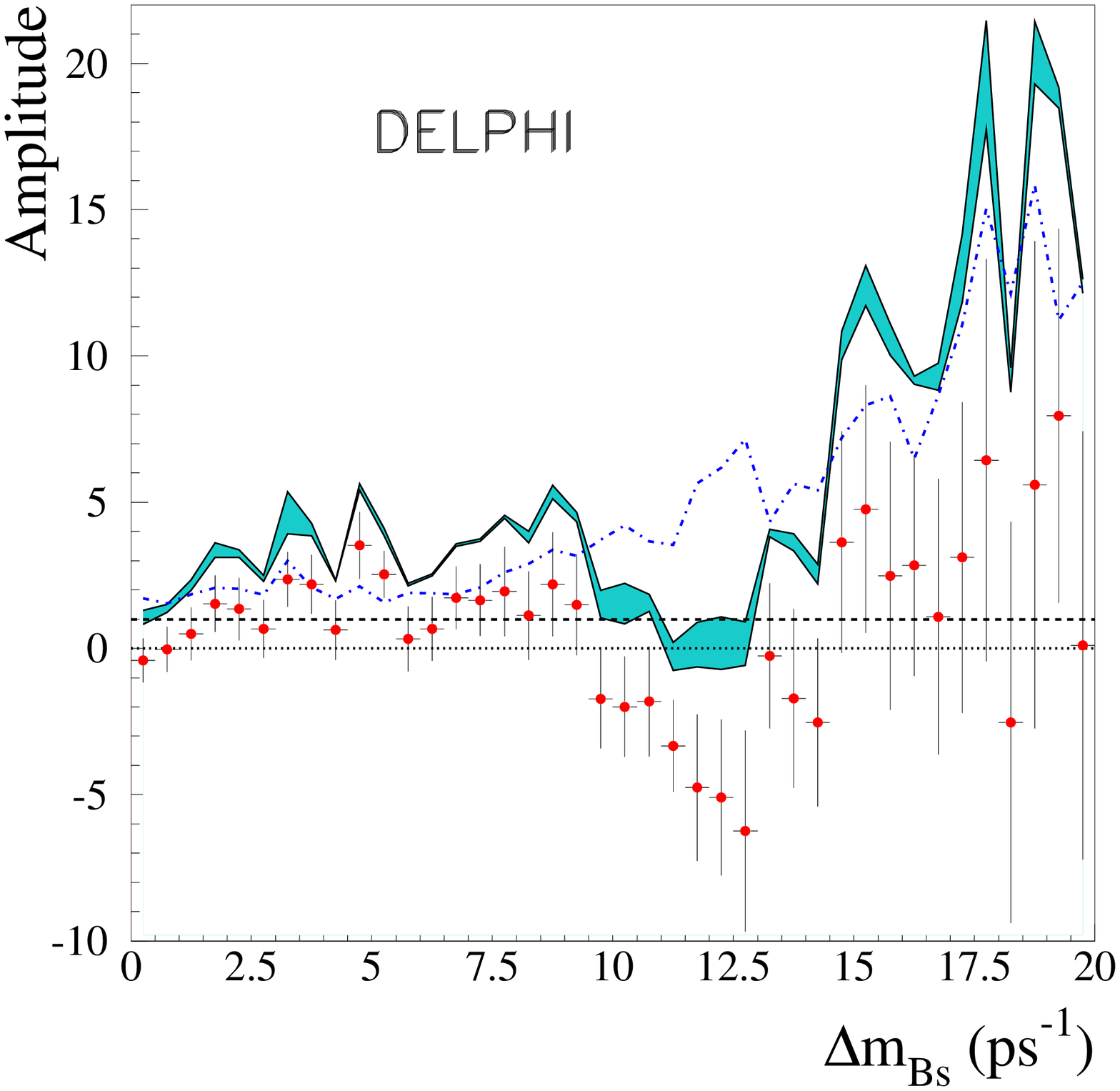,width=7.5cm}}
\caption {\it 
Exclusive $\Bs$ analysis: {variation of the oscillation  amplitude $A$ as a function of $\Delta m_{\Bs}$.
The lower continuous line corresponds to $ A + 1.645~\sigma(A)$ where $\sigma(A)$ includes statistical uncertainties only, 
while
the shaded area shows the contribution from systematics.
The dashed-dotted line corresponds to the sensitivity curve. The lines at A=0 and A=1 are also given.The points with error bars are real data. }}
\label{fig:bsexcl_dms}
\end{center}
\end{figure}

\begin{figure}
\begin{center}
\begin{tabular}{cc}
{\epsfig{file=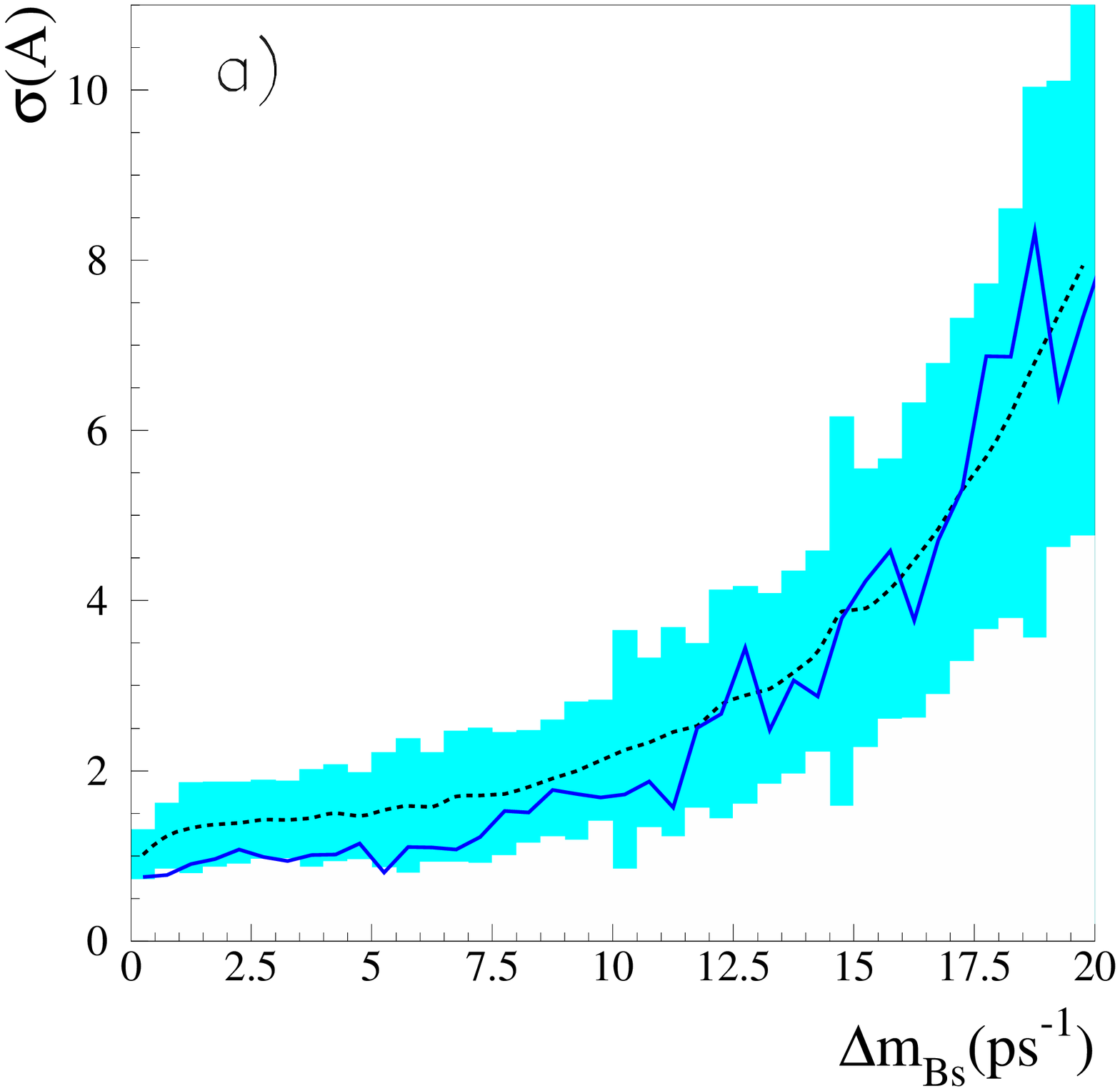,width=7.cm}}
{\epsfig{file=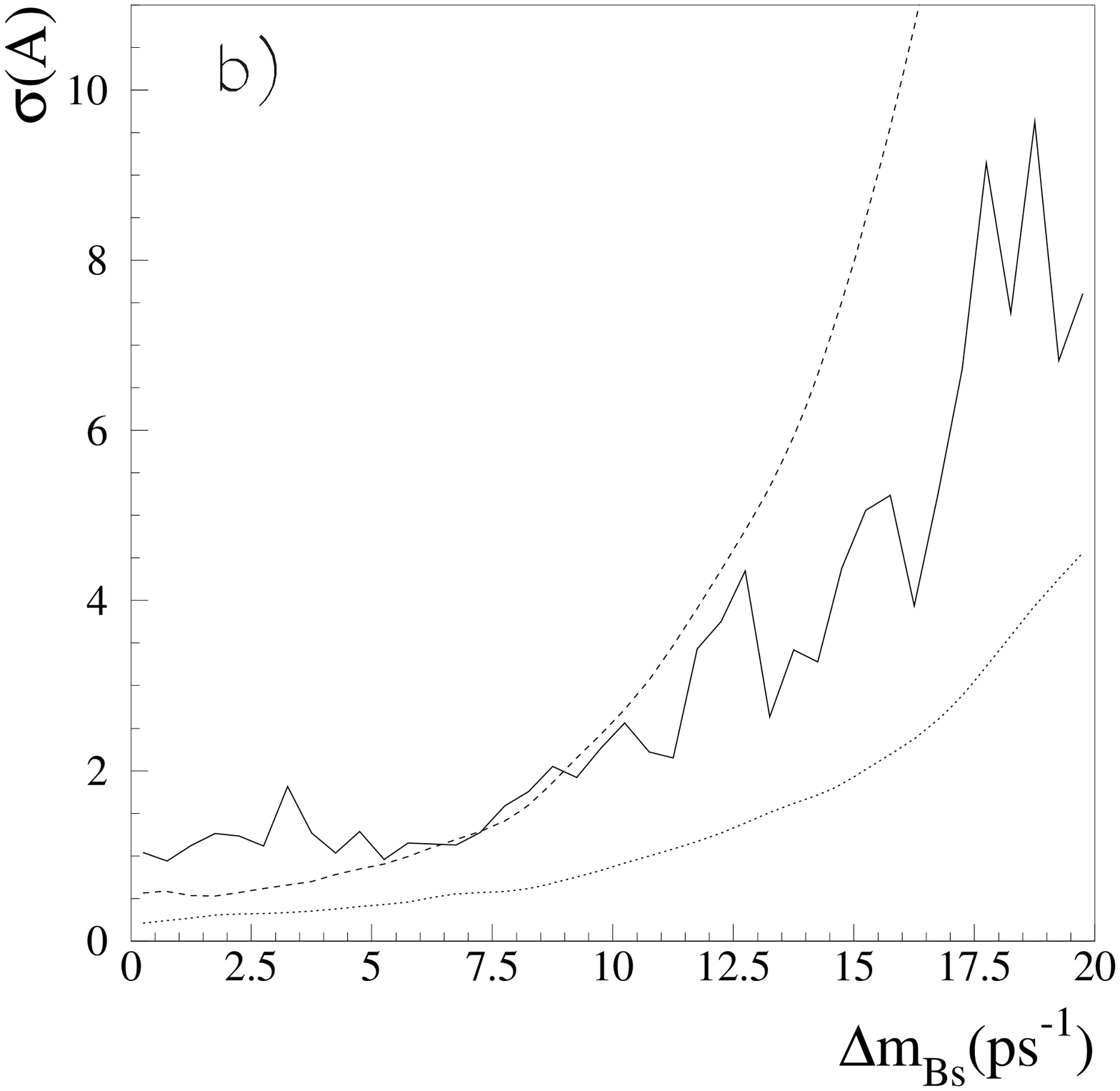,width=7.cm}}
\end{tabular}
\caption{ \it{ $\sigma(A)$ as a function of $\Delta m_{\Bs}$. 
a) Exclusive $\Bs$ analysis:
the full curve shows the result from the data. 
The dashed curve shows the average result from ${\mathrm 100}$ toy experiments 
with the same statistics as in data and the shaded area gives the $\pm2 
\sigma(A) $ region around this average. The  systematic effects are not included.
b) Comparison of the $\sigma(A)$  as function of $\dms$  for three analyses:
the full curve shows the result from the  exclusive $\Bs$ analysis,
the dashed and dotted  curves show the result from 
 $\rm{D}_{s}^{\pm}{\mathrm h}^{\mp}$ 
and $\rm{D}_{s}^{\pm}\ell^{\mp}$  analyses, respectively. }}
\label{fig:bsexcl_err}
\end{center}
\end{figure}

\newpage
\begin{figure}
\begin{center}
\epsfig{file=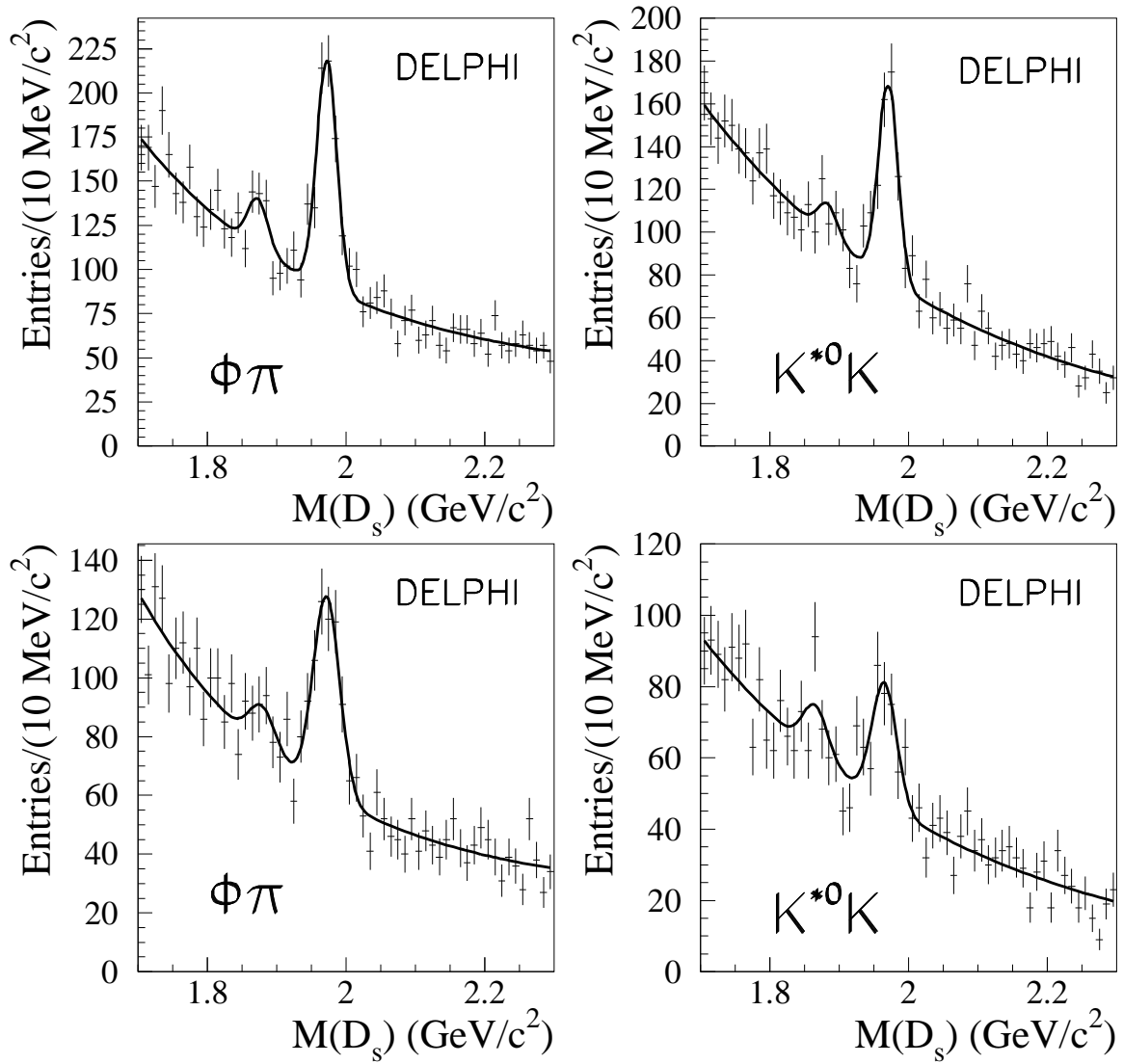,width=18cm,bbllx=0pt,bburx=595pt,bblly=110pt,bbury=580} 
\caption{ \it Invariant mass distributions for $\Ds$ candidates selected 
in $\phi \pi^-$ and ${\mathrm K}^{*0} \mathrm{K}^-$ decay channels. 
The upper and lower plots refer to data samples
registered in ${\mathrm 1994}$-${\mathrm 1995}$ and ${\mathrm 1992}$-${\mathrm 1993}$, respectively.
 The selected $\Ds$ candidates are accompanied by a hadron of opposite  charge
(or by several hadrons), measured in the same event hemisphere. 
The curves show the fits described in the text.}
\label{fig:dsh1}
\end{center}
\end{figure}

\newpage
\begin{figure}
\begin{center}
{\epsfig{file=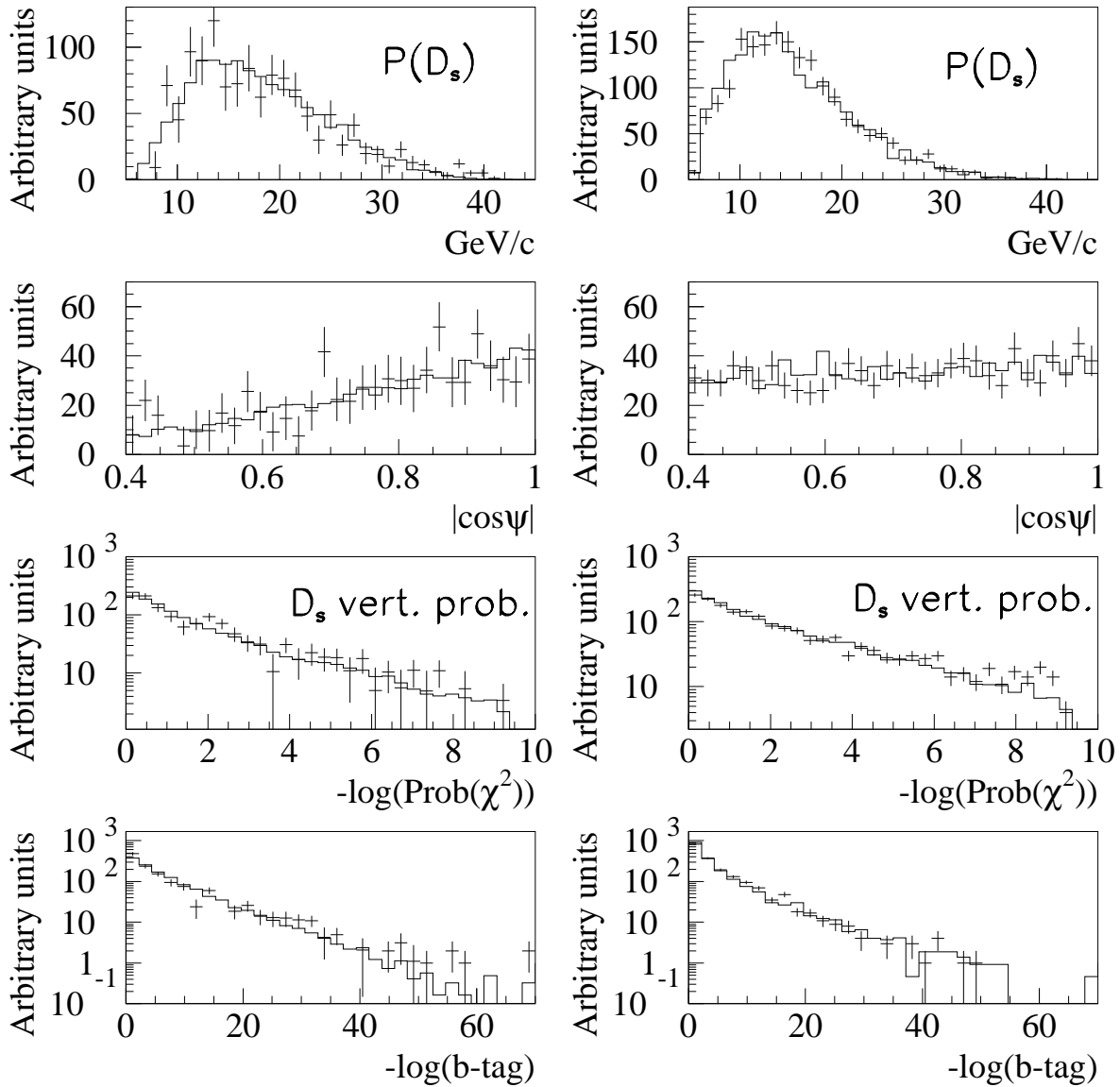,width=18.cm}}
\caption{ \it Distributions of the variables used to increase the 
${\mathrm{B_s}}$ purity. The distributions on the left
are for events selected in the signal region after having subtracted the corresponding distributions of background 
events, which have been obtained using events situated in the side-bands of the ${\mathrm{D_s}}$ signal.
 The corresponding 
distributions for background events are shown on the right. The points with error bars correspond to the data and the 
histograms are simulated events. For the  $|\cos\psi|$ distribution,  only the $\phi \pi^{+}$ decay mode is shown
because the cut on this variable was set at  ${\mathrm 0.6}$ 
for ${\mathrm K}^{*0} \mathrm{K}^-$  and at  ${\mathrm 0.4}$ for $\phi \pi^{+}$ channel.
In addition, the $\Ds$ mass was used as the fifth discriminating variable: the signal has a Gaussian 
and the background has an exponential distribution.  }
\label{fig:dsh_disc}
\end{center}
\end{figure}

\newpage
\begin{figure}
\begin{center}
\epsfig{file=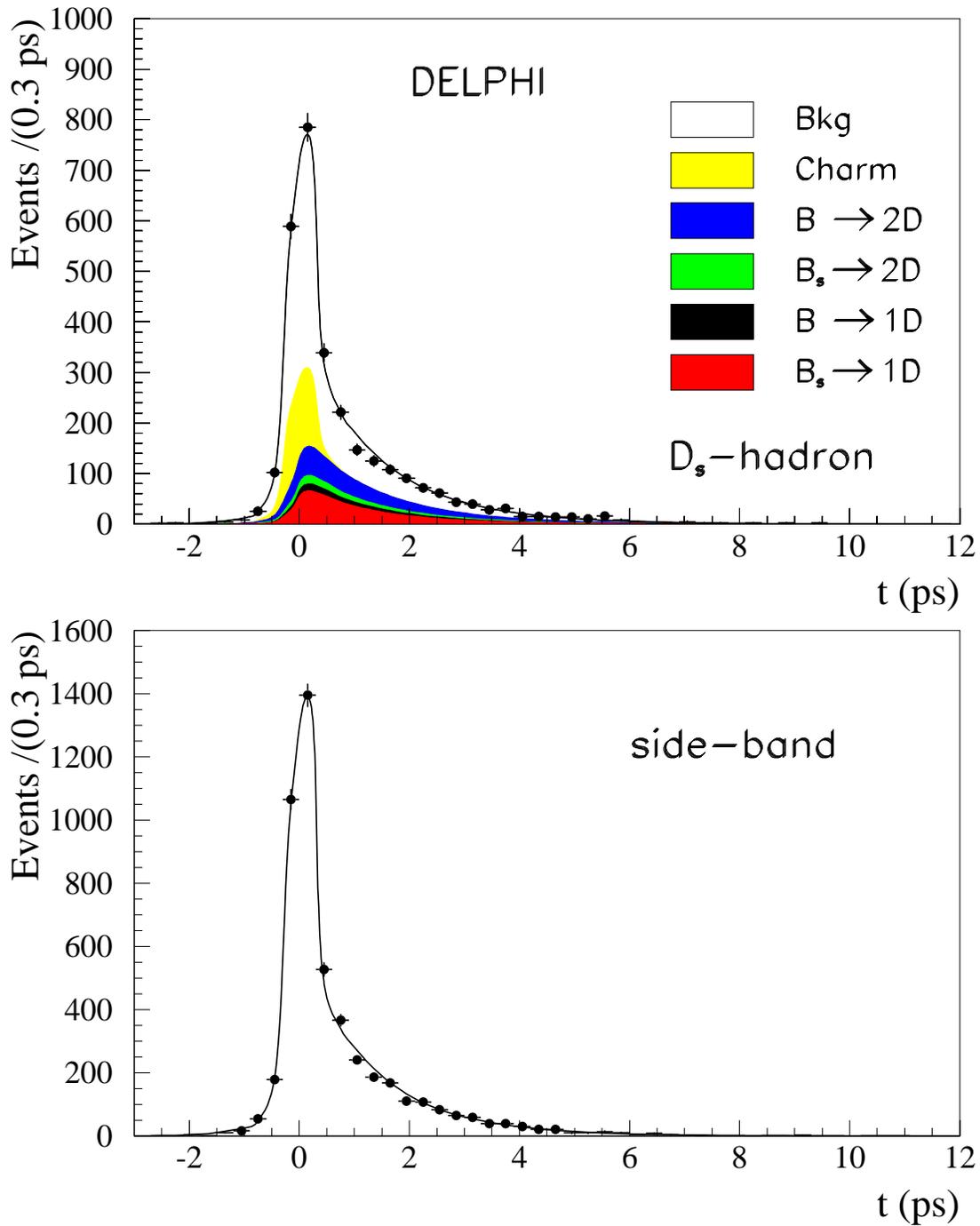,width=17cm,bbllx=0pt,bburx=595pt,bblly=0pt,bbury=580} 
\vspace{-2.5cm}
\caption{  \it  $\rm{D}_{s}^{\pm}{\mathrm h}^{\mp}$ analysis. 
Upper plot: Proper time distribution for events in the signal mass region.
          The points show the data and the shaded regions correspond to the different
          contributions to the selected events. The curve shows the result of the fit
           described in the text.
          Lower plot:  the same as the upper plot but for events 
          situated in the $\Ds$ mass side-band.  } 
\label{fig:dsh_tau}
\end{center}
\end{figure}

\newpage
\begin{figure}[ph]
\begin{center}
\epsfig{figure=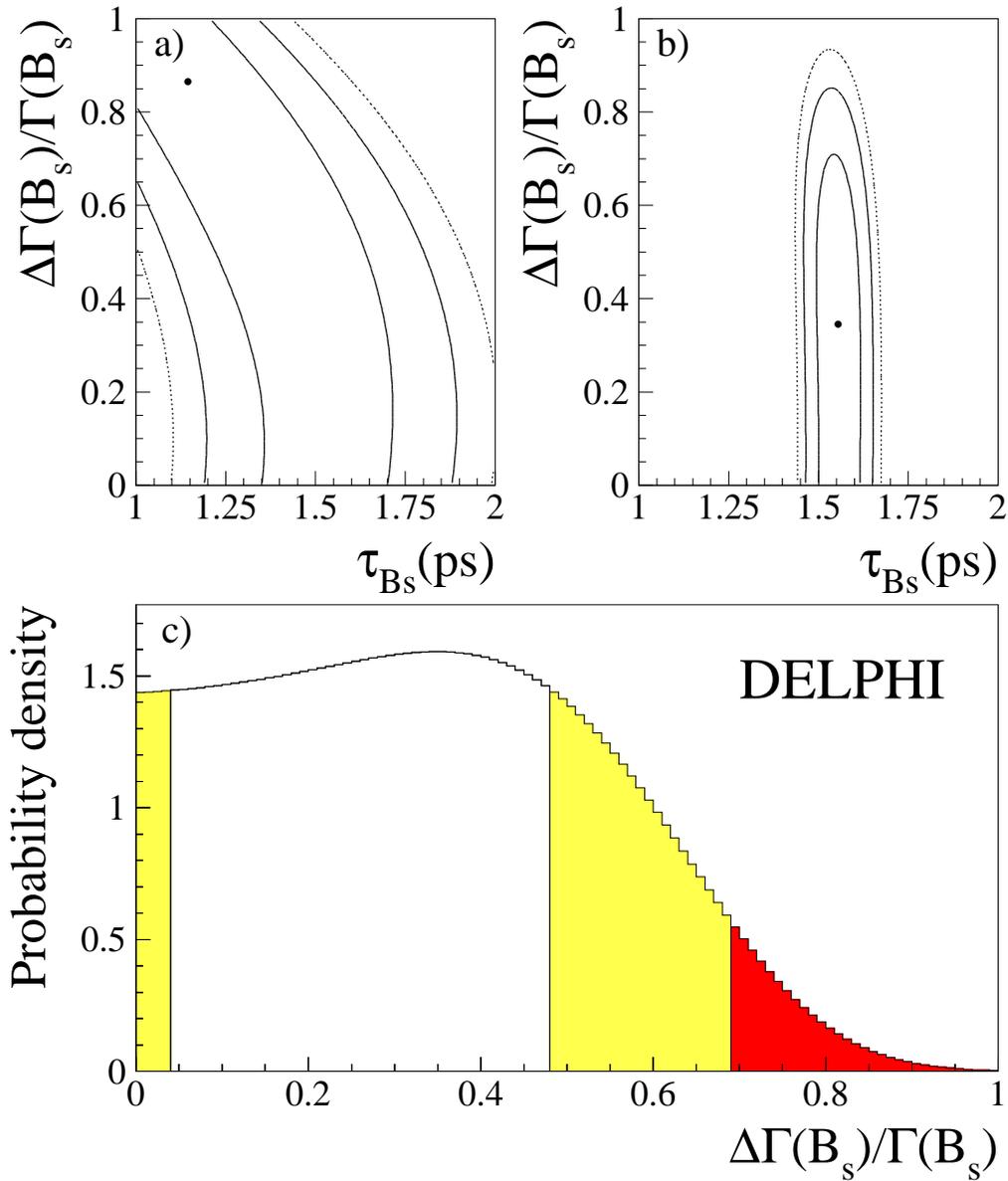,height=16cm}
\caption{ \it  $\rm{D}_{s}^{\pm}{\mathrm h}^{\mp}$ analysis: 
           68\%, 95\%, 99\%~C.L.  contours of the negative log-likelihood 
             in the plane $\dgbs - \tbs$ a) without and b) with $\tau_{\Bs}=\tau_{\Bd}$
constraint.
 The point indicates the minimum.
          c) Probability density distribution for 
          $\Delta\Gamma_{\mathrm{B_s}}/\Gamma_{\mathrm{B_s}}$;
             the two lightly shaded regions at the ${\mathrm 68}$\%~C.L. 
          and dark one  at the  ${\mathrm 95}$\%~C.L. are also shown.}
\label{fig:dsh_dg}
\end{center}
\end{figure}

\newpage
\begin{figure}
\begin{center}
\begin{tabular}{cc}
{\epsfig{file=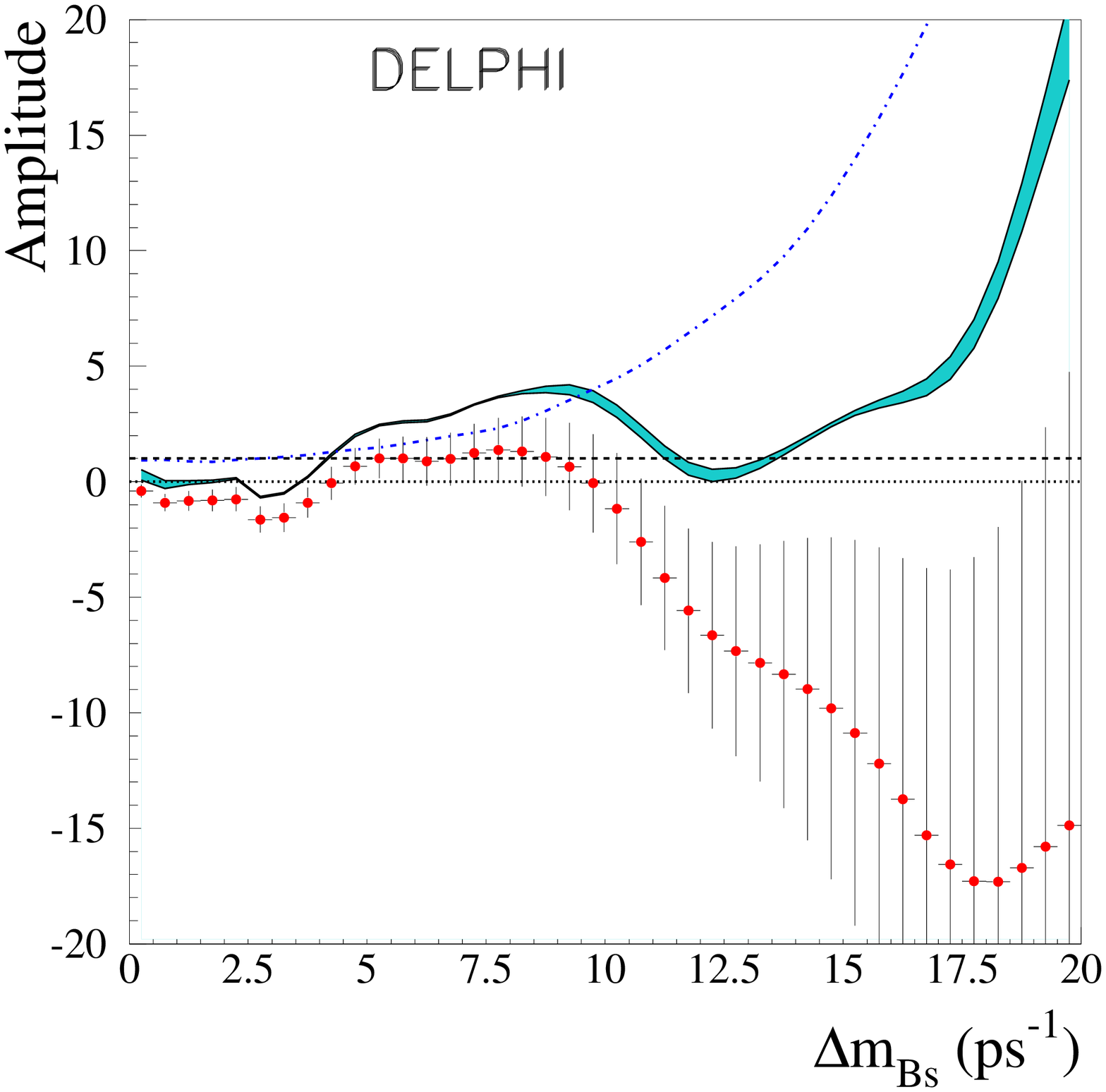,width=7cm}}
{\epsfig{file=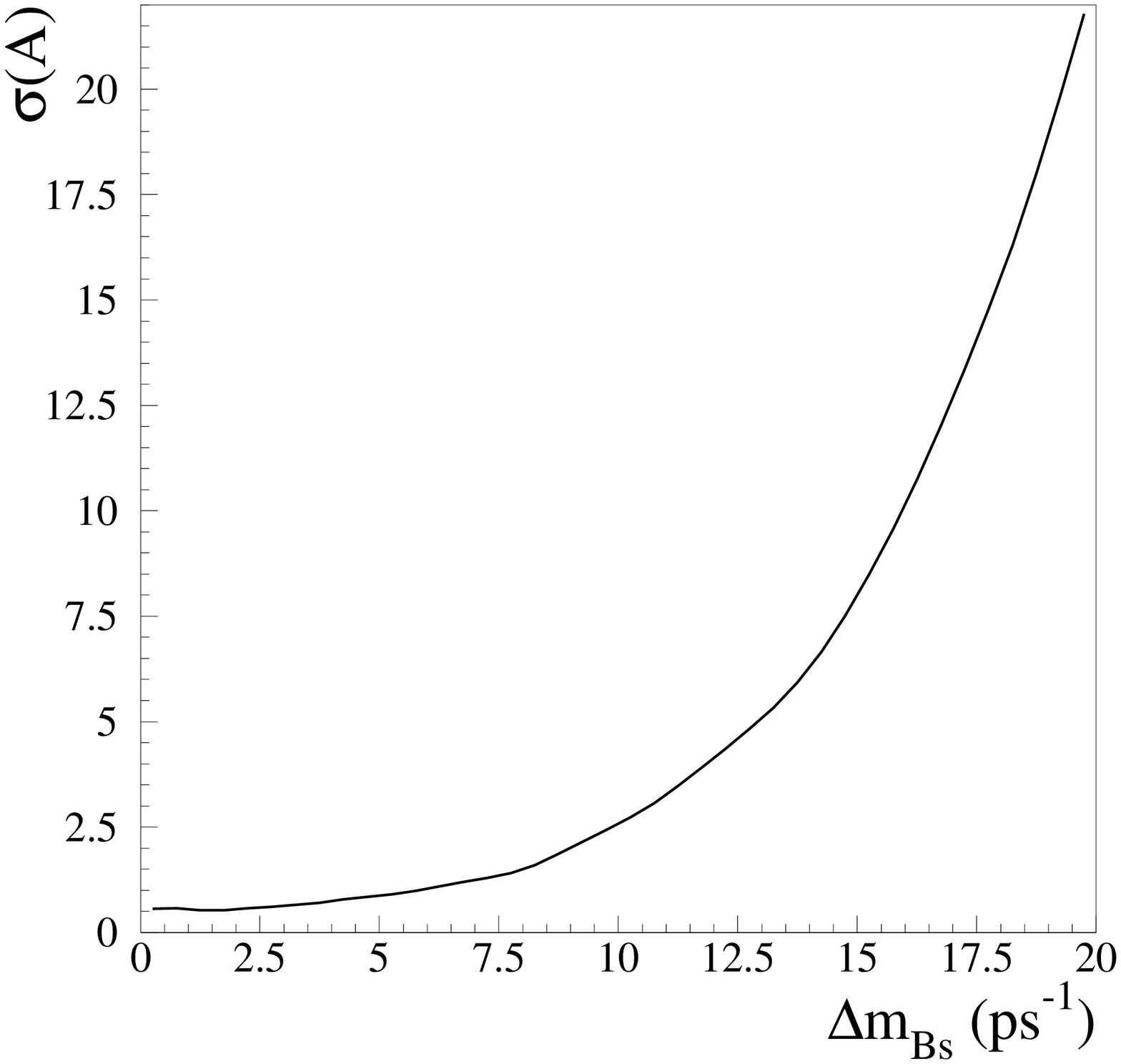,width=7cm}}
\end{tabular}
\caption {\it $\rm{D}_{s}^{\pm}{\mathrm h}^{\mp}$ analysis. 
{Left plot: variation of the oscillation  amplitude $A$ as a function of $\Delta m_{\Bs}$.
The lower continuous line corresponds to $ A + 1.645~\sigma(A)$ where
 $\sigma(A)$ includes statistical uncertainties only, while the shaded area shows the contribution from systematics. 
The dashed-dotted line corresponds to the sensitivity curve. The lines at A=0 and A=1 are also given.
The points with error bars are real data.
Right plot: variation  of the error on the amplitude as a function of $\Delta m_{\Bs}$, including systematic uncertainties.}}
\label{fig:dsh_dms}

\begin{tabular}{cc}
{\epsfig{file=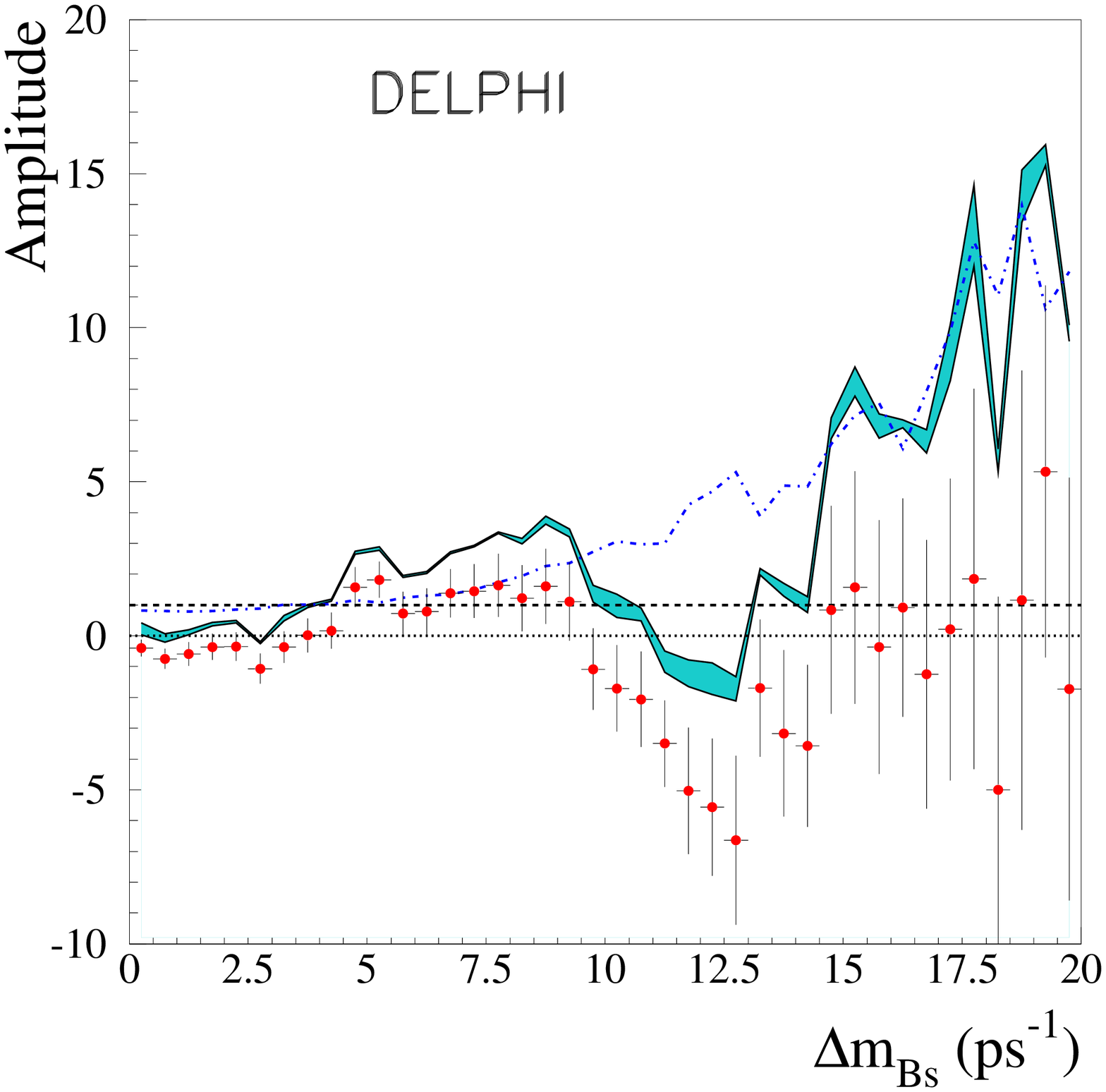,width=7cm}}
{\epsfig{file=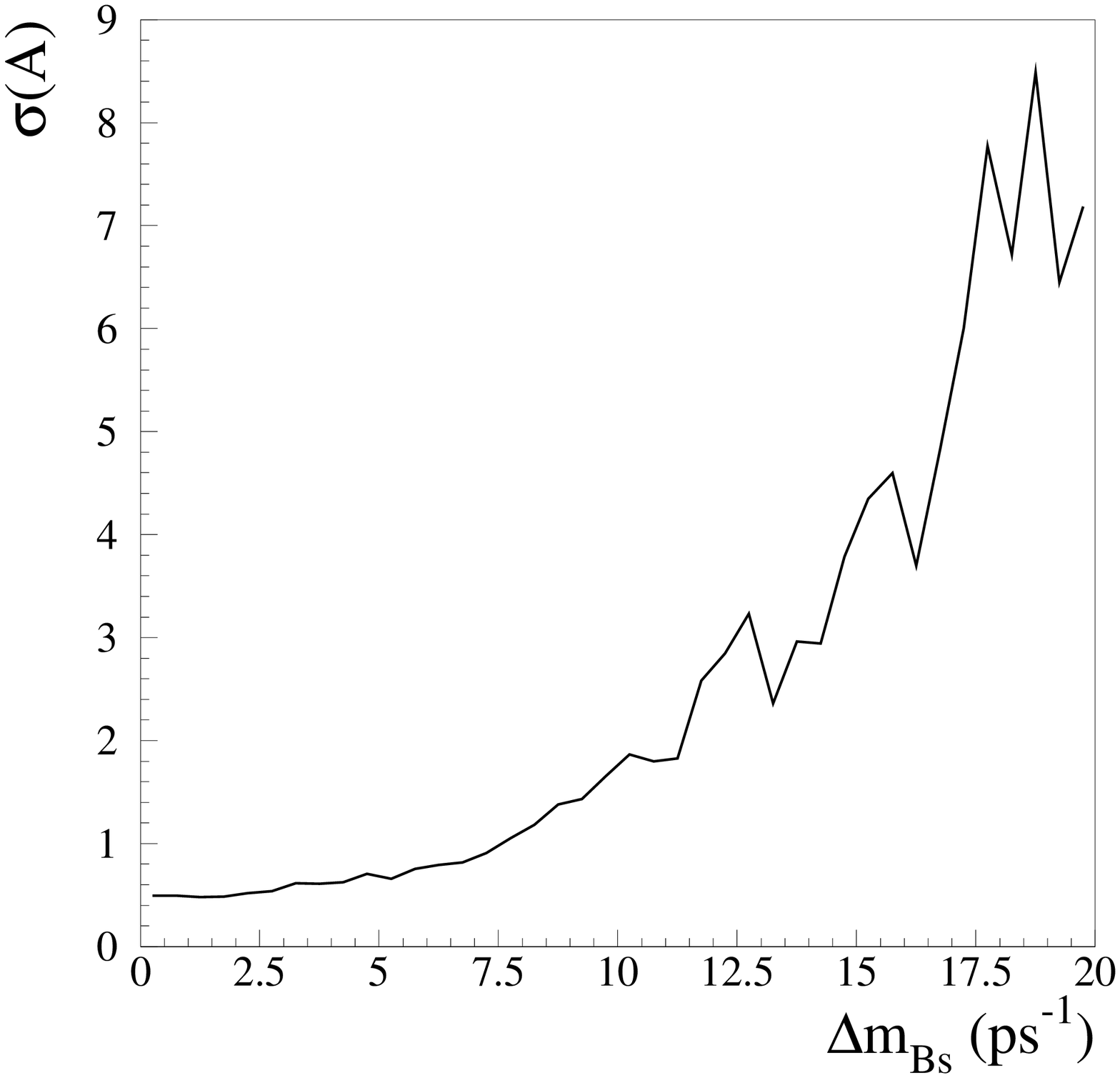,width=7cm}}
\end{tabular}
\caption {\it 
Combination of the $\rm{D}_{s}^{\pm}{\mathrm h}^{\mp}$ and exclusive $\Bs$ analyses. {Left plot: 
variation of the oscillation  amplitude $A$ as a 
function of $\Delta m_{\Bs}$. The lower continuous line corresponds to $ A + 1.645~\sigma(A)$ where
 $\sigma(A)$ includes statistical uncertainties only, while the shaded area shows the contribution from 
systematics. The dashed-dotted line corresponds to the sensitivity curve. The lines at A=0 and A=1 are also given.The points with error bars are real data.
 Right plot: variation  of the error on the amplitude as a function of $\Delta m_{\Bs}$ including systematic uncertainties. It should be noted
that the  $\rm{D}_{s}^{\pm}{\mathrm h}^{\mp}$ analysis dominates at low values  
of $\Delta m_{\Bs}$ and the  exclusive $\Bs$ analysis dominates at large values of $\Delta m_{\Bs}$.  }}
\label{fig:bsdsh_dms}
\end{center}
\end{figure}

\newpage
\begin{figure}[ph]
\begin{center}
\epsfig{figure=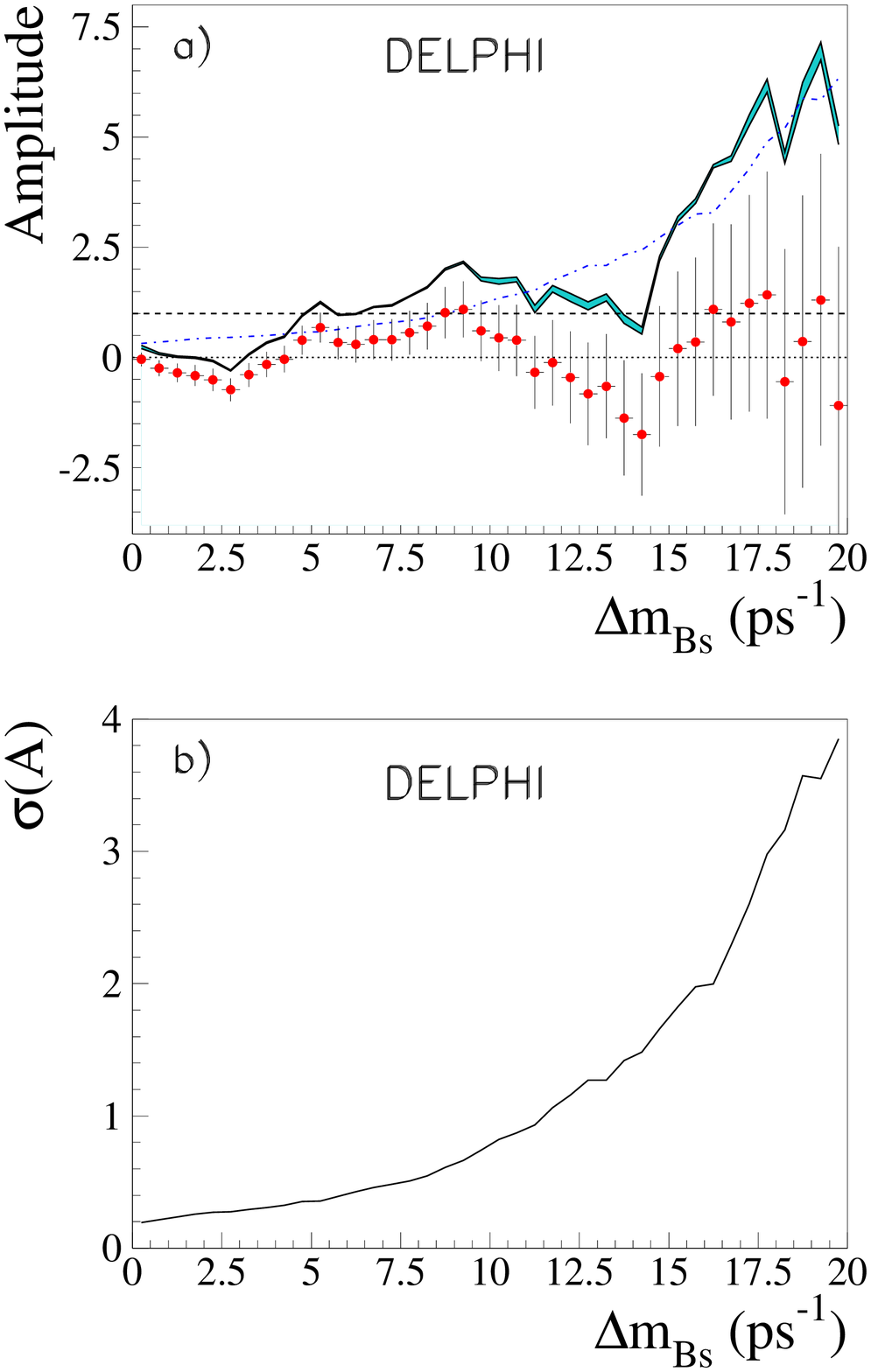,width=12cm}
\caption {\it 
Combined DELPHI analysis:
a) variation of the oscillation  amplitude $A$ as a function of $\Delta m_{\Bs}$.
The lower continuous line corresponds to $ A + 1.645~\sigma(A)$ where $\sigma(A)$ includes 
statistical uncertainties only, while the shaded area shows the contribution from systematics. 
The dashed-dotted line corresponds 
to the sensitivity curve. The lines at A=0 and A=1 are also given.  
The points with error bars are real data.
b) variation  of the error on the amplitude as a function of $\Delta m_{\Bs}$.}
\label{fig:delphi_dms}
\end{center}
\end{figure}


\begin{thebibliography}{ref99}

\bibitem{ref:bbd} M. Beneke, G. Buchalla and I. Dunietz,  Phys. Rev. {\bf D54} (1996) 4419.

\bibitem{ref:wolf}   L. Wolfenstein, Phys. Rev. Lett. {\bf 51} (1983) 1945.

\bibitem{ref:dmd_teo} G. Altarelli and P.J. Franzini,  Z. Phys. {\bf C37} (1988) 271;\\
                      P.J. Franzini, Phys Rep. {\bf 173} (1989) 1.

\bibitem{ref:bubu}   A. J. Buras, Acta Phys. Polon. {\bf B26} (1995) 755.

\bibitem{ref:bbag1}  A. Abada et al., Nucl. Phys. {\bf B376} (1992) 172.

\bibitem{ref:bello}  P.  Paganini, F. Parodi, P. Roudeau and A. Stocchi, 
                     Physica Scripta {\bf 58}  (1998)  556; \\
             F. Parodi, P. Roudeau and A. Stocchi, Nuovo Cimento {\bf 112A} (1999) N.7.

\bibitem{ref:osciwg}
G. Blaylock. Dec 1999. 
Presented at 19th International Symposium on Lepton and Photon Interactions at High-Energies (LP 99), Stanford, California,
9-14 Aug 1999. 
e-Print Archive: hep-ex/9912038 

\bibitem{ALDMS_LQ}
 D. Buskulic et al.,  ALEPH Coll., Eur. Phys. J. {\bf C7} (1999) 553. 

\bibitem{OPALDMS}
 G. Abbiendi et al.,   OPAL Coll., 
Eur. Phys. J. {\bf C11} (1999) 587. 

\bibitem{ref:dsl} P. Abreu et al., DELPHI Coll., 
``Measurement of the ${\mathrm B^0_s}$ Lifetime and Study of ${\mathrm B^0_s}-\overline{\mathrm B^0_s}$ Oscillations  
using D$_{\mathrm s}-\ell$ Events", 
CERN-EP/2000-043, submitted to Eur. Phys. J. C.

\bibitem{ADMSDSL}
  D. Buskulic et al.,  ALEPH Coll., Phys. Lett. {\bf B377} (1996) 205.            

\bibitem{ref:dsh_al} D. Buskulic et al., ALEPH Coll., Eur. Phys. J. {\bf C4} (1998) 367.

\bibitem{ref:amplitude} H.G. Moser and A. Roussarie, Nucl. Instr. Meth. {\bf A384} (1997) 491.

\bibitem{ref:bigi}
I. Bigi, Phys. Rep. {\bf 289} (1997) 1; \\
M. Neubert, CERN-TH/98-2 (1998), 
invited talk at the International Europhysics Conference on High Energy Physics (HEP 97),
Jerusalem, Israel (19-26 Aug 1997),
ed. D. Lellouch, G. Mikenberg, E. Rabinovici (Springer, 1998).

\bibitem{ref:bbd_new} M. Beneke et al., Phys. Lett. {\bf B459} (1999) 631.            

\bibitem{ref:perfo} P. Abreu et al., DELPHI Coll., Nucl. Instr. Meth. {\bf A378} (1996) 57; \\
E.G. Anassontzis et al., Nucl. Instr. Meth. {\bf A323}  (1992) 351.

\bibitem{Delphi:VD} V. Chabaud et al., Nucl. Instr. Meth. {\bf A368} (1996) 314.

\bibitem{delphi:Zshape} P. Abreu et al., DELPHI Coll., Nucl. Phys. {\bf B418} (1994) 403.

\bibitem{ref:luclus} T. Sj\"ostrand, Comp. Phys. Commun. {\bf 82} (1994) 74.

\bibitem{ref:idid}  M. Battaglia and P.M. Kluit, Nucl. Instr. Meth. {\bf A443} (1999) 252. 


\bibitem{btag} G. Borisov and C. Mariotti, Nucl. Instr. Meth. {\bf A372} (1996) 181.

\bibitem{ref:tuning} P. Abreu et al, DELPHI Coll., Z. Phys. {\bf C73} (1996) 11.

\bibitem{ref:isgw} N. Isgur, et al, Phys. Rev. {\bf D39} (1989) 799.

\bibitem{ref:b-en-2-corps-factorisation}  A. Deandrea et al., Phys. Lett. {\bf B318} (1993) 549.

\bibitem{ref:book} 
Review of Particle Physics, Eur. Phys. J. {\bf C3} (1998) 1.

\bibitem{ref:Neubert1997uc} M. Neubert and B. Stech, in
A.J. Buras and M. Lindner (ed.): Heavy Flavours 2nd edition, p. 294-344.
World Scientific, Singapore (1998).

\bibitem{ref:bspapertau} P. Abreu et al., DELPHI Coll.,  Z. Phys. {\bf C71} (1996) 11.

\bibitem{ref:oldbs}   W. Adam et al., DELPHI Coll., Phys. Lett. {\bf B414} (1997) 382.



\bibitem{ref:mf} P. Abreu et al., DELPHI Coll., Z. Phys.  {\bf C68} (1995) 353.


\bibitem{lepds}   
G. Alexander et al.,  OPAL Coll., Z. Phys. {\bf C72} (1996) 1; \\
D. Buskulic et al., ALEPH Coll., Phys. Lett. {\bf B388} (1996) 648; \\
P. Abreu et al.,  DELPHI Coll., Eur. Phys. J. {\bf C12} (2000) 225.
\bibitem{cleodl}
H. Albrecht et al., ARGUS Coll., Z. Phys. {\bf C54} (1992) 1;\\
D.Gibaut et al., CLEO Coll., Phys. Rev. {\bf D53} (1996) 4734. \\
\bibitem{ref:book1} X.Fu et al.,CLEO Coll., Preprint CLEO-Conf 95-11. \\
\bibitem{ref:book2}  P.Abreu et al.,DELPHI Coll., Phys. Lett. {\bf B426} (1998) 193. \\

\bibitem{ref:aleksan} R. Aleksan, A. Le Yaouanc, 
                      L. Oliver, O. Pene and J.C. Raynal, 
                      Phys.Lett. {\bf B316} (1993) 567.


\end{thebibliography}
\end{document}